\documentclass[
reprint,
nofootinbib,
 amsmath,amssymb,
 aps,
]{revtex4-2}

\usepackage[T1]{fontenc}
\usepackage{graphicx}
\usepackage{dcolumn}
\usepackage{bm}
\usepackage[colorlinks=true,linktocpage=true,linkcolor=blue,citecolor=blue]{hyperref}
\usepackage[dvipsnames]{xcolor}
\usepackage[normalem]{ulem}

\begin{document}

\title{Bayesian Constraints on Pre-Equilibrium Jet Quenching\\ and Predictions for Oxygen Collisions}

\author{Daniel Pablos}
\email{pablosdaniel@uniovi.es}
\affiliation{IGFAE, Universidade de Santiago de Compostela, E-15782 Galicia-Spain}
\affiliation{Departamento de F\'isica, Universidad de Oviedo, Avda. Federico Garc\'ia Lorca 18, 33007 Oviedo, Spain}
\affiliation{Instituto Universitario de Ciencias y Tecnolog\'ias Espaciales de Asturias (ICTEA), Calle de la Independencia 13, 33004 Oviedo, Spain}

\author{Adam Takacs}
\email{takacs@thphys.uni-heidelberg.de}
\affiliation{Institute for Theoretical Physics, Heidelberg University, Philosophenweg 16, 69120 Heidelberg, Germany}

\begin{abstract}
The contrast between the as-yet unmeasurable energy-loss effects in proton-nucleus collisions and the striking magnitude of the so-called high-momentum flow coefficients challenges our understanding of jet quenching mechanisms in large nucleus-nucleus collisions when applied to smaller systems. Intermediate-sized, light ion collisions will offer key insight into the system-size dependence of the interplay between jet energy loss and jet flow effects. To make quantitative predictions, we extend a semi-analytic jet quenching framework by coupling it to state-of-the-art event-by-event hydrodynamics and, for the first time, incorporate pre-equilibrium energy loss via the hydrodynamic attractor. A Bayesian analysis shows that an early-time onset of energy loss is compatible with RHIC and LHC measurements of jet suppression and jet elliptic flow in large systems, as well as hadron suppression, with the exception of hadron elliptic flow. Using these constraints, we predict both hadron and jet quenching observables in oxygen-oxygen collisions, finding sizable energy loss that exceeds the no-quenching baseline. 
\end{abstract}

\maketitle

\section{Introduction}
\label{sec:introduction}

Understanding the emergence of collective behavior in small collision systems, such as $pp$ and $pA$ collisions, remains one of the central challenges for both the high-energy physics and heavy-ion communities since their discovery at the LHC more than a decade ago~\cite{CMS:2010ifv,ATLAS:2012cix,ALICE:2012eyl,CMS:2012qk} (for recent reviews, see~\cite{Nagle:2018nvi,Noronha:2024dtq,Grosse-Oetringhaus:2024bwr}). While the dynamical origin of this collectivity is still under investigation, hydrodynamic models, originally thought to be valid only for large systems such as $AA$ collisions, have proven quantitatively successful in describing several low-momentum observables in small systems as well~\cite{Weller:2017tsr,PHENIX:2018lia,ALICE:2019zfl,Nijs:2020roc} (a non-exhaustive list of alternative explanations includes~\cite{Bierlich:2018xfw,Kurkela:2021ctp,Bierlich:2024lmb,Torres:2024rgw,Soudi:2024slz,JETSCAPE:2024dgu}). 

This hydrodynamic picture requires the presence of a deconfined quark-gluon plasma (QGP), implying final-state interactions and consequently jet quenching~\cite{Armesto:2015ioy,Connors:2017ptx,Cunqueiro:2021wls}. In $AA$ collisions, jet modifications are firmly established, and their dependence on the medium size is consistent with expectations from parton energy-loss calculations performed using perturbative QCD or holographic techniques~\cite{Casalderrey-Solana:2007knd,dEnterria:2009xfs,Mehtar-Tani:2013pia,Cao:2024pxc}. However, in $pA$ collisions, jet quenching remains elusive, possibly due to the short path lengths involved~\cite{ATLAS:2014cpa,ALICE:2016faw,ALICE:2021wct,ALICE:2023ama}. Despite the absence of measurable energy loss, significant azimuthal anisotropies have been observed in high-$p_T$ particles in $pA$ systems~\cite{ATLAS:2019vcm,ATLAS:2023bmp,CMS:2025kzg}. In $AA$ collisions, this anisotropy is understood in terms of differences in energy loss due to the different path lengths traversed as a function of the jet orientation with respect to the reaction plane of the collision~\cite{Gyulassy:2000gk,Wang:2000fq}, adding further puzzling elements to the picture we attempt to draw for small systems.

Recent LHC runs on OO and NeNe collisions have provided a valuable bridge between small ($pp$ and $pA$) and large (such as PbPb and AuAu) collision systems, thereby reducing uncertainties related to short path lengths and centrality determination~\cite{Citron:2018lsq,Brewer:2021kiv,Loizides:2017sqq,Park:2025mbt}. In these smaller systems, fluctuations and out-of-equilibrium dynamics are believed to play an increasingly critical role in achieving quantitative precision~\cite{Heinz:2013th,Gale:2013da,Mantysaari:2017cni,Romatschke:2017ejr}. 

In this work, we extend our semi-analytic jet quenching framework~\cite{Mehtar-Tani:2021fud,Takacs:2021bpv,Mehtar-Tani:2024jtd} by incorporating these elements to enable reliable predictions for the upcoming experimental results on the potential existence of jet quenching effects in OO collisions. We account for medium fluctuations by coupling the framework to state-of-the-art event-by-event hydrodynamic simulations of the expanding fireball within which jets are produced and subsequently traverse. Before hydrodynamics is believed to be applicable, we incorporate quenching effects during the pre-equilibrium phase via the hydrodynamic attractor extracted from QCD kinetic theory, representing the first time this is done in a realistic model for jet quenching. 

We benchmark our model against a broad set of $AA$ data, including both LHC and RHIC energies and collision systems, spanning wide ranges in centrality, jet cone size and transverse momentum. Using Bayesian parameter estimation, we achieve a simultaneous and consistent description of jet suppression $R^{\rm jet}_{AA}(p_T)$ and elliptic anisotropy $v_2^{\rm jet}(p_T)$ across all available data. We show that the combination of $R^{\rm jet}_{AA}$ and $v^{\rm jet}_2$ provides constraints on energy-loss effects in the pre-equilibrium phase. The resulting posterior distributions can then be used to predict jet and hadron observables in other collision systems, including OO as we do in the present work. 

\section{Framework}
\label{sec:Framework}
Our jet quenching framework builds on previous results from \cite{Mehtar-Tani:2021fud,Takacs:2021bpv,Mehtar-Tani:2024jtd}, which we briefly summarize here. We introduce two key new components: (i) the coupling to event-by-event fluctuating hydrodynamics, and (ii) energy loss in the pre-equilibrium phase using the hydrodynamic attractor.

Our starting point is the vacuum jet cross section, $\sigma^{pp}_i(p_T, R)$, computed using \texttt{MadGraph+Pythia}~\cite{Alwall:2014hca,Bierlich:2022pfr,Skands:2014pea} at NLO+LL accuracy. We reconstructed jets and assigned their flavor ($i=q/g$) with the \texttt{Flavor-kt}~\cite{Banfi:2006hf}. Nuclear PDF (nPDF) effects are incorporated using \texttt{EPPS21}~\cite{Eskola:2021nhw} within the \texttt{LHAPDF}~\cite{Buckley:2014ana} framework, yielding the modified cross section $\tilde\sigma^{pp}(p_T, R)$. The charged hadron spectrum is evaluated by convoluting this jet spectrum with the \texttt{NPC23}~\cite{Gao:2024dbv} fragmentation function (FF).

The quenched jet cross section is described via the convolution of the nPDF-modified jet cross section with the energy-loss probability distribution:
\begin{equation}
    \begin{split}
        \sigma^{AA}(p_T,R) &= \sum_{i=q,g}\int d\varepsilon P_i(\varepsilon,R,p_T)\tilde\sigma^{pp}_i(p_T+\varepsilon,R)\,.
    \end{split}
\end{equation}
Energy is lost predominantly due to medium-induced emissions escaping the jet cone $R$, as encoded in $P_i(\varepsilon,R,p_T)$. The emission rate is computed using the Improved Opacity Expansion in a static medium, which captures both single hard and multiple soft scatterings, and also includes finite path length corrections~\cite{Mehtar-Tani:2019tvy,Barata:2020rdn}\footnote{The static medium rates can mimic the rates of expanding medium in the multiple soft scattering limit~\cite{Adhya:2019qse,Caucal:2020xad}.}.
Medium properties affect these rates through LO Hard Thermal Loop results, namely the (bare) jet quenching parameter $\hat q_0=g^2_{\rm med}N_cm_D^2T/(4\pi)$, and the Debye screening mass $m_D^2=3g_{\rm med}^2T^2/2$, where $T$ is the local temperature of the hydrodynamic medium. Here, $g_{\rm med}$ is the (fixed, i.e. non-running) strong coupling for jet-medium interactions, representing the first of the two parameters we will estimate using Bayesian inference. We assume factorization between induced collinear emissions and transverse momentum broadening: semi-hard emissions undergo Gaussian broadening, while hard emissions inherit their transverse momentum from the Coulomb tail of the scattering potential. Since the softest emissions thermalize rapidly~\cite{Blaizot:2015lma}, they should be considered part of the medium and contribute to hydrodynamic wakes~\cite{Casalderrey-Solana:2004fdk,Chesler:2007an}, a manifestation of medium response to the jet passage. We approximate these physics by assuming that the thermalized energy is distributed uniformly around the jet hemisphere. Elastic energy loss is also included and constrained via the Einstein relation $\hat e=\hat q/(4T)$, and is assumed to thermalize in a similar fashion. Here, $\hat{q}$ accounts for the logarithmic corrections of $\hat{q}_0$ (see later). 

Energetic jets consist of multiple partons produced by the high-virtuality DGLAP evolution that occurs at the early stages of the in-medium jet propagation~\cite{Caucal:2018dla}. These partons represent an additional source of energy loss. We account for this by resumming these multiple sources via a factorization between early vacuum-like and medium-induced emissions~\cite{Mehtar-Tani:2017web}. Importantly, we also include color-coherence, and -resolution~\cite{Mehtar-Tani:2010ebp,Mehtar-Tani:2011hma,Casalderrey-Solana:2011ule} effects (referred to as \emph{coherent energy loss}) by incorporating the non-local angular scale below which the medium cannot resolve individual partons, $\theta_c \approx 1/\sqrt{\hat{q}L^3}$, with $L$ being the traversed medium length. Detailed formulas are provided in~\cite{Mehtar-Tani:2024jtd}.

Jet propagation occurs within an event-by-event fluctuating 2+1D hydrodynamic background. We employ hydro profiles from the comprehensive multi-stage framework \texttt{IP-Glasma+JIMWLK+MUSIC+UrQMD}, which successfully describes a wide range of bulk observables in heavy-ion collisions~\cite{Mantysaari:2025tcg}. This state-of-the-art framework evolves the bulk starting from the initial energy density deposited at specific hotspots determined by the \texttt{IPSat}~\cite{Kowalski:2003hm}. Jet probes are placed precisely onto these hotspots, whose number is proportional to the number of binary collisions $N_{\rm coll}$, and their initial orientation is uniformly distributed in the azimuthal angle $\phi$. Straight jet trajectories are sampled from their production point until they exit the deconfined medium at $T<T_c$, with $T_c$ marks the crossover transition to the hadron gas phase. The medium and jet-medium interaction quantities needed for the computation of radiative and elastic energy loss are then obtained by averaging along each jet trajectory, as extracted from the hydrodynamic profiles. For example, we evaluate the Debye mass as $m_D^2\propto (\int {\rm d}x_F T^2(x))/L$ for a given trajectory, where $x_F$ is the distance traversed in the fluid rest frame, thereby accounting for different flow velocities, and $L$ is the integral of ${\rm d}x_F$. The expressions for other relevant quantities can be found in Eq.~(29) in~\cite{Mehtar-Tani:2024jtd}. We neglect further quenching below the freeze-out temperature $T_c$. For more details on the embedding of our jet quenching framework in the realistic heavy-ion environment, see Ref.~\cite{Mehtar-Tani:2024jtd}, where event-averaged hydro profiles were used instead. Event classification into centrality classes is based on the final charged particle multiplicity of the bulk, from which one also extracts the soft event-plane angles $\Psi_n$, necessary for computing the jet flow coefficients. Thus, our framework predicts the quenched jet spectrum for various parameters $\sigma^{AA\%}(p_T, \eta, \phi, R)$.

The hydrodynamic description of the bulk begins at a finite time $\tau_{\rm hyd} = 0.4$ fm. This parameter was fixed when the \texttt{IP-Glasma+JIMWLK+MUSIC+UrQMD} framework was tuned to LHC and RHIC data. The initial condition of this nearly ideal fluid is reached through the Glasma evolution of saturated gluon fields. The Glasma evolution should be followed by a kinetic evolution that drives the system toward hydrodynamics; however, this is not the setup used in the current bulk evolution framework. For reviews of this approach to thermalization, see~\cite{Schlichting:2019abc,Berges:2020fwq}. Quenching effects have traditionally been neglected before $\tau_{\rm hyd}$, largely due to the absence of reliable energy-loss calculations in the non-equilibrated phases. While an earlier start could be mimicked by adjusting the jet-medium coupling, combined observables such as high-$p_T$ $R_{AA}$ and $v_2$ are known to be sensitive to the actual value of the quenching onset time~\cite{Andres:2019eus,Stojku:2020wkh}. The need to address jet quenching in the earliest stages has been emphasized in recent works where the pre-hydro phase is modeled either by Glasma~\cite{Ipp:2020nfu,Avramescu:2023qvv} or by QCD effective kinetic theory (EKT)~\cite{Boguslavski:2023alu,Boguslavski:2023waw}. Despite differences in approach, these studies consistently conclude that the broadening per unit length in the short pre-equilibrium phase is considerably larger than in the equilibrated hydro phase.

In the present work, we extend our energy loss framework to the kinetic pre-equilibrium stage as follows. The out-of-equilibrium evolution of the local energy density has been extensively studied in EKT~\cite{Kurkela:2018wud,Kurkela:2018vqr,Mazeliauskas:2018yef,Kurkela:2019set,Almaalol:2020rnu,Soloviev:2021lhs}, and it is well captured by the hydrodynamic attractor. We use the attractor to extrapolate the effective local temperature for times $\tau$ earlier than $\tau_{\rm hyd}$ as~\cite{Giacalone:2019ldn,Garcia-Montero:2023lrd}:
\begin{equation}
    (\tau^{1/3}T)_{\rm eff}^4=(\tau^{1/3}T)_{\rm hyd}^4\cdot\mathcal E(\tilde\omega)\,.
\end{equation}
where $\mathcal{E}(\tilde\omega)$ is the attractor function, $\tilde\omega = \tau T_{\rm eff} / (4\pi\eta/s)$, and we implicitly used the the three flavor conformal equation of state of the EKT. We account for viscous corrections as
\begin{equation}
    (\tau^{1/3}T)_{\rm hyd}=\tau^{1/3}_{\rm hyd}\left(T(x)+\frac{2}{3}\frac{\eta/s}{\tau_{\rm hyd}}\right)\,,
\end{equation}
with $\eta/s = 0.12$, where $T(x)$ is the local initial temperature. We assume that our energy-loss model can be applied without any additional modification by using these extrapolated pre-equilibrium temperatures. Before $\tau_{\rm min}$ , and for the Glasma phase, no quenching is applied, and we take this time as the second parameter of our Bayesian inference. 

Finally, fluid velocity fields also evolve during the kinetic pre-equilibrium phase, as initial spatial anisotropies begin to generate flow~\cite{Berges:2020fwq}. This non-equilibrium evolution is well approximated by the universal pre-flow extracted from EKT~\cite{Keegan:2016cpi}. We incorporate it by scaling the transverse velocity linearly with time:
\begin{equation}
    \begin{split}
        v_{{\rm eff},T}(\tau,\vec x) &= \tfrac{\tau}{\tau_{\rm hyd}}v_T(\tau_{\rm hyd},\vec x)\,,\\
        v_{{\rm eff},z}(\tau) &= \tanh\eta_s\,,
    \end{split}
\end{equation}
where the longitudinal velocity scales with the space-time rapidity $\eta_s$.

\section{Results}
\label{sec:results}
We evaluate the quenched inclusive jet spectrum $\sigma^{AA\%}(p_T,\eta,\phi,R)$ for various collision systems and centralities. Our focus is on the simultaneous description of $R^{\rm jet}_{AA}$ and $v^{\rm jet}_n$ observables, which we evaluate using
\begin{equation}
\label{eq:obs}
    \begin{split}
        R^{\rm jet}_{AA} =& \sigma^{AA\%}(p_T,R)/\sigma^{pp}(p_T,R)\,,\\
        v^{\rm jet}_n =& \frac{\int d\eta d\phi \cos(n(\phi-\Psi_n))\sigma^{AA\%}(p_T,\eta,\phi,R)}{\sigma^{AA}(p_T,R)}\,.
    \end{split}
\end{equation}
where $\sigma(p_T,R)=\int d\eta d\phi\sigma(p_T,\eta,\phi,R)$ with the appropriate experimental cuts. We use the event plane method to compute $v_n^{\rm jet}$, following the procedure in the ATLAS measurement to which we compare our model~\cite{ATLAS:2021ktw}.\footnote{We also compared our results using the event plane method against the more experimentally common scalar product method, in which one correlates the jet momenta with the soft sector (see e.g.~\cite{Noronha-Hostler:2016eow}), and found very little change in $v_2$, below $4\%$, consistent with what was shown by ATLAS~\cite{ATLAS:2018ezv}.}
We evaluate the hadron spectrum as
\begin{equation}
    \sigma(p^h_T,\eta,\phi)=\sum_{i=q,g}\int^1_x\frac{{\rm d}z}{z}\sigma_i(\tfrac{p^h_T}{z},R,\eta,\phi)\,D_i^h(z,\mu_F^2)\,,
\end{equation}
where $x=2p_T^h\cosh(\eta)/\sqrt{s}$, and we used the jet scale $\mu_F=p_T^hR/z$. We averaged over $(u,d,s)$ FFs and neglected the others. This simple convolution neglects medium effects in fragmentation, and thus ignores, for example, the impact of the angular redistribution of particles due to broadening.

Our work resembles previous studies of $R_{AA}$ and $v_2$~\cite{Betz:2014cza,Noronha-Hostler:2016eow,Andres:2019eus,Andres:2022bql,Arleo:2022shs,Faraday:2024qtl, Zigic:2018smz,Zigic:2018ovr,Zigic:2019sth,Stojku:2020wkh,Zigic:2021rku,Zigic:2022xks,Karmakar:2023ity,Karmakar:2024jak,Zhao:2021vmu,Barreto:2022ulg,He:2022evt}. In contrast to these works, we focus on jets instead of hadrons, since jets are infrared- and collinear-safe, and have small non-perturbative corrections in vacuum~\cite{Ellis:318585,Salam:2010zt,Luisoni:2015xha,Marzani:2019hun}. We use state-of-the-art hydrodynamics and energy-loss formulas, including elastic energy loss as well as the effects of medium response. To our knowledge, this is the first model implementation of jet energy loss in the pre-equilibrium phase carried out in a physically well-motivated manner.

\begin{figure}
    \centering
     \includegraphics[width=\linewidth]{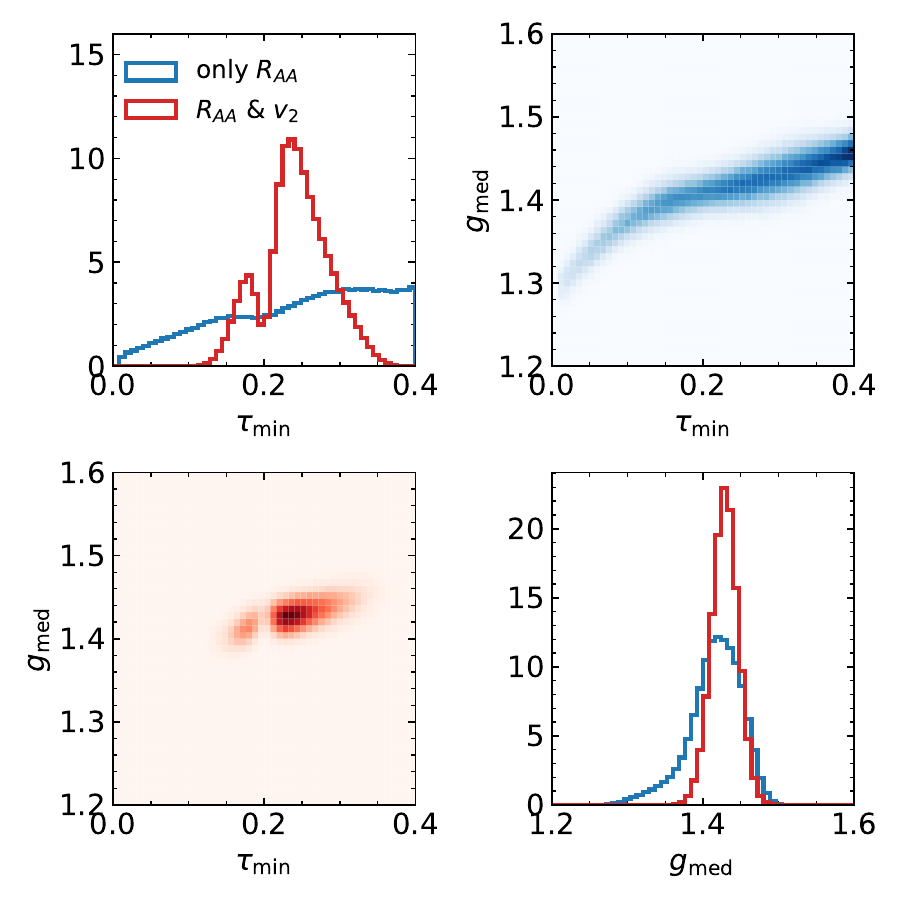} 
     \caption{Parameter posterior probability distributions and correlations. Red distributions include $R^{\rm jet}_{AA}$ and $v_2^{\rm jet}$ data, while blue only includes $R_{AA}^{\rm jet}$.}
     \label{fig:ParameterPosterior}
\end{figure}

\begin{figure*}
    \centering
     \includegraphics[width=0.47\linewidth,page=1]{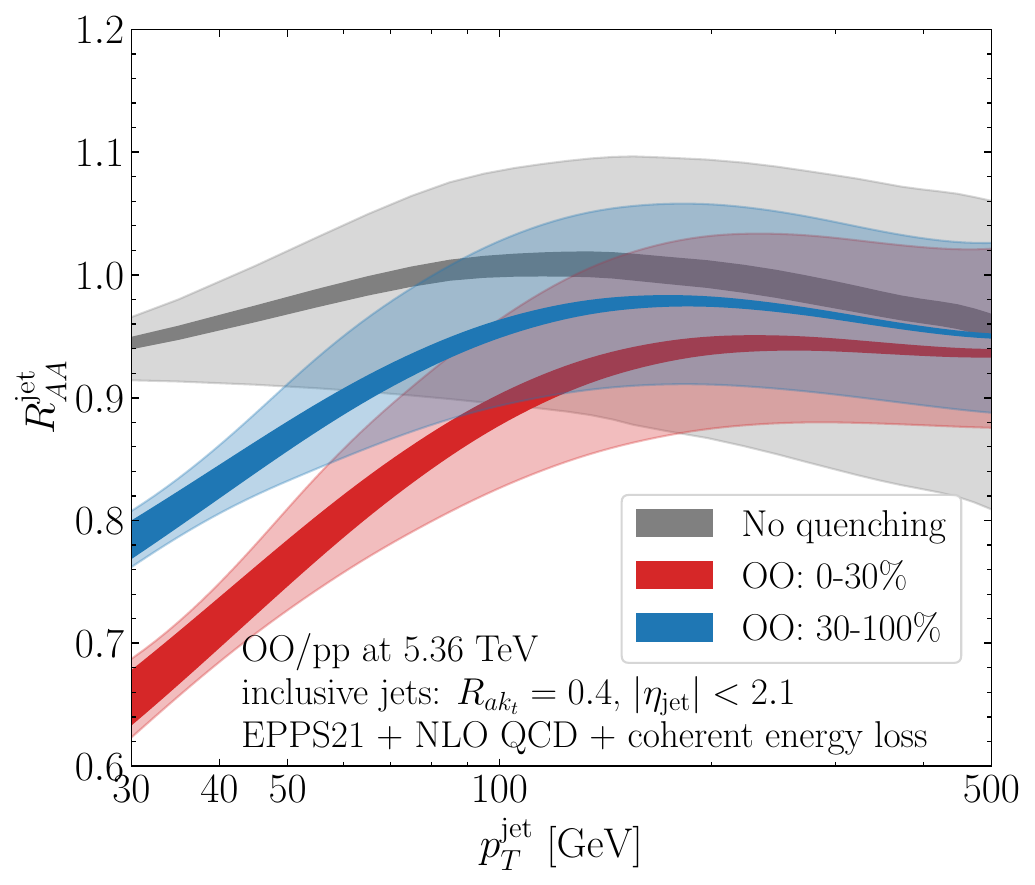} 
     \includegraphics[width=0.47\linewidth,page=3]{figs/plot_Oxygen_predictions.pdf} 
     \caption{Nuclear modification for jets and charged hadrons in OO collisions for two different centrality selections. Dark bands represent the 68\% confidence interval for quenching parameters, and light bands represent the 68\% nPDF uncertainties. }
     \label{fig:OO_RAA}
\end{figure*}

\begin{figure*}
    \centering
     \includegraphics[width=0.47\linewidth, page=2]{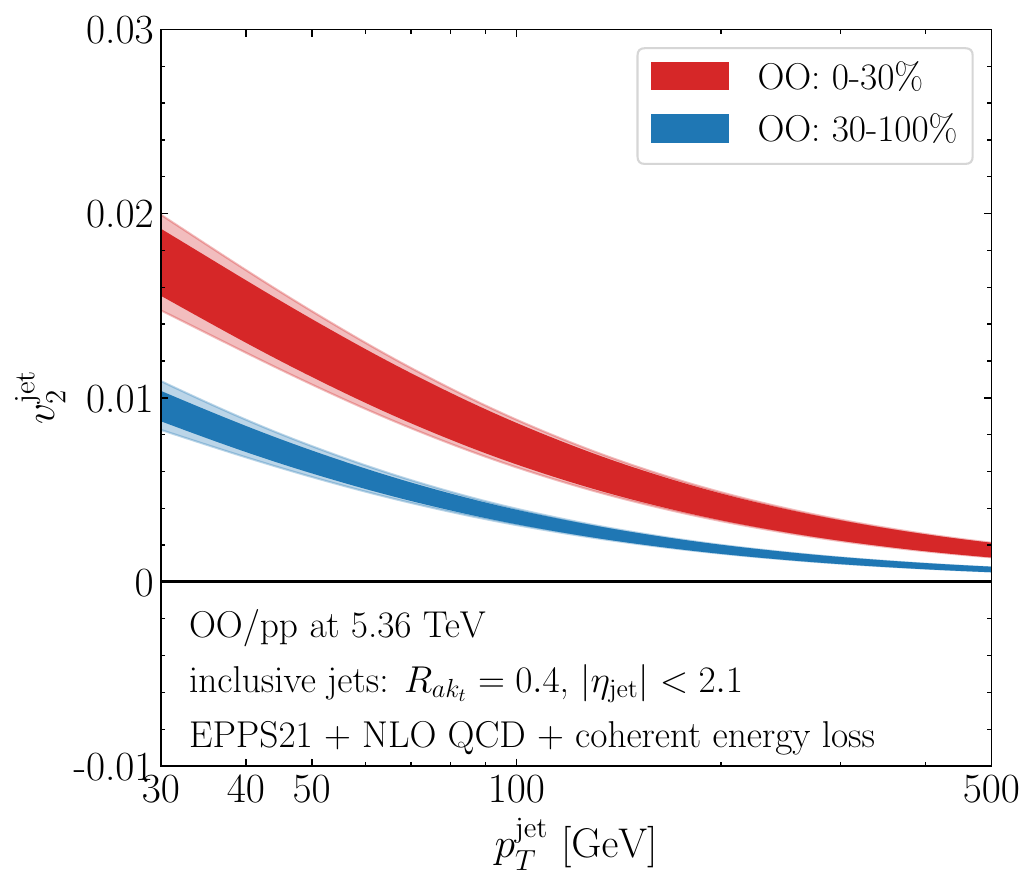} 
     \includegraphics[width=0.47\linewidth, page=4]{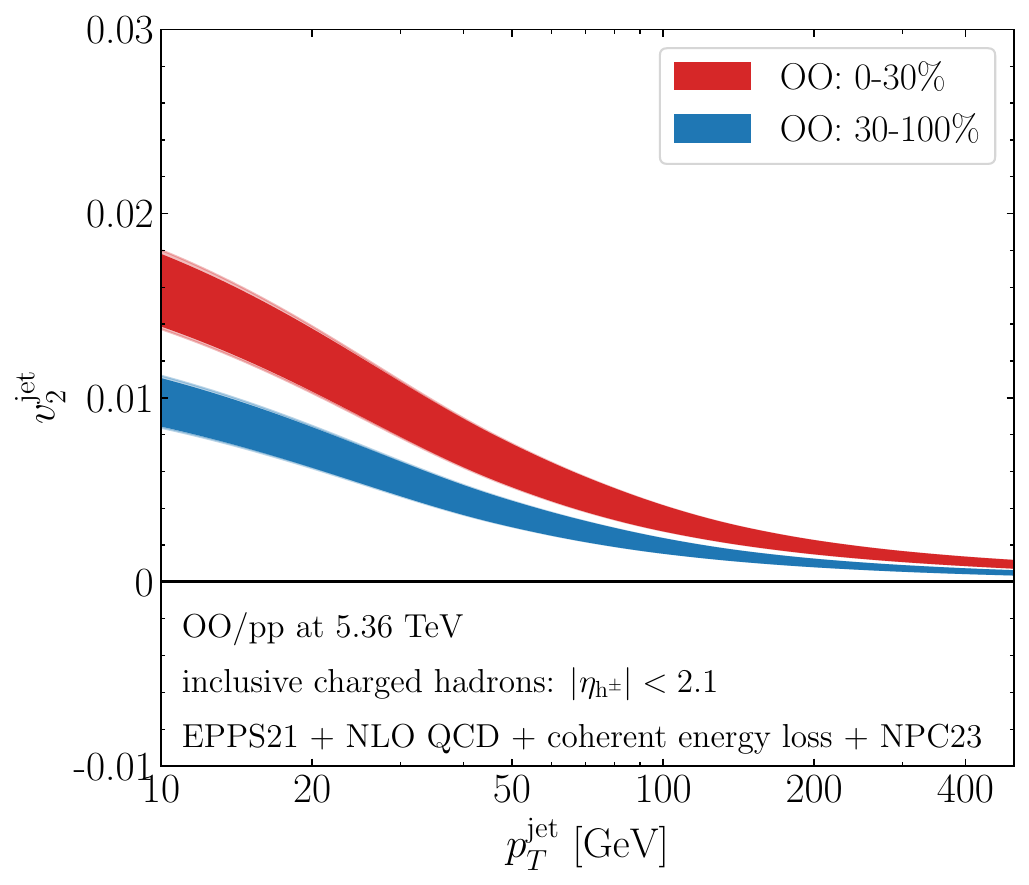} 
     \caption{Charged hadron and jet elliptic flow in oxygen collisions for two different centrality selections. Dark bands represent the 68\% confidence interval for quenching parameters, while light bands represent the 68\% nPDF uncertainties.}
     \label{fig:OO_V2}
\end{figure*}

We employ Bayesian inference, implemented via the \texttt{JETSCAPE/STAT} toolkit~\cite{JETSCAPE/STAT_GitHub,JETSCAPE:2021ehl,JETSCAPE:2024cqe}, to constrain the jet-medium strong coupling $g_{\rm med}$ and the initial quenching time in the pre-equilibrium phase $\tau_{\rm min}$. We used flat priors in the intervals $1.2\leq g_{\rm med}\leq1.7$, and $0.01\leq \tau_{\rm min}\,[{\rm fm}]\leq 0.4$. We do not consider the scenario when $\tau_{\min}>\tau_{\rm hyd}$ as we believe delaying quenching beyond hydrodynamics is unphysical. Our data points include $R^{\rm jet}_{AA}$ at LHC energies for different jet-cones and centrality selections~\cite{ATLAS:2018gwx,ALICE:2019qyj,CMS:2021vui}, as well as jet $v_2^{\rm jet}$~\cite{ATLAS:2021ktw}, up to 40-50\% centrality. The inference incorporates statistical and systematic uncertainties of the data, and we estimate the full correlation matrix by employing a finite correlation length of 0.2~\cite{JETSCAPE:2021ehl,Soltz:2024gkm}. Luminosity and centrality uncertainties are added to the systematics. We neglect inherent model uncertainties in the inference such as scale and nPDF uncertainties, and we will present them separately for the OO predictions.

Figure.~\ref{fig:ParameterPosterior} shows the parameter posterior for two different scenarios. Fitting only $R^{\rm jet}_{AA}$ (blue) results in little constraint on $\tau_{\rm min}$. This is expected, as an earlier starting time can be compensated by a relatively smaller medium coupling to produce the same energy loss. Including $v_2^{\rm jet}$ constrains $\tau_{\rm min}\approx 0.24$ fm (red). This inference illustrates the role of pre-equilibrium effects in $v^{\rm jet}_2$.\footnote{It would be interesting to include $\tau_{\rm hyd}$ in the inference, since it could affect our results. This would require the simultaneous tuning of the soft and hard sectors, a task that we defer to future work.}

Before moving to our predictions for oxygen collisions, Figs.~\ref{fig:PosteriorPredictiveDistribution_1}-\ref{fig:PosteriorPredictiveDistribution_2} in App.~\ref{app:Bayesian} show the posterior predictive distributions. We observe excellent overall agreement with all data. Small tensions can be found in $v_2^{\rm jet}$ at low $p_T$, which motivates further improvements of the energy loss framework. For completeness, Figs.~\ref{fig:PosteriorPredictiveDistribution_1}-\ref{fig:PosteriorPredictiveDistribution_2} in App.~\ref{app:Bayesian} also show two additional inference results on $g_{\rm med}$ with two limiting scenarios: early pre-equilibrium with fixed $\tau_{\min}=0.08$ fm, and no pre-equilibrium with $\tau_{\min}=0.4$ fm. The best fit (whose parameters are shown in Fig.~\ref{fig:ParameterPosterior}) outperforms these two scenarios, although differences are not very significant. The extracted bare jet transport coefficient is approximately $\hat{q}_0/T^3 \approx 1.7$, while after logarithmic corrections the effective value is $\hat{q}/T^3=\hat{q}_0/T^3 \log(Q_s^2/\mu_*^2)\approx 7$, where $Q_s^2= \hat{q}_0 L$, and $\mu_*^2=m_D^2 e^{-2+2\gamma_E}/4$. We note that this extracted $\hat q$ is consistent with other recent extractions using Bayesian inference in the \texttt{JETSCAPE} framework~\cite{JETSCAPE:2024cqe}.

Figures~\ref{fig:PosteriorPredictiveDistribution_hadrons_1}-\ref{fig:PosteriorPredictiveDistribution_hadrons_2} in App.~\ref{app:Bayesian} provide predictions for jets at RHIC energies~\cite{STAR:2020xiv} and charged hadrons at LHC energies~\cite{CMS:2016xef,ALICE:2018vuu,ATLAS:2018ezv,ATLAS:2022kqu} using the extracted parameters. Even though our framework was designed for jets, it also provides a good description of the charged hadron $R_{AA}$ without including them in the inference. We also see, however, that our framework underestimates hadron elliptic flow at low $p_T$. This failure is consistent with previous findings using pQCD energy loss~\cite{Andres:2019eus,Stojku:2020wkh}. It has been pointed out that a rather late quenching time of around $\approx 1$ fm would resolve the disagreement~\cite{Stojku:2020wkh}. Alternatively, the absence of this issue for jets signals the need for model improvements which could be more relevant for (mid-$p_T$) hadrons than for jets. Again, for completeness, we show predictions with the previous two limiting scenarios (early pre-eq. and no pre-eq. energy loss), resulting as before in subtle differences.

\paragraph*{OO predictions:}
The recent OO run at the LHC has attracted considerable attention, with numerous dedicated studies on, for example, baseline analyses~\cite{Brewer:2021tyv,Paakkinen:2021jjp,Gebhard:2024flv,Mazeliauskas:2025clt} and hadron quenching predictions~\cite{Katz:2019qwv,Huss:2020dwe,Huss:2020whe,Zakharov:2021uza,Xie:2022fak,Ke:2022gkq,Vitev:2023nti,Ogrodnik:2024qug,vanderSchee:2025hoe}. Compared to these studies, our framework incorporates additional improvements, with an emphasis on jet observables. For our OO predictions of $R_{AA}$ and $v_2$ we have set $\sqrt{s_{NN}}=5.36$ TeV and $|\eta|<2.1$. We defer to future work the computation of higher-order flow harmonics, along with the corresponding predictions for NeNe collisions.

Figure~\ref{fig:OO_RAA} shows our jet and charged hadron predictions for OO collisions. The left panel shows the inclusive jet suppression, where the no-quenching baseline, $\tilde\sigma^{pp}/\sigma^{pp}$, is indicated with gray bands. The dark band represents NLO scale uncertainties, while the light band is the 68\% confidence of the nPDF. Below 50 GeV for jets and 30 GeV for hadrons, we underestimate nPDF uncertainties because of the relatively high momentum cutoff we chose for the NLO cross-sections~\cite{Gebhard:2024flv,Mazeliauskas:2025clt}. This cutoff does not affect the central value. Quenching effects are shown with red and blue for different centrality selections. Dark bands are the 68\% confidence of our Bayesian parameter estimation, and on top of these, 68\% nPDF uncertainties are shown with light bands. Peripheral collisions (30-100\%) are less suppressed than central ones (0-30\%), as expected, and clear quenching signals are present towards lower jet $p_T$ for both centrality selections. The right panel in Fig.~\ref{fig:OO_RAA} shows the charged hadron suppression, which presents a similar pattern to jets. 

Figure~\ref{fig:OO_V2} shows the inclusive jet and charged hadron elliptic flow. There is relatively smaller sensitivity to the nPDF, as $v_2$ is an observable that is differential in the transverse plane, using solely the medium spectra, so the impact of the nPDF largely cancels in Eq.~\eqref{eq:obs}. Semi-inclusive observables can present analogous advantages~\cite{Gebhard:2024flv}. We observe that, in contrast to large collision systems, the elliptic flow coefficient $v_2$ increases with centrality. This is a consequence of the more prominent role played by the fluctuations of the initial nuclear shape in small collision systems~\cite{Katz:2019qwv,Rybczynski:2019adt,Huss:2020whe}. Jet distributions are not shown for very low $p_T$, where our framework is not expected to apply.

Due to the relatively small path lengths involved in OO, the typical value of the coherence angle is relatively large, $\theta_c\approx 0.5$, and so $R=0.4$ jets are unresolved and lose energy as single color charges. Therefore, hadron and jet suppression in our framework can be understood by the mere $p_T$ shift resulting from hadronization. In contrast, in larger systems (with smaller values of $\theta_c$) both $R_{AA}$ and $v_2$ receive additional contributions due to resolved jet substructure fluctuations~\cite{Casalderrey-Solana:2018wrw,Mehtar-Tani:2024jtd}.

Finally, Fig.~\ref{fig:OO_early_vs_late} in App.~\ref{app:Oxygen_early_vs_late} shows our OO predictions for minimum bias with the different limiting cases introduced previously. Only subtle differences are observed for earlier vs. later quenching for the ranges of $\tau_{\rm min}$ explored.

\section{Conclusions}
\label{sec:conclusion}

In this work, we have integrated our semi-analytical jet quenching model with a state-of-the-art description of heavy-ion collisions, \texttt{IP-Glasma+JIMWLK+MUSIC+UrQMD}. We incorporated energy loss in the pre-equilibrium phase via the hydrodynamic attractor for the first time. Our framework provides a simultaneous description of jet $R_{AA}$ and $v_2$ in agreement with data in $AA$ collisions, offering a strong validation of the approach. We show that a joint description of these observables places constraints on the onset time of jet-medium interactions, which is around 0.2 fm, deep in the pre-equilibrium phase. Then, we applied these parameters to predict jet and hadron quenching in other collision systems. We saw that our framework successfully predicts hadron suppression but underestimates hadron elliptic flow towards lower $p_T$. Our results provide a complementary perspective to previous phenomenological studies. By incorporating early-time energy loss, we achieve a simultaneous description of jet $R_{AA}$ and $v_2$, as well as hadron $R_{AA}$ and $v_2$ at high-$p_T$. This resolves the, in our opinion, unphysical delay of quenching invoked in earlier approaches (e.g.~\cite{Andres:2019eus,Stojku:2020wkh}) to reproduce hadron $R_{AA}$ and $v_2$. The successful description of jet observables supports the onset of quenching at early times, whereas the remaining discrepancy in hadron $v_2$ at moderate transverse momentum indicates the necessity of additional medium effects beyond longitudinal energy loss. 

Our constrained framework has been used to make realistic predictions for the intermediate-sized OO collisions. We obtained sizable energy loss both for jets and charged hadrons, in quantitative agreement with the results from CMS~\cite{CMS:2025bta}, where our predictions are also presented. Even though we have shown that pre-equilibrium effects play a moderate role in our model, we anticipate that including forthcoming data on high-$p_T$ flow coefficients will result in further constraints.  Furthermore, in this work we have provided predictions for elliptic anisotropies and different centrality selections, both for hadrons and jets. Due to the small path lengths in OO, unlike in PbPb or AuAu, jet constituents are barely resolved (jets are fully coherent), and thus we obtain similar energy loss for hadrons and jets (modulo a $p_T$ shift in fragmentation). This outcome constitutes a prediction of our framework which can be confronted with future experimental results on jet suppression in light ions. All of our data and plots are available at~\cite{Takacs_Bayesian_constraints_on_2026}.

As an outlook, our jet quenching framework could benefit from an improved description of medium-induced emissions, elastic scatterings, and medium response in several ways. It is important to note that, at very early times, local color fields are saturated and highly anisotropic, a scenario which our energy-loss framework, built on local equilibrium, cannot accommodate. For these initial stages, it would be desirable to incorporate recent developments in energy-loss calculations for anisotropic and flowing media~\cite{Sadofyev:2021ohn,Fu:2022idl,Barata:2022krd,Andres:2022ndd,Hauksson:2023tze,Kuzmin:2023hko,Adhya:2024nwx}, correlated color fields~\cite{Barata:2024xwy}, and out of-equilibrium plasma~\cite{Altenburger:2025iqa,Lindenbauer:2025ctw}, so as to extend quenching for those earliest times where the attractor solution is not meant to apply.

\begin{acknowledgments}
We are grateful to Bj\"orn Schenke and Chun Shen for sharing the hydrodynamic profiles and to Jannis Gebhard for sharing oxygen events. We greatly appreciate discussions with Aleksas Mazeliauskas, Yacine Mehtar-Tani, and Konrad Tywoniuk. The work of A.T. is supported by DFG through Emmy Noether Programme (project number 496831614) and through CRC 1225 ISOQUANT (project number 27381115). D.P. is funded by the European Union's Horizon 2020 research and innovation program under the Marie Sklodowska-Curie grant agreement No 101155036 (AntScat), by the European Research Council project ERC-2018-ADG-835105 YoctoLHC, by the Spanish Research State Agency under project PID2020-119632GB-I00, by Xunta de Galicia (CIGUS Network of Research Centres) and the European Union, and by Unidad de Excelencia Mar\'ia de Maetzu under project CEX2023-001318-M. D.P. also acknowledges support from the the Ram\'on y
Cajal fellowship RYC2023-044989-I.
\end{acknowledgments}

\appendix
\section{Bayesian parameter estimation}\label{app:Bayesian}
We have already introduced our Bayesian framework in Sec.~\ref{sec:Framework}, and here we extend this discussion by providing additional details. In this appendix, PbPb and AuAu cross-sections $\tilde\sigma^{AA}(p_T,\eta,R)$ are computed using the microjet evolution as employed in~\cite{Mehtar-Tani:2021fud,Mehtar-Tani:2024jtd}, with initial conditions from LO \texttt{Pythia} at $R_0=1$. Our framework is built on the \texttt{JETSCAPE/STAT} toolkit~\cite{JETSCAPE/STAT_GitHub,JETSCAPE:2021ehl,JETSCAPE:2024cqe}. We evaluated our model for 24 design points, equally distributed in the parameter space. To focus on primary model features that are robust against statistical uncertainty, we reduced our data to its 10 principal components, and trained a Gaussian Process Emulator to interpolate between the predictions. The resulting parameter posterior is shown in the main text in Fig.~\ref{fig:ParameterPosterior}. Furthermore, we provide two alternative inferences for $g_{\rm med}$: one with fixed $\tau_{\min}=0.08$ fm, (and extracted $\langle g_{\rm med}\rangle=1.385$) referred to as ``early preeq'', and another with $\tau_{\min}=0.4$ fm, (extracted $\langle g_{\rm med}\rangle=1.473$), referred to as ``no preeq''.

\begin{figure*}
    \centering
     \includegraphics[width=0.28\textwidth,page=1]{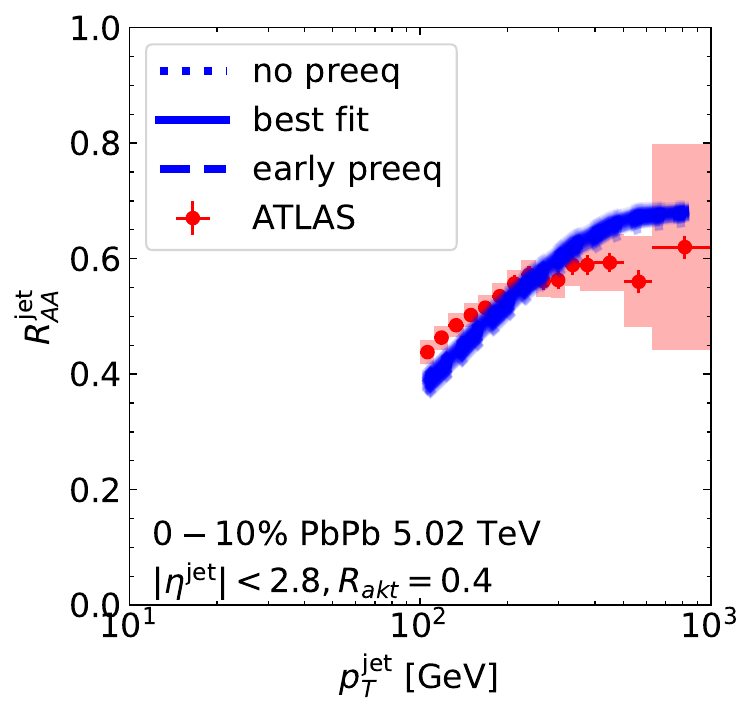}
     \includegraphics[width=0.28\textwidth,page=2]{figs/plot_Bayesian_posteriors.pdf}
     \includegraphics[width=0.28\textwidth,page=3]{figs/plot_Bayesian_posteriors.pdf}
     \includegraphics[width=0.28\textwidth,page=4]{figs/plot_Bayesian_posteriors.pdf}
     \includegraphics[width=0.28\textwidth,page=5]{figs/plot_Bayesian_posteriors.pdf}\\
     \includegraphics[width=0.28\textwidth,page=6]{figs/plot_Bayesian_posteriors.pdf}
     \includegraphics[width=0.28\textwidth,page=7]{figs/plot_Bayesian_posteriors.pdf}
     \includegraphics[width=0.28\textwidth,page=8]{figs/plot_Bayesian_posteriors.pdf}
     \includegraphics[width=0.28\textwidth,page=9]{figs/plot_Bayesian_posteriors.pdf}
     \caption{Posterior predictive distributions, the fitting performance of our model. Jet $R_{AA}$ and jet $v_2$ data from ATLAS.}
     \label{fig:PosteriorPredictiveDistribution_1}
\end{figure*}

\begin{figure*}
    \centering
     \includegraphics[width=0.28\textwidth,page=12]{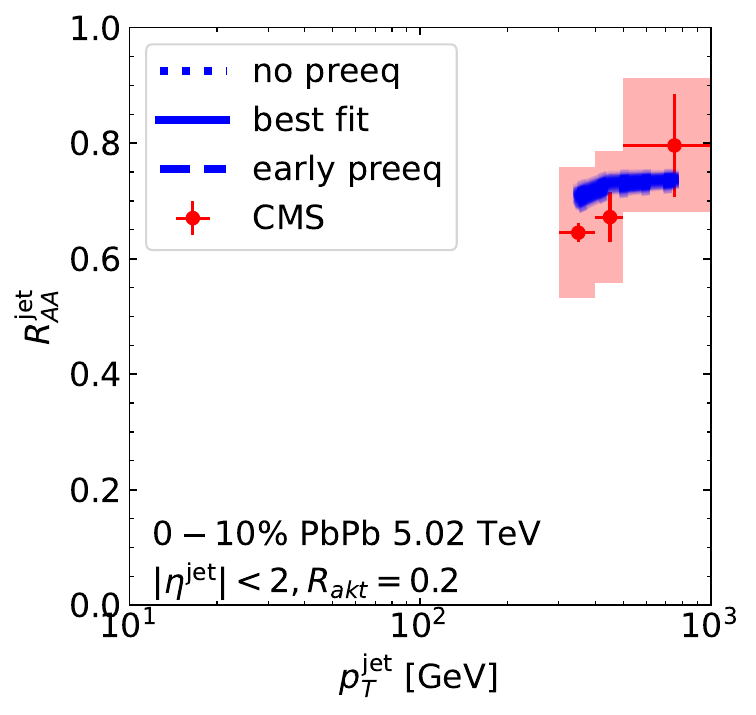}
     \includegraphics[width=0.28\textwidth,page=13]{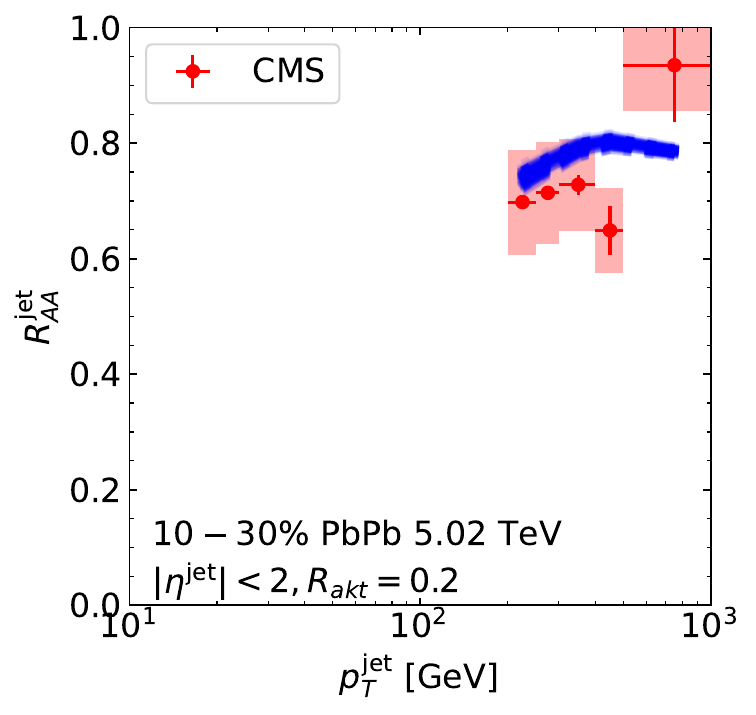}
     \includegraphics[width=0.28\textwidth,page=14]{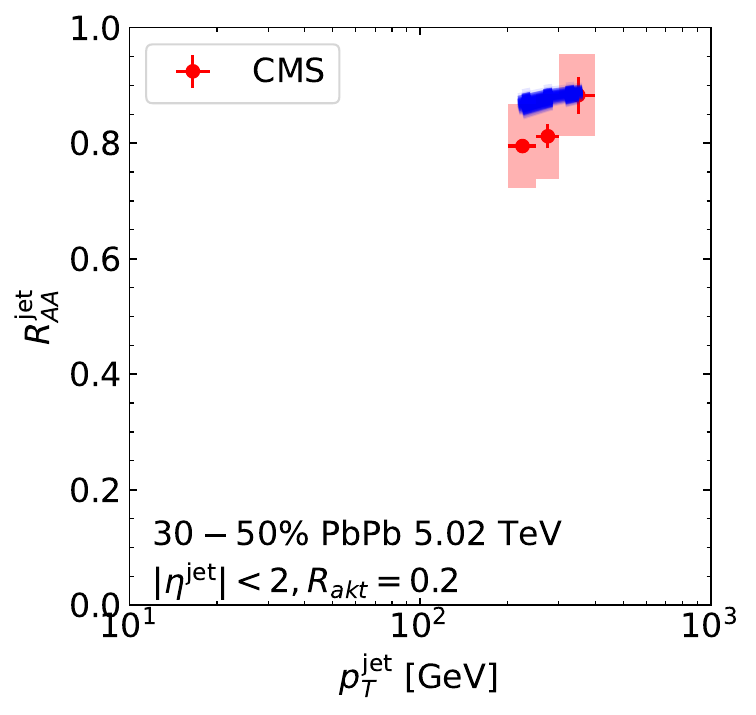}\\  
     \includegraphics[width=0.28\textwidth,page=15]{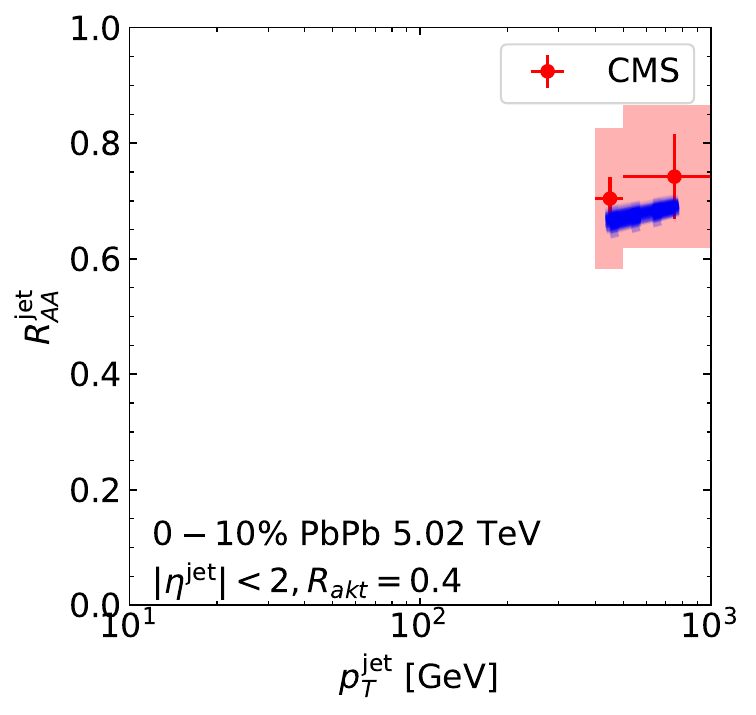}
     \includegraphics[width=0.28\textwidth,page=16]{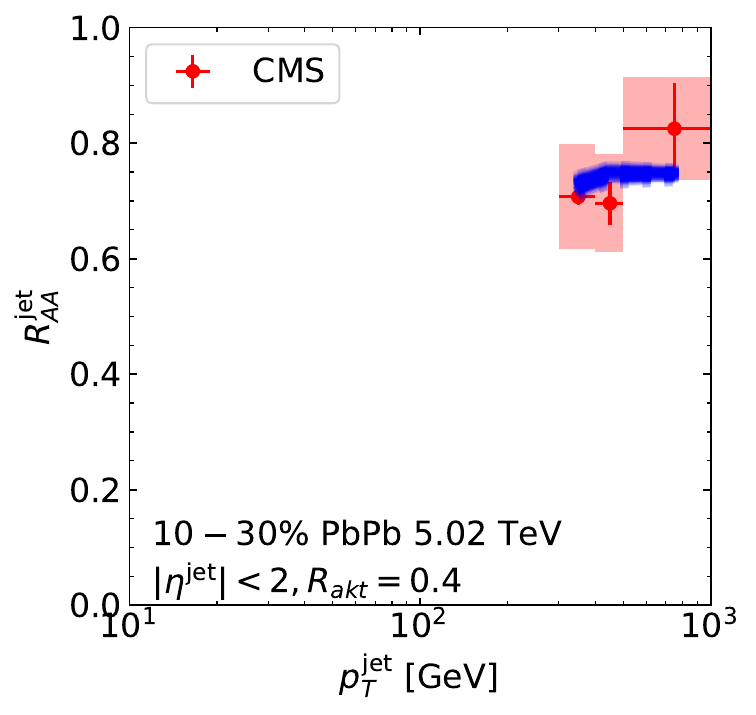}
     \includegraphics[width=0.28\textwidth,page=17]{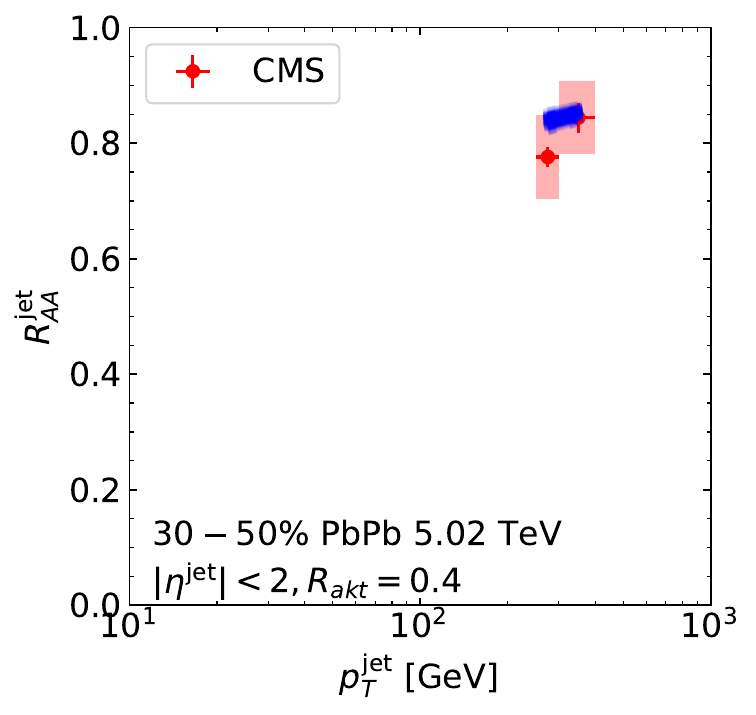}\\
     \includegraphics[width=0.28\textwidth,page=18]{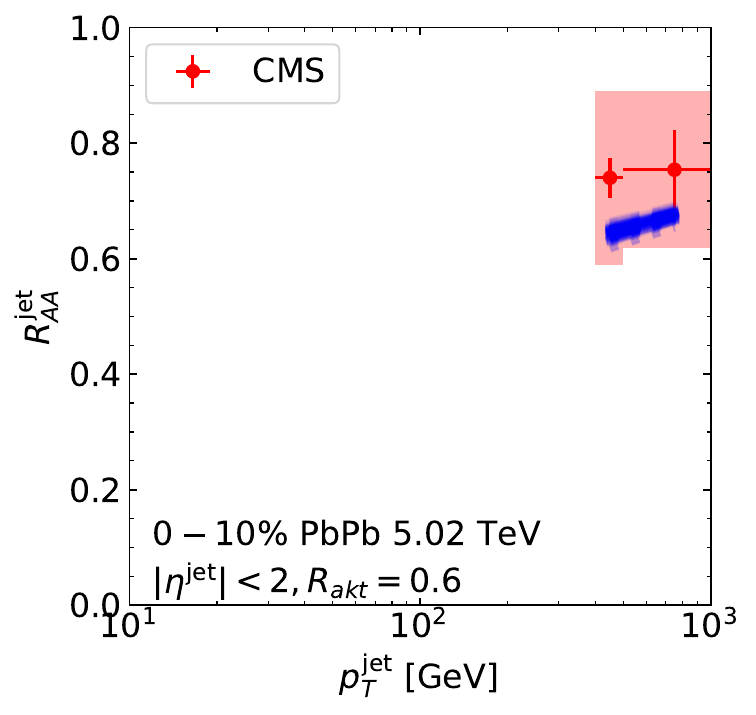}
     \includegraphics[width=0.28\textwidth,page=19]{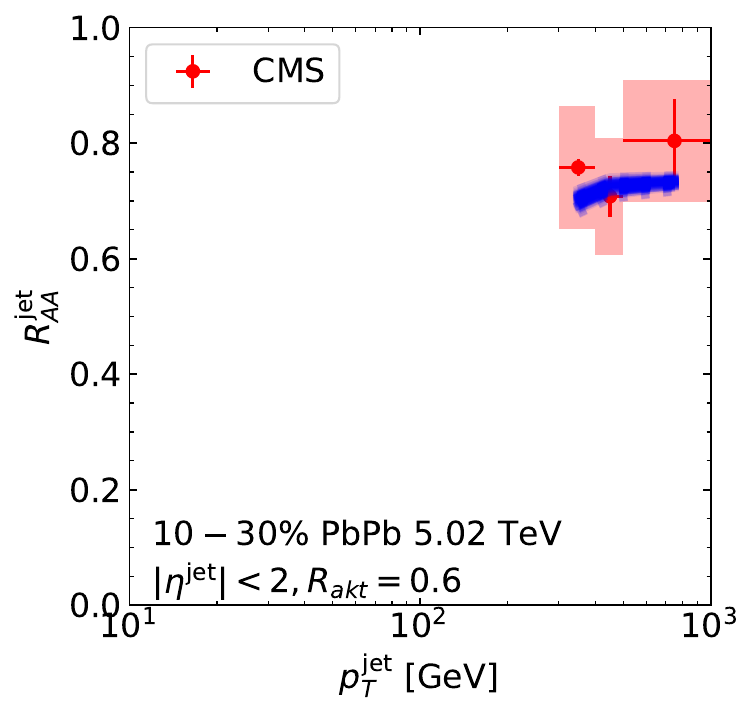}
     \includegraphics[width=0.28\textwidth,page=20]{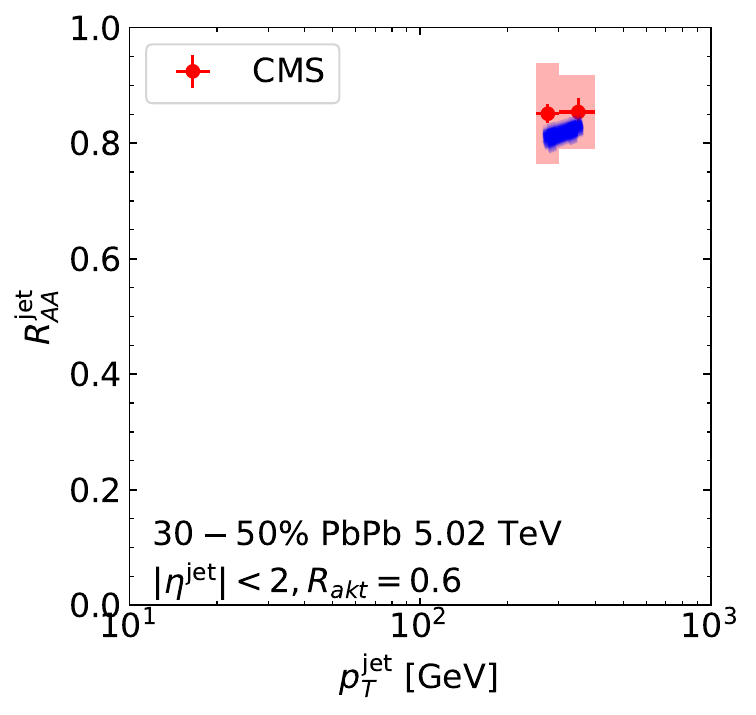}\\
     \includegraphics[width=0.28\textwidth,page=10]{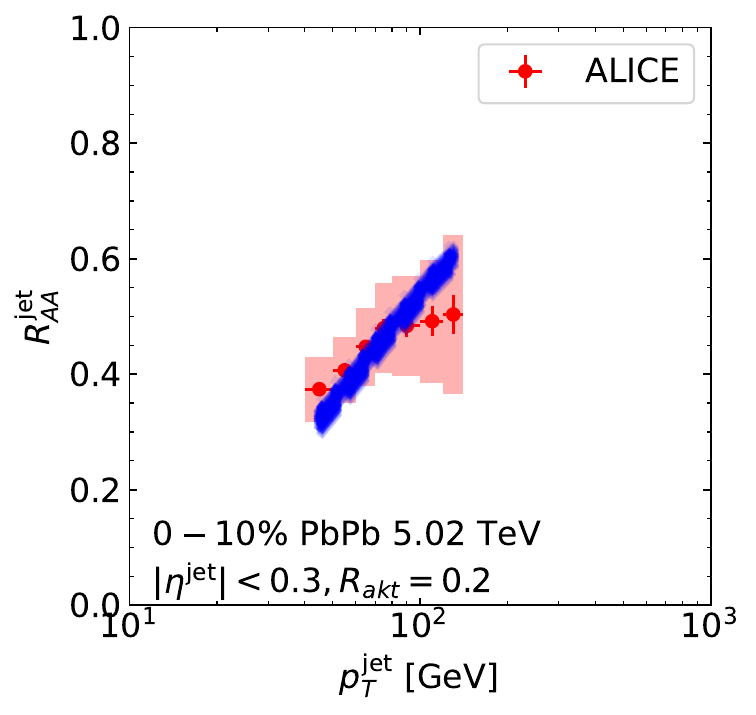}
     \includegraphics[width=0.28\textwidth,page=11]{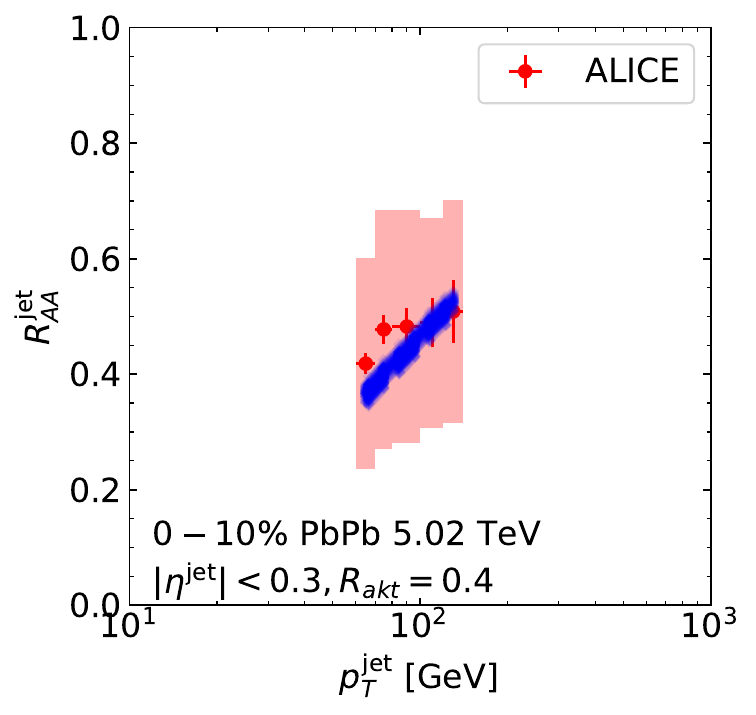}
     \caption{Posterior predictive distributions, the fitting performance of our model. Jet $R_{AA}$ from ALICE and CMS.}
     \label{fig:PosteriorPredictiveDistribution_2}
\end{figure*}

Figures~\ref{fig:PosteriorPredictiveDistribution_1}-\ref{fig:PosteriorPredictiveDistribution_2} show the corresponding posterior predictive distributions. These jet data were used to constrain the model parameters. Different rows correspond to different measurements. We observe good agreement with all data for both $R^{\rm jet}_{AA}$ and $v^{\rm jet}_2$, including all collision energies, centralities, rapidities, jet cone sizes, and momenta. There is a slight tension in $v^{\rm jet}_2$ for the lowest $p_T$ bins. This tension could be do to missing ingredients in the description of energy loss towards lower $p_T$. Dotted and dashed lines show the two alternative fits, where $\tau_{\min}$ is kept fixed. There are subtle differences between different fits, with the fixed earlier quenching reducing the jet $v_2$, as expected. 

\begin{figure*}
    \centering
     \includegraphics[width=0.28\textwidth,page=21]{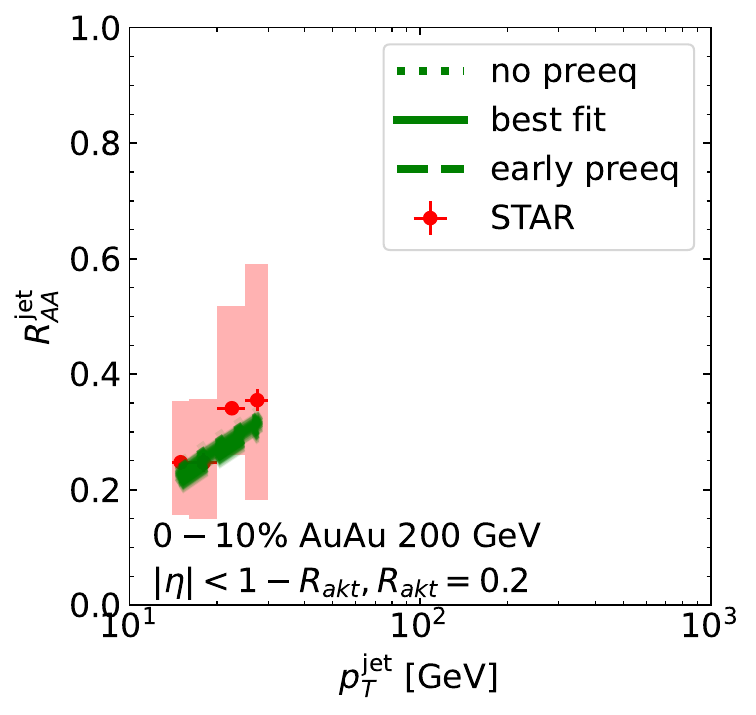}
     \includegraphics[width=0.28\textwidth,page=22]{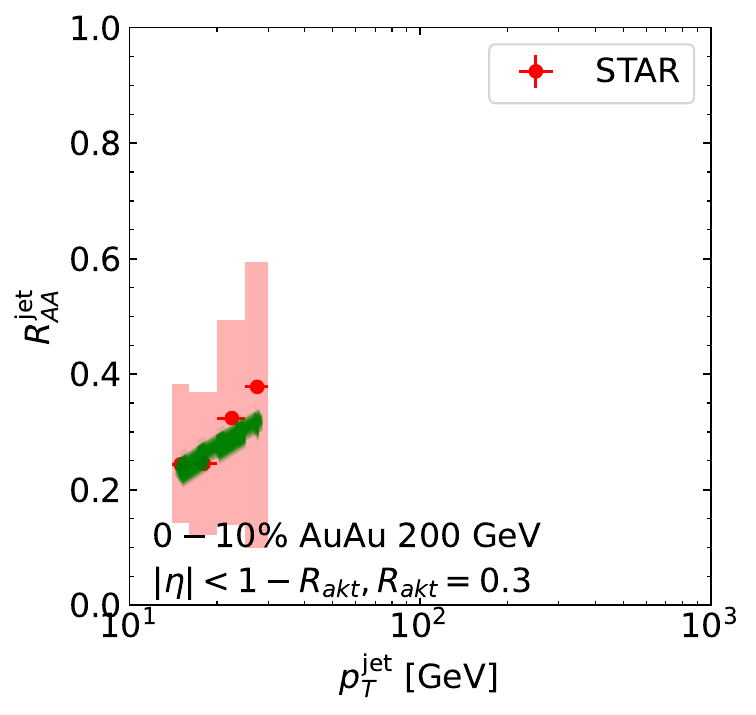}
     \includegraphics[width=0.28\textwidth,page=23]{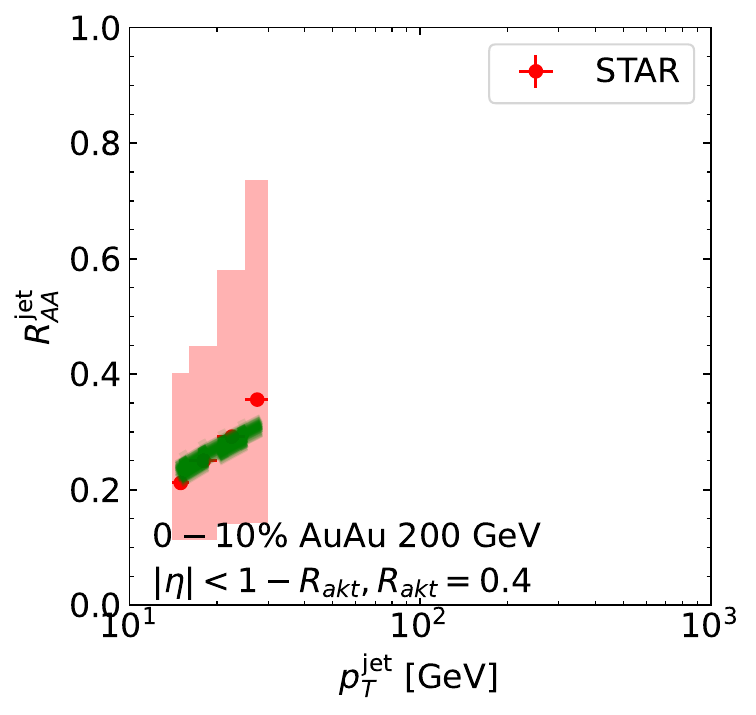}\\
     \includegraphics[width=0.28\textwidth,page=24]{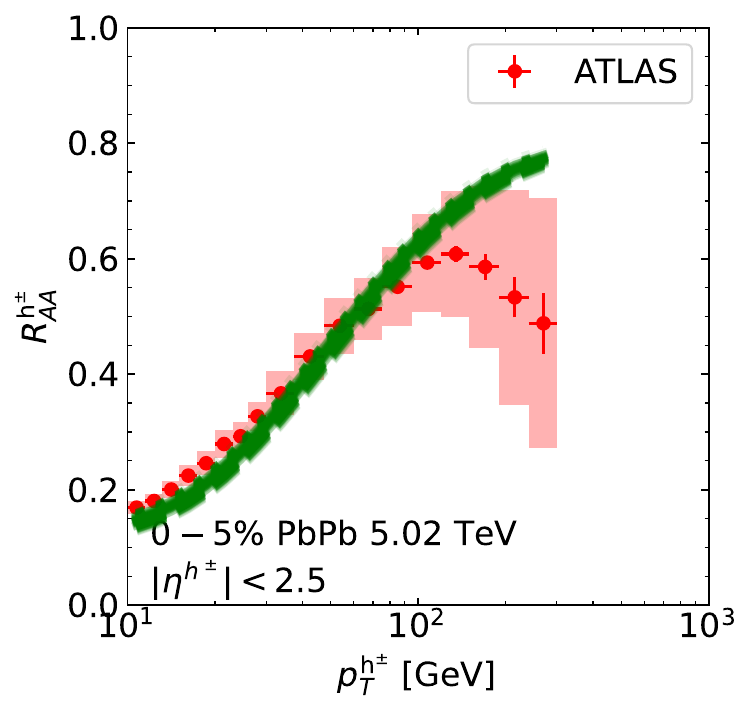}
     \includegraphics[width=0.28\textwidth,page=25]{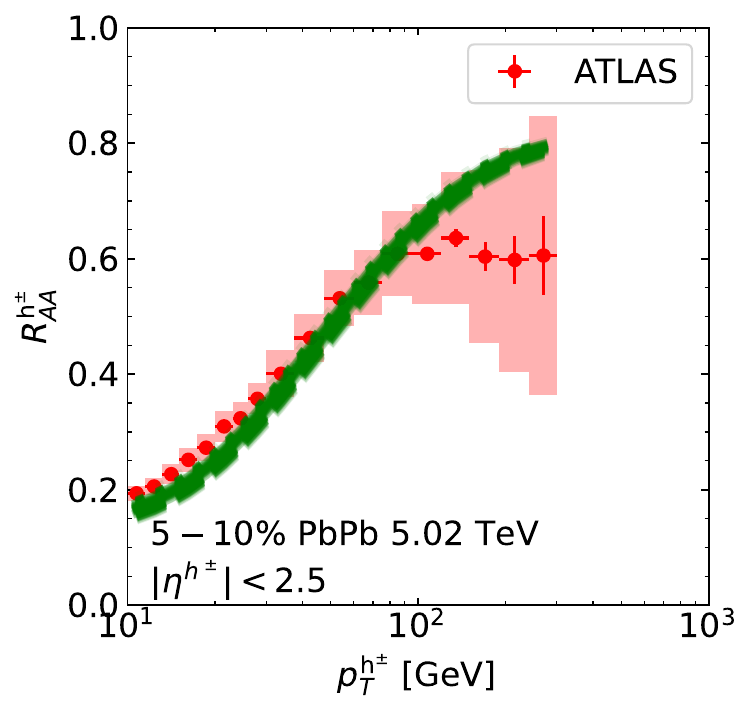}
     \includegraphics[width=0.28\textwidth,page=26]{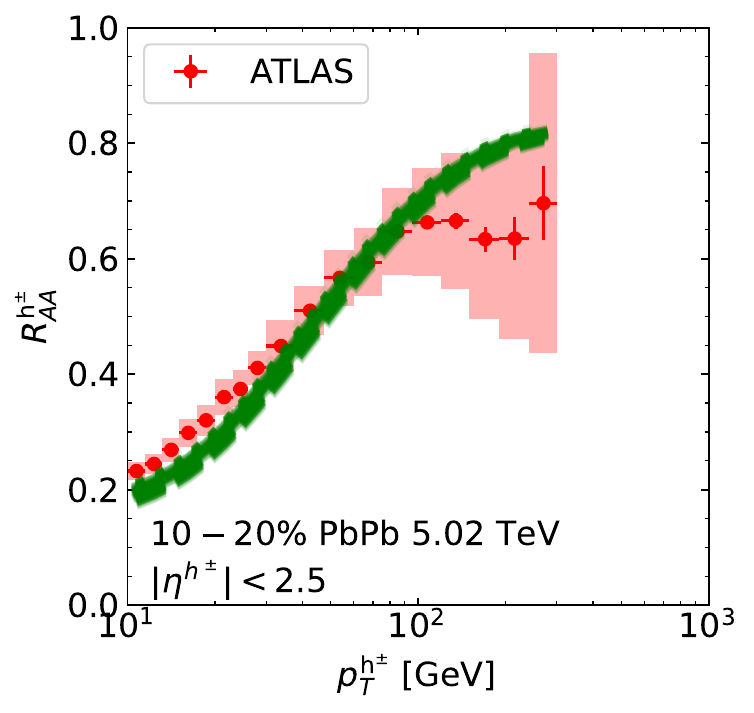}
     \includegraphics[width=0.28\textwidth,page=27]{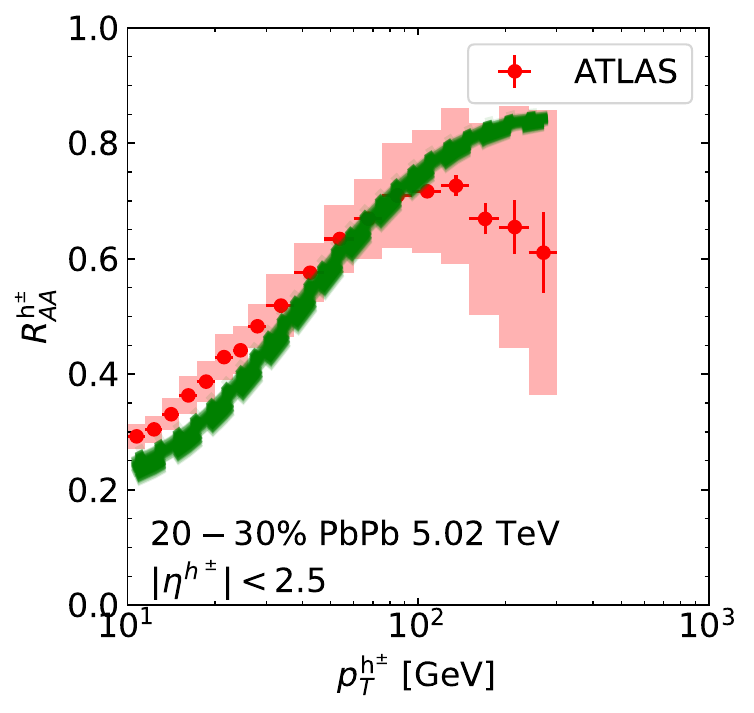}
     \includegraphics[width=0.28\textwidth,page=28]{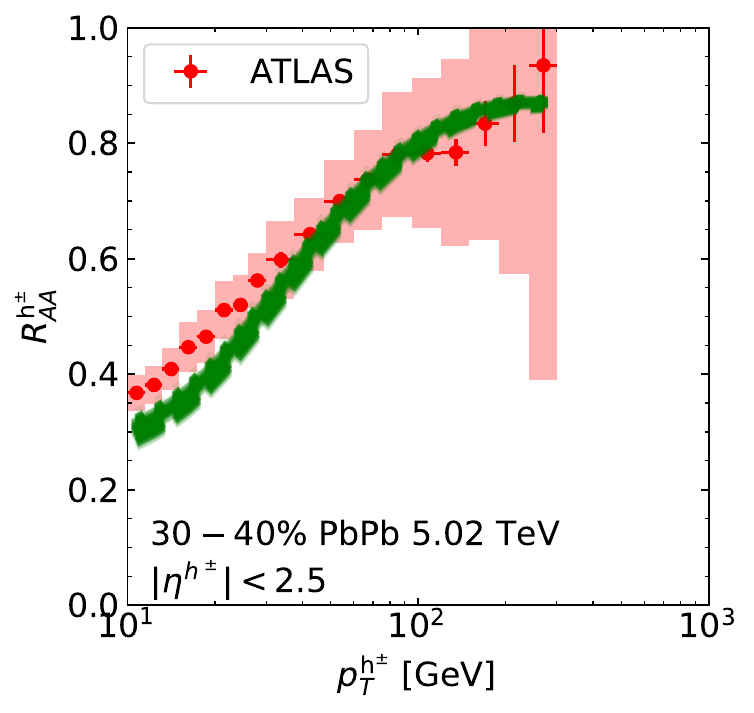}
     \includegraphics[width=0.28\textwidth,page=29]{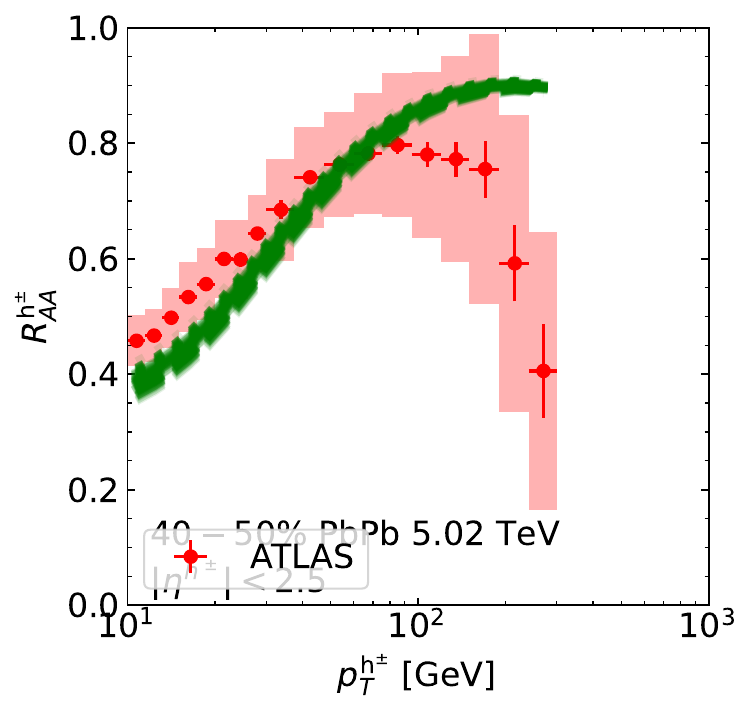}\\
     \includegraphics[width=0.28\textwidth,page=30]{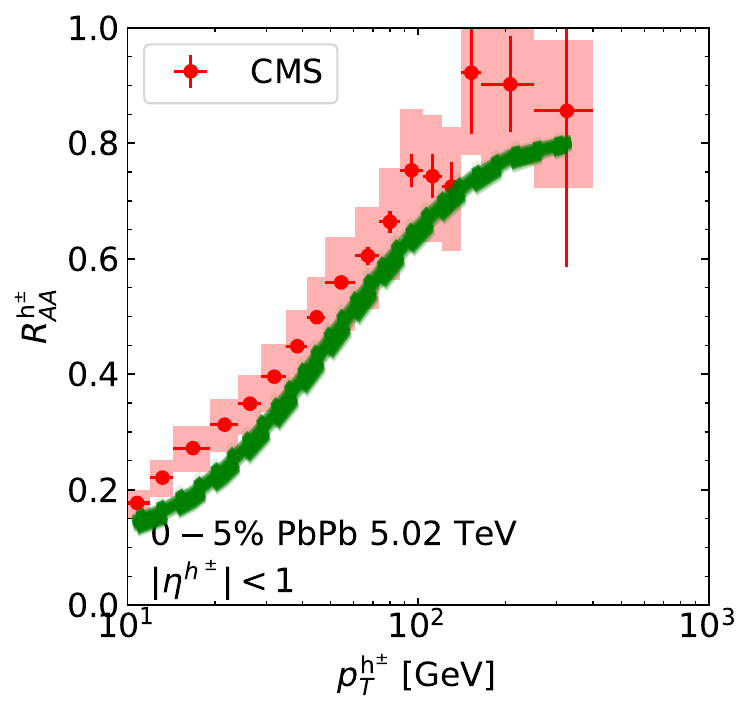}
     \includegraphics[width=0.28\textwidth,page=31]{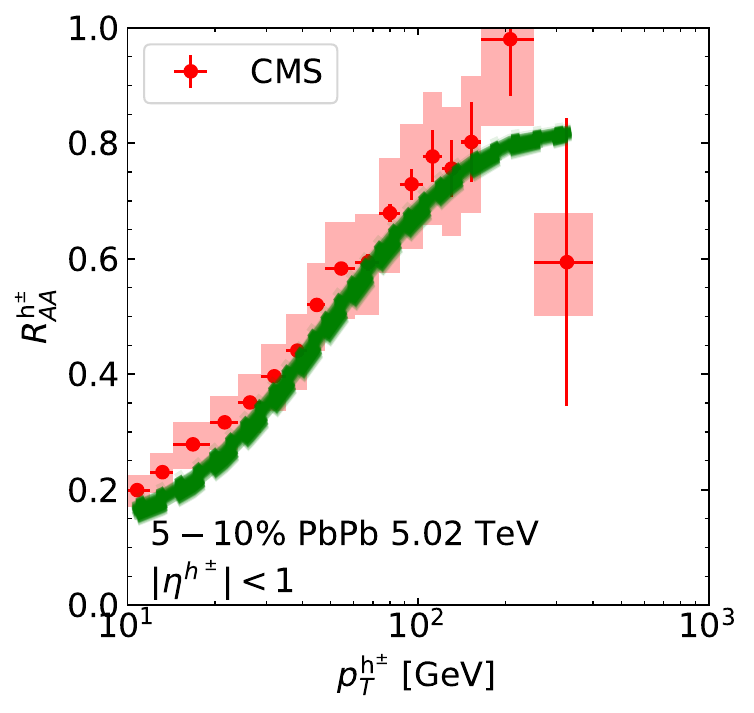}
     \includegraphics[width=0.28\textwidth,page=32]{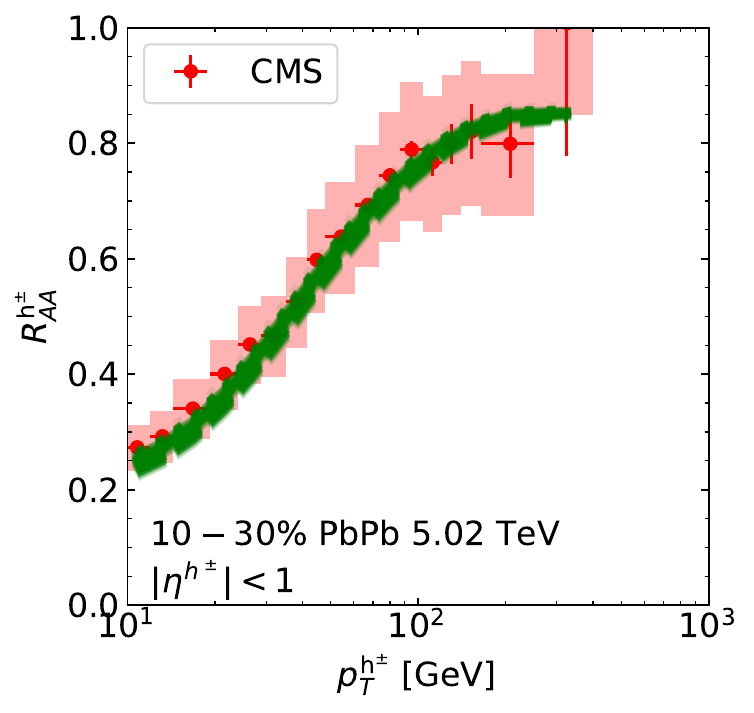}
     \includegraphics[width=0.28\textwidth,page=33]{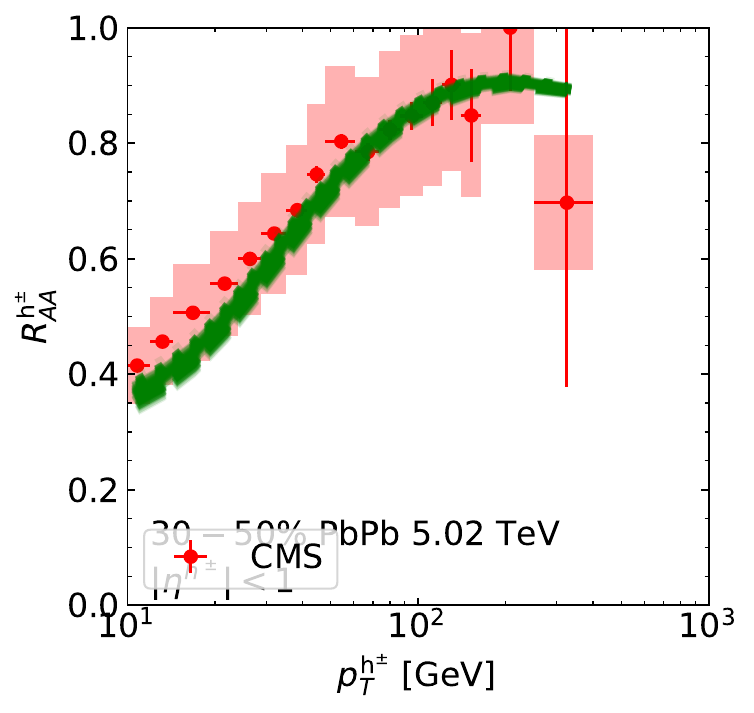}\\
     \caption{Posterior predictive distributions for RHIC energy and for hadrons (not included in the inference). Data shown for RHIC jets and charged hadrons across different LHC experiments.}
     \label{fig:PosteriorPredictiveDistribution_hadrons_1}
\end{figure*}

\begin{figure*}
    \centering
     \includegraphics[width=0.28\textwidth,page=34]{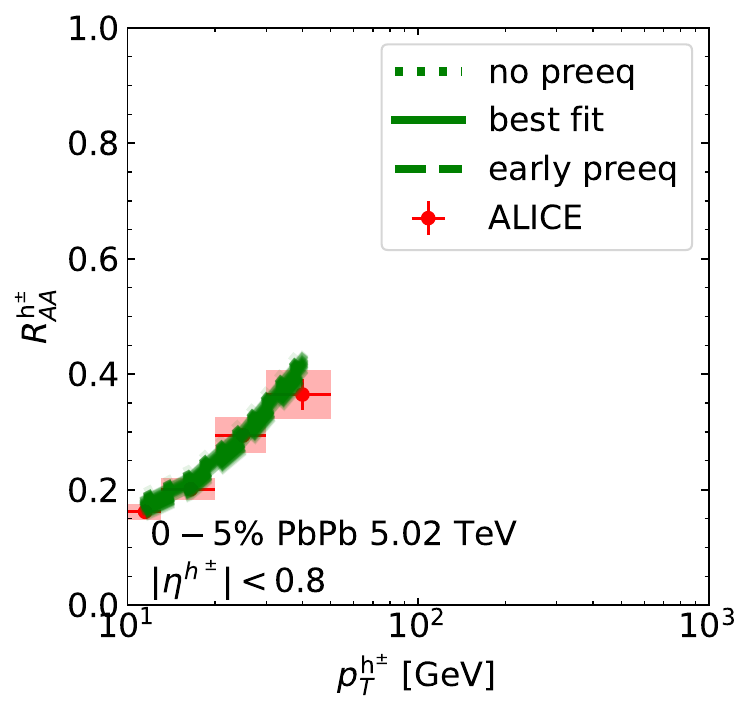}
     \includegraphics[width=0.28\textwidth,page=35]{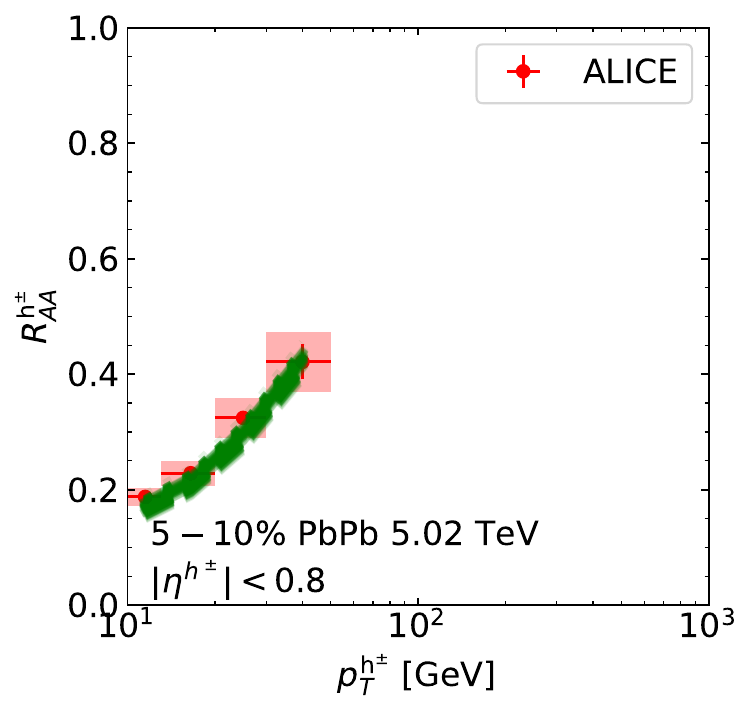}
     \includegraphics[width=0.28\textwidth,page=36]{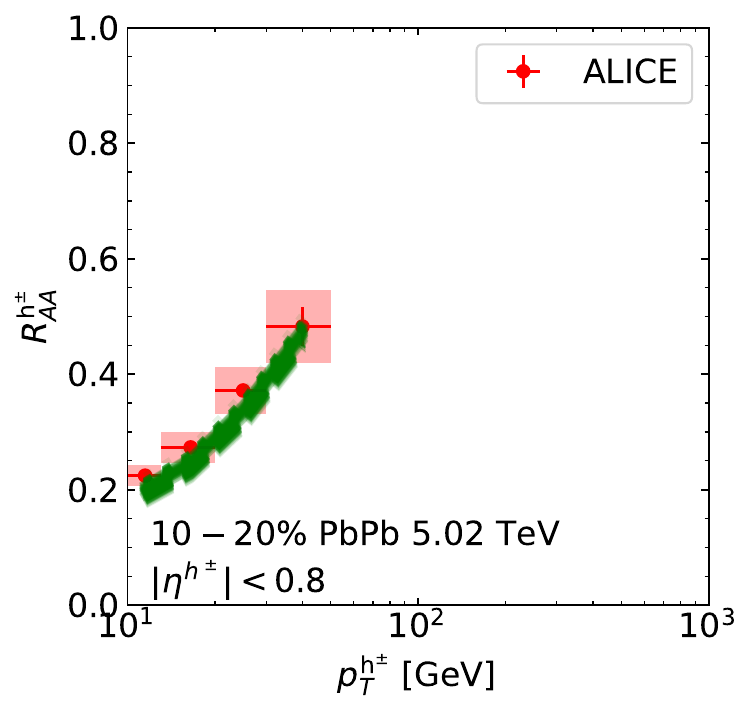}
     \includegraphics[width=0.28\textwidth,page=37]{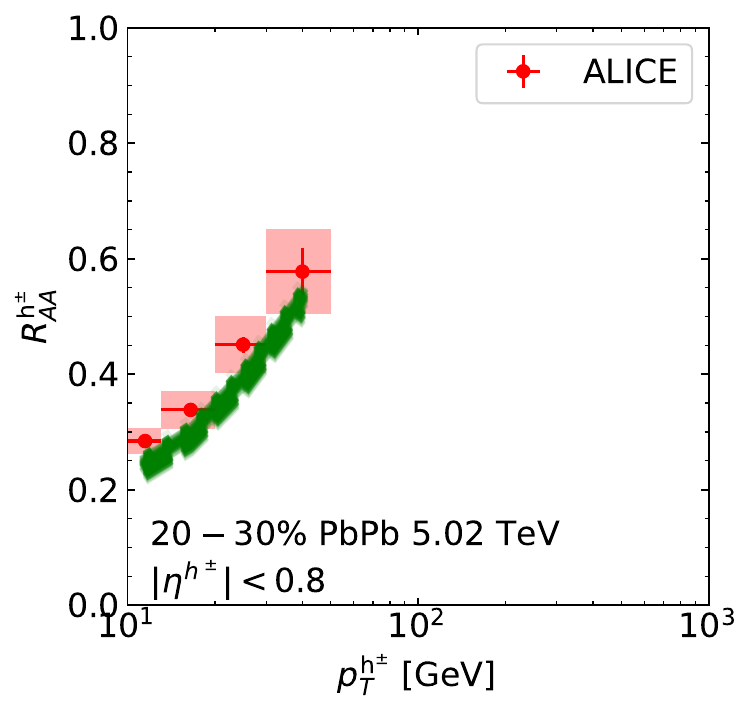}
     \includegraphics[width=0.28\textwidth,page=38]{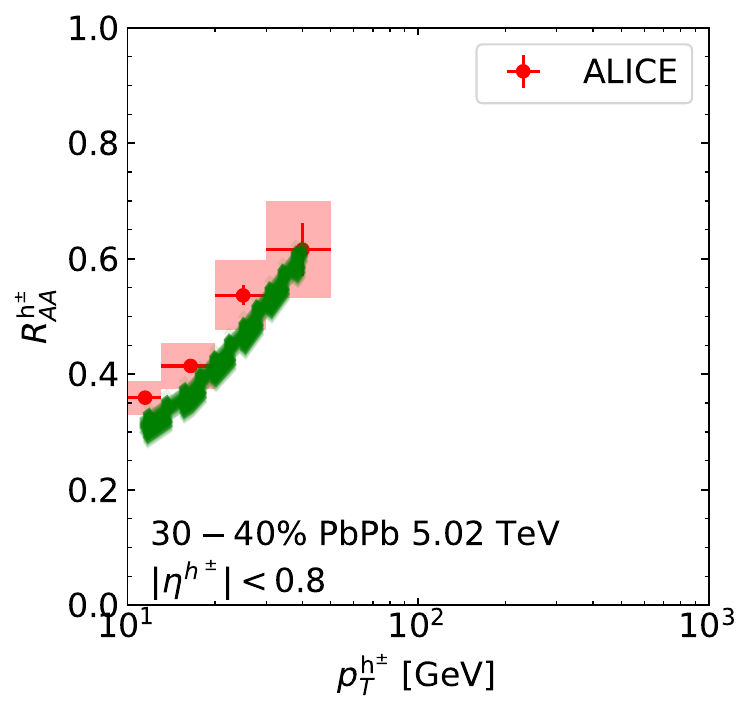}
     \includegraphics[width=0.28\textwidth,page=39]{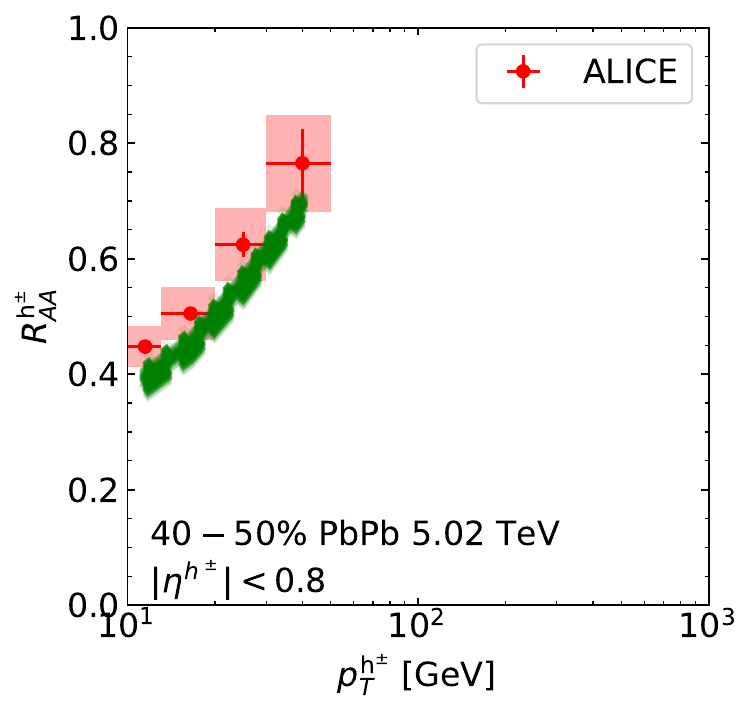}\\
     \includegraphics[width=0.28\textwidth,page=40]{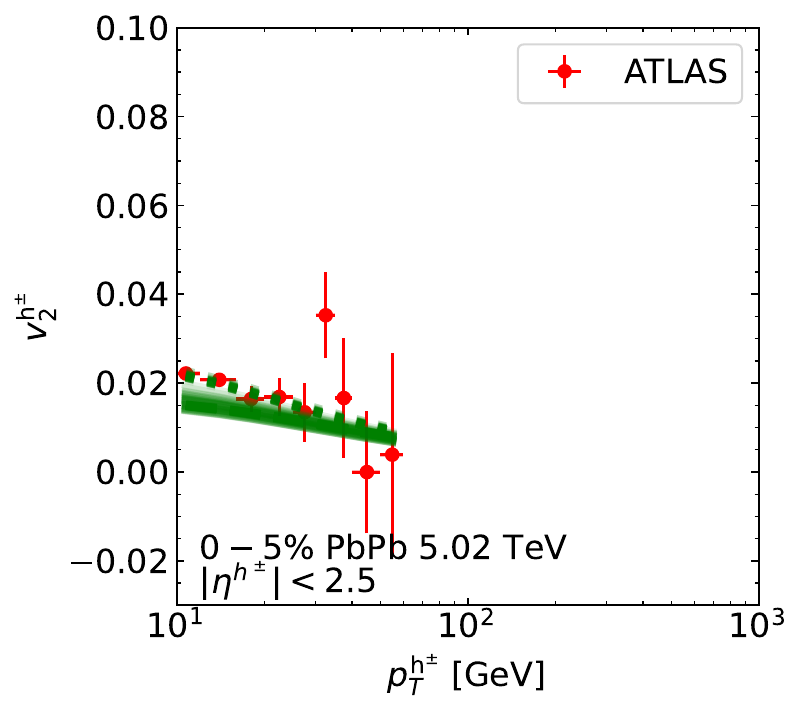}
     \includegraphics[width=0.28\textwidth,page=41]{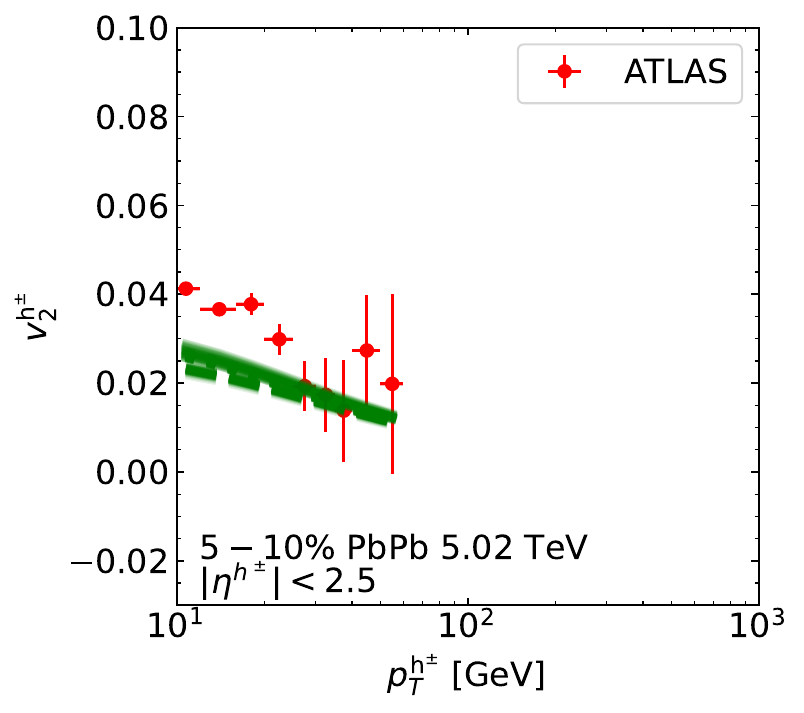}
     \includegraphics[width=0.28\textwidth,page=42]{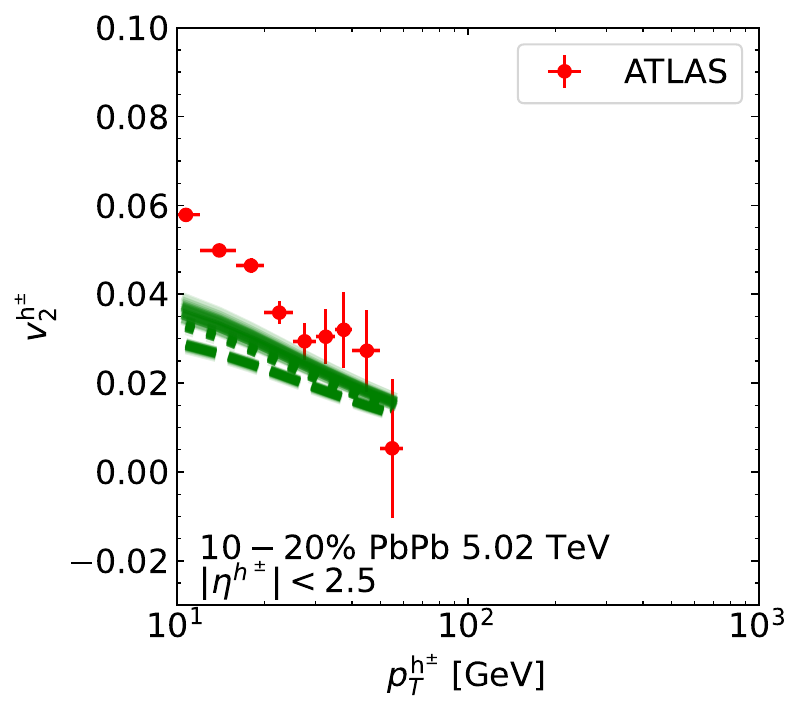}
     \includegraphics[width=0.28\textwidth,page=43]{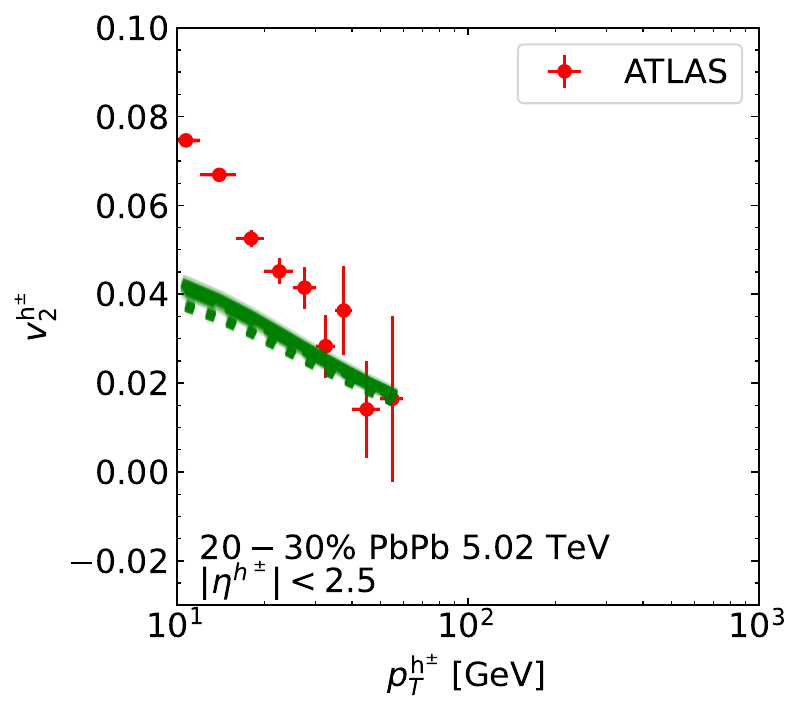}
     \includegraphics[width=0.28\textwidth,page=44]{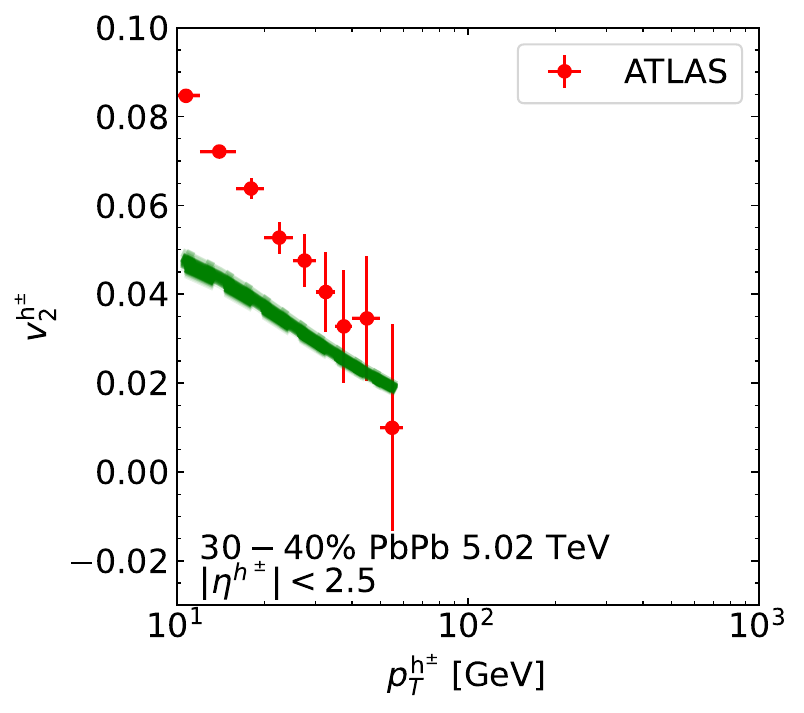}
     \includegraphics[width=0.28\textwidth,page=45]{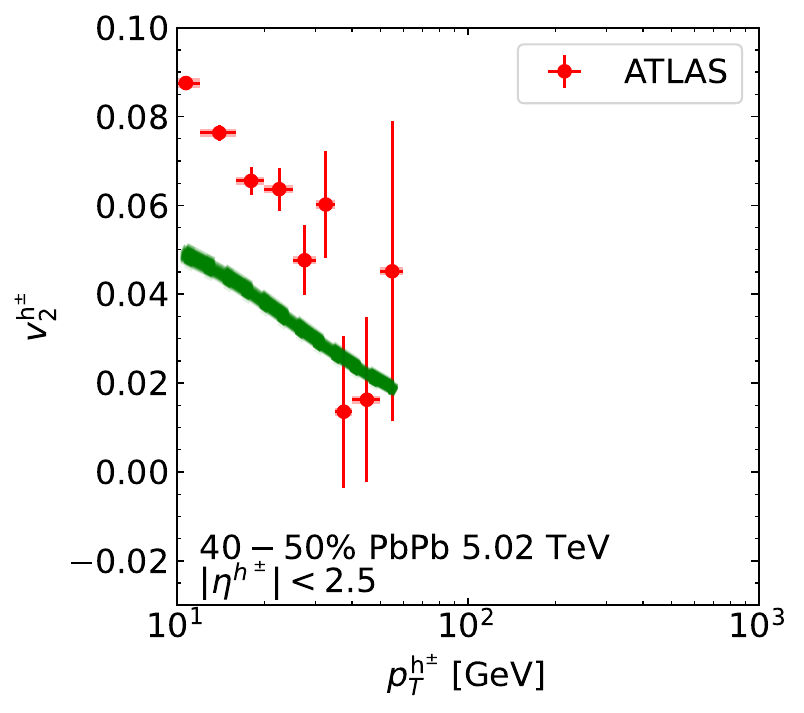}
     \caption{Posterior predictive distributions for RHIC energy and for hadrons (not included in the inference). Data shown for RHIC jets and charged hadrons across different LHC experiments.}
     \label{fig:PosteriorPredictiveDistribution_hadrons_2}
\end{figure*}

Finally, we also compare our results to jets at RHIC energies and charged hadron observables, shown in Figs.~\ref{fig:PosteriorPredictiveDistribution_hadrons_1}-\ref{fig:PosteriorPredictiveDistribution_hadrons_2}. As our framework was designed for jets, we did not include these data in the inference. We see an overall good agreement with both RHIC energies and charged hadron $R_{AA}$. Interestingly, hadron $R_{AA}$ seems to favor early quenching, as indicated by the dashed lines. There is a clear underestimation of the hadron elliptic flow at low momenta, consistent with previous findings from other theory collaborations, likely linked to the (milder) tension with low $p_T$ jets mentioned above. 

\section{OO predictions without pre-equilibrium quenching}
\label{app:Oxygen_early_vs_late}
\begin{figure*}
    \centering
     \includegraphics[width=0.47\linewidth]{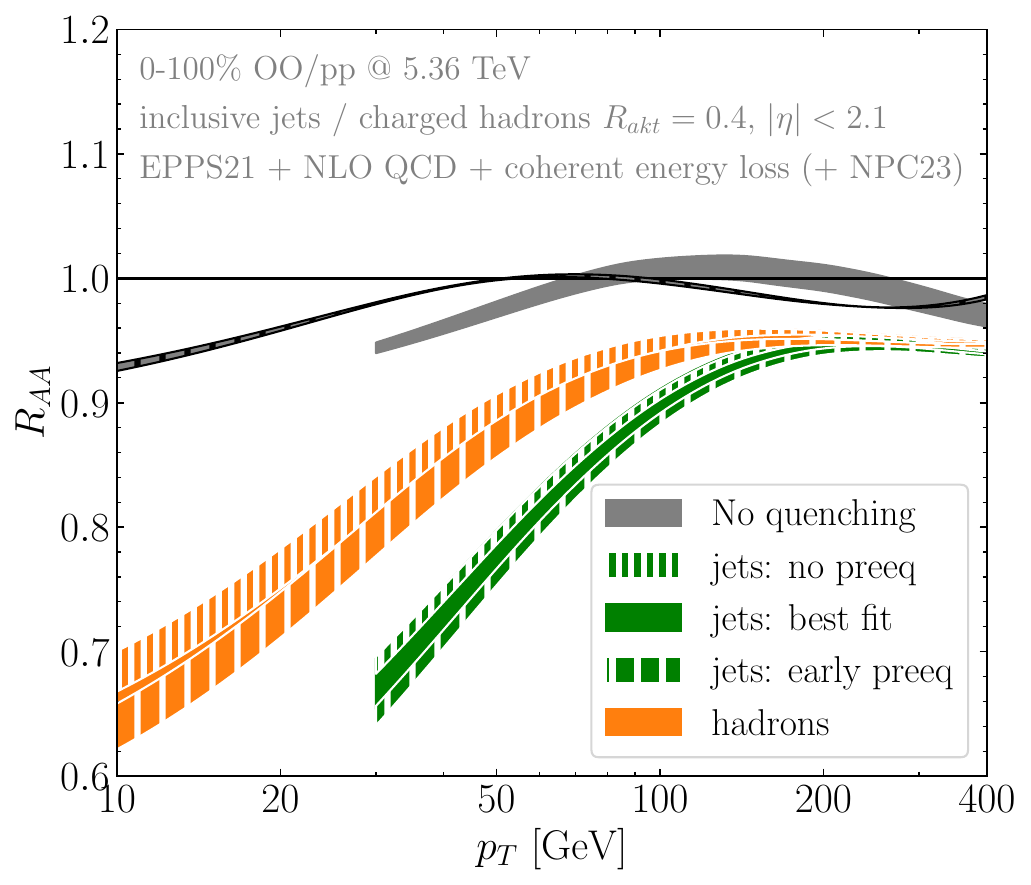} 
     \includegraphics[width=0.47\linewidth]{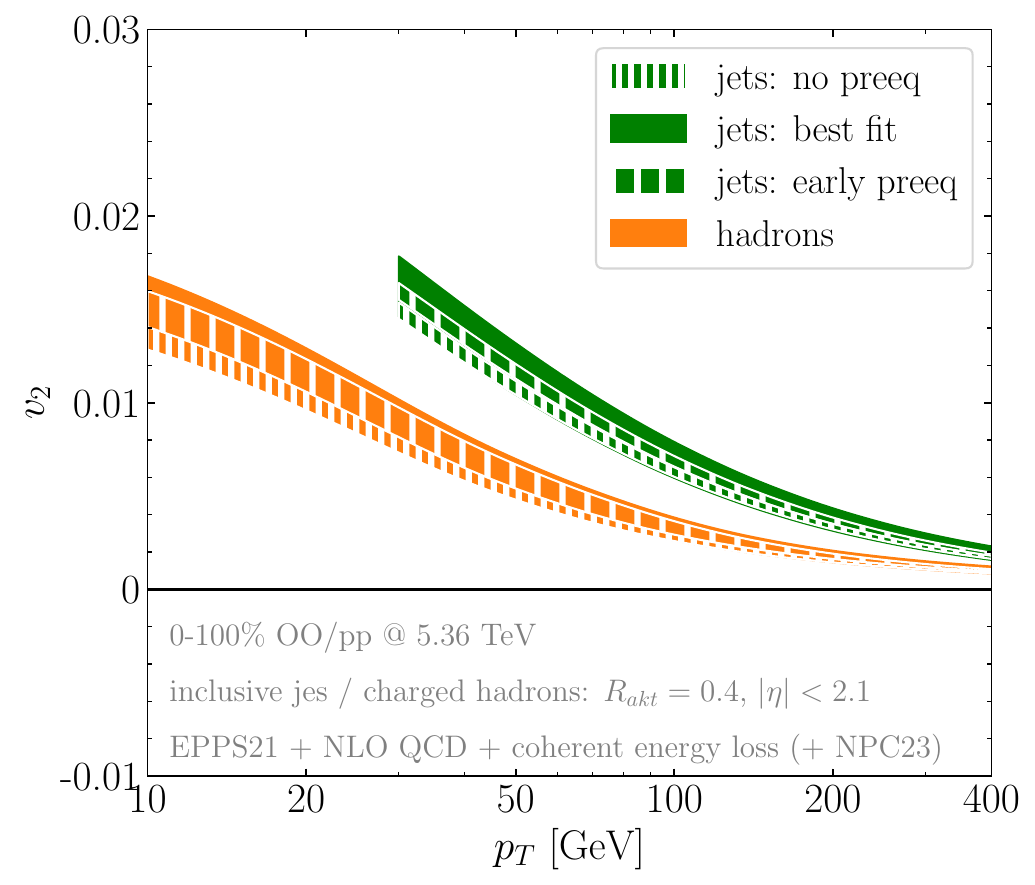} 
     \caption{Nuclear modification and elliptic flow for jets and charged hadrons in minimum bias OO collisions. Different lines corresponds to different limiting scenarios in the pre-equilibrium quenching. The bands represent the 68\% confidence interval for quenching parameters. }
     \label{fig:OO_early_vs_late}
\end{figure*}

Figure~\ref{fig:OO_early_vs_late} shows the nuclear modification and elliptic flow for minimum bias OO collisions. Curves represent the limiting scenarios tested in App.~\ref{app:Bayesian} regarding pre-equilibrium quenching: ``early preeq'' denotes $\tau_{\min}=0.08$ fm, ``no preeq'' $\tau_{\min}=0.4$ fm, with their corresponding inferred $g_{\rm med}$. The best fit denotes our default setting shown in Fig.~\ref{fig:ParameterPosterior}. For simplicity, we don't show nPDF uncertainties. Slight differences are present in the different limiting scenarios. Interestingly, an earlier quenching can result in a relatively larger $v_2$, in contrast to larger systems, likely reflecting the more important role of fluctuations over average geometry in smaller systems.

\bibliography{refs_vac.bib,refs_med.bib,refs_Dani.bib}

\begin{thebibliography}{141}%
\makeatletter
\providecommand \@ifxundefined [1]{%
 \@ifx{#1\undefined}
}%
\providecommand \@ifnum [1]{%
 \ifnum #1\expandafter \@firstoftwo
 \else \expandafter \@secondoftwo
 \fi
}%
\providecommand \@ifx [1]{%
 \ifx #1\expandafter \@firstoftwo
 \else \expandafter \@secondoftwo
 \fi
}%
\providecommand \natexlab [1]{#1}%
\providecommand \enquote  [1]{``#1''}%
\providecommand \bibnamefont  [1]{#1}%
\providecommand \bibfnamefont [1]{#1}%
\providecommand \citenamefont [1]{#1}%
\providecommand \href@noop [0]{\@secondoftwo}%
\providecommand \href [0]{\begingroup \@sanitize@url \@href}%
\providecommand \@href[1]{\@@startlink{#1}\@@href}%
\providecommand \@@href[1]{\endgroup#1\@@endlink}%
\providecommand \@sanitize@url [0]{\catcode `\\12\catcode `\$12\catcode
  `\&12\catcode `\#12\catcode `\^12\catcode `\_12\catcode `\%12\relax}%
\providecommand \@@startlink[1]{}%
\providecommand \@@endlink[0]{}%
\providecommand \url  [0]{\begingroup\@sanitize@url \@url }%
\providecommand \@url [1]{\endgroup\@href {#1}{\urlprefix }}%
\providecommand \urlprefix  [0]{URL }%
\providecommand \Eprint [0]{\href }%
\providecommand \doibase [0]{https://doi.org/}%
\providecommand \selectlanguage [0]{\@gobble}%
\providecommand \bibinfo  [0]{\@secondoftwo}%
\providecommand \bibfield  [0]{\@secondoftwo}%
\providecommand \translation [1]{[#1]}%
\providecommand \BibitemOpen [0]{}%
\providecommand \bibitemStop [0]{}%
\providecommand \bibitemNoStop [0]{.\EOS\space}%
\providecommand \EOS [0]{\spacefactor3000\relax}%
\providecommand \BibitemShut  [1]{\csname bibitem#1\endcsname}%
\let\auto@bib@innerbib\@empty
\bibitem [{\citenamefont {Khachatryan}\ \emph {et~al.}(2010)\citenamefont
  {Khachatryan} \emph {et~al.}}]{CMS:2010ifv}%
  \BibitemOpen
  \bibfield  {author} {\bibinfo {author} {\bibfnamefont {V.}~\bibnamefont
  {Khachatryan}} \emph {et~al.} (\bibinfo {collaboration} {CMS}),\ }\bibfield
  {title} {\bibinfo {title} {{Observation of Long-Range Near-Side Angular
  Correlations in Proton-Proton Collisions at the LHC}},\ }\href
  {https://doi.org/10.1007/JHEP09(2010)091} {\bibfield  {journal} {\bibinfo
  {journal} {JHEP}\ }\textbf {\bibinfo {volume} {09}},\ \bibinfo {pages}
  {091}},\ \Eprint {https://arxiv.org/abs/1009.4122} {arXiv:1009.4122 [hep-ex]}
  \BibitemShut {NoStop}%
\bibitem [{\citenamefont {Aad}\ \emph {et~al.}(2013)\citenamefont {Aad} \emph
  {et~al.}}]{ATLAS:2012cix}%
  \BibitemOpen
  \bibfield  {author} {\bibinfo {author} {\bibfnamefont {G.}~\bibnamefont
  {Aad}} \emph {et~al.} (\bibinfo {collaboration} {ATLAS}),\ }\bibfield
  {title} {\bibinfo {title} {{Observation of Associated Near-Side and Away-Side
  Long-Range Correlations in $\sqrt{s_{NN}}$=5.02 TeV Proton-Lead Collisions
  with the ATLAS Detector}},\ }\href
  {https://doi.org/10.1103/PhysRevLett.110.182302} {\bibfield  {journal}
  {\bibinfo  {journal} {Phys. Rev. Lett.}\ }\textbf {\bibinfo {volume} {110}},\
  \bibinfo {pages} {182302} (\bibinfo {year} {2013})},\ \Eprint
  {https://arxiv.org/abs/1212.5198} {arXiv:1212.5198 [hep-ex]} \BibitemShut
  {NoStop}%
\bibitem [{\citenamefont {Abelev}\ \emph {et~al.}(2013)\citenamefont {Abelev}
  \emph {et~al.}}]{ALICE:2012eyl}%
  \BibitemOpen
  \bibfield  {author} {\bibinfo {author} {\bibfnamefont {B.}~\bibnamefont
  {Abelev}} \emph {et~al.} (\bibinfo {collaboration} {ALICE}),\ }\bibfield
  {title} {\bibinfo {title} {{Long-range angular correlations on the near and
  away side in $p$-Pb collisions at $\sqrt{s_{NN}}=5.02$ TeV}},\ }\href
  {https://doi.org/10.1016/j.physletb.2013.01.012} {\bibfield  {journal}
  {\bibinfo  {journal} {Phys. Lett. B}\ }\textbf {\bibinfo {volume} {719}},\
  \bibinfo {pages} {29} (\bibinfo {year} {2013})},\ \Eprint
  {https://arxiv.org/abs/1212.2001} {arXiv:1212.2001 [nucl-ex]} \BibitemShut
  {NoStop}%
\bibitem [{\citenamefont {Chatrchyan}\ \emph {et~al.}(2013)\citenamefont
  {Chatrchyan} \emph {et~al.}}]{CMS:2012qk}%
  \BibitemOpen
  \bibfield  {author} {\bibinfo {author} {\bibfnamefont {S.}~\bibnamefont
  {Chatrchyan}} \emph {et~al.} (\bibinfo {collaboration} {CMS}),\ }\bibfield
  {title} {\bibinfo {title} {{Observation of Long-Range Near-Side Angular
  Correlations in Proton-Lead Collisions at the LHC}},\ }\href
  {https://doi.org/10.1016/j.physletb.2012.11.025} {\bibfield  {journal}
  {\bibinfo  {journal} {Phys. Lett. B}\ }\textbf {\bibinfo {volume} {718}},\
  \bibinfo {pages} {795} (\bibinfo {year} {2013})},\ \Eprint
  {https://arxiv.org/abs/1210.5482} {arXiv:1210.5482 [nucl-ex]} \BibitemShut
  {NoStop}%
\bibitem [{\citenamefont {Nagle}\ and\ \citenamefont
  {Zajc}(2018)}]{Nagle:2018nvi}%
  \BibitemOpen
  \bibfield  {author} {\bibinfo {author} {\bibfnamefont {J.~L.}\ \bibnamefont
  {Nagle}}\ and\ \bibinfo {author} {\bibfnamefont {W.~A.}\ \bibnamefont
  {Zajc}},\ }\bibfield  {title} {\bibinfo {title} {{Small System Collectivity
  in Relativistic Hadronic and Nuclear Collisions}},\ }\href
  {https://doi.org/10.1146/annurev-nucl-101916-123209} {\bibfield  {journal}
  {\bibinfo  {journal} {Ann. Rev. Nucl. Part. Sci.}\ }\textbf {\bibinfo
  {volume} {68}},\ \bibinfo {pages} {211} (\bibinfo {year} {2018})},\ \Eprint
  {https://arxiv.org/abs/1801.03477} {arXiv:1801.03477 [nucl-ex]} \BibitemShut
  {NoStop}%
\bibitem [{\citenamefont {Noronha}\ \emph {et~al.}(2024)\citenamefont
  {Noronha}, \citenamefont {Schenke}, \citenamefont {Shen},\ and\ \citenamefont
  {Zhao}}]{Noronha:2024dtq}%
  \BibitemOpen
  \bibfield  {author} {\bibinfo {author} {\bibfnamefont {J.}~\bibnamefont
  {Noronha}}, \bibinfo {author} {\bibfnamefont {B.}~\bibnamefont {Schenke}},
  \bibinfo {author} {\bibfnamefont {C.}~\bibnamefont {Shen}},\ and\ \bibinfo
  {author} {\bibfnamefont {W.}~\bibnamefont {Zhao}},\ }\bibfield  {title}
  {\bibinfo {title} {{Progress and challenges in small systems}},\ }\href
  {https://doi.org/10.1142/9789811294679_0004} {\bibfield  {journal} {\bibinfo
  {journal} {Int. J. Mod. Phys. E}\ }\textbf {\bibinfo {volume} {33}},\
  \bibinfo {pages} {2430005} (\bibinfo {year} {2024})},\ \Eprint
  {https://arxiv.org/abs/2401.09208} {arXiv:2401.09208 [nucl-th]} \BibitemShut
  {NoStop}%
\bibitem [{\citenamefont {Grosse-Oetringhaus}\ and\ \citenamefont
  {Wiedemann}(2024)}]{Grosse-Oetringhaus:2024bwr}%
  \BibitemOpen
  \bibfield  {author} {\bibinfo {author} {\bibfnamefont {J.~F.}\ \bibnamefont
  {Grosse-Oetringhaus}}\ and\ \bibinfo {author} {\bibfnamefont {U.~A.}\
  \bibnamefont {Wiedemann}},\ }\bibfield  {title} {\bibinfo {title} {{A Decade
  of Collectivity in Small Systems}},\ }\href@noop {} {\  (\bibinfo {year}
  {2024})},\ \Eprint {https://arxiv.org/abs/2407.07484} {arXiv:2407.07484
  [hep-ex]} \BibitemShut {NoStop}%
\bibitem [{\citenamefont {Weller}\ and\ \citenamefont
  {Romatschke}(2017)}]{Weller:2017tsr}%
  \BibitemOpen
  \bibfield  {author} {\bibinfo {author} {\bibfnamefont {R.~D.}\ \bibnamefont
  {Weller}}\ and\ \bibinfo {author} {\bibfnamefont {P.}~\bibnamefont
  {Romatschke}},\ }\bibfield  {title} {\bibinfo {title} {{One fluid to rule
  them all: viscous hydrodynamic description of event-by-event central p+p,
  p+Pb and Pb+Pb collisions at $\sqrt{s}=5.02$ TeV}},\ }\href
  {https://doi.org/10.1016/j.physletb.2017.09.077} {\bibfield  {journal}
  {\bibinfo  {journal} {Phys. Lett. B}\ }\textbf {\bibinfo {volume} {774}},\
  \bibinfo {pages} {351} (\bibinfo {year} {2017})},\ \Eprint
  {https://arxiv.org/abs/1701.07145} {arXiv:1701.07145 [nucl-th]} \BibitemShut
  {NoStop}%
\bibitem [{\citenamefont {Aidala}\ \emph {et~al.}(2019)\citenamefont {Aidala}
  \emph {et~al.}}]{PHENIX:2018lia}%
  \BibitemOpen
  \bibfield  {author} {\bibinfo {author} {\bibfnamefont {C.}~\bibnamefont
  {Aidala}} \emph {et~al.} (\bibinfo {collaboration} {PHENIX}),\ }\bibfield
  {title} {\bibinfo {title} {{Creation of quark{\textendash}gluon plasma
  droplets with three distinct geometries}},\ }\href
  {https://doi.org/10.1038/s41567-018-0360-0} {\bibfield  {journal} {\bibinfo
  {journal} {Nature Phys.}\ }\textbf {\bibinfo {volume} {15}},\ \bibinfo
  {pages} {214} (\bibinfo {year} {2019})},\ \Eprint
  {https://arxiv.org/abs/1805.02973} {arXiv:1805.02973 [nucl-ex]} \BibitemShut
  {NoStop}%
\bibitem [{\citenamefont {Acharya}\ \emph {et~al.}(2019)\citenamefont {Acharya}
  \emph {et~al.}}]{ALICE:2019zfl}%
  \BibitemOpen
  \bibfield  {author} {\bibinfo {author} {\bibfnamefont {S.}~\bibnamefont
  {Acharya}} \emph {et~al.} (\bibinfo {collaboration} {ALICE}),\ }\bibfield
  {title} {\bibinfo {title} {{Investigations of Anisotropic Flow Using
  Multiparticle Azimuthal Correlations in pp, p-Pb, Xe-Xe, and Pb-Pb Collisions
  at the LHC}},\ }\href {https://doi.org/10.1103/PhysRevLett.123.142301}
  {\bibfield  {journal} {\bibinfo  {journal} {Phys. Rev. Lett.}\ }\textbf
  {\bibinfo {volume} {123}},\ \bibinfo {pages} {142301} (\bibinfo {year}
  {2019})},\ \Eprint {https://arxiv.org/abs/1903.01790} {arXiv:1903.01790
  [nucl-ex]} \BibitemShut {NoStop}%
\bibitem [{\citenamefont {Nijs}\ \emph {et~al.}(2021)\citenamefont {Nijs},
  \citenamefont {van~der Schee}, \citenamefont {G{\"u}rsoy},\ and\
  \citenamefont {Snellings}}]{Nijs:2020roc}%
  \BibitemOpen
  \bibfield  {author} {\bibinfo {author} {\bibfnamefont {G.}~\bibnamefont
  {Nijs}}, \bibinfo {author} {\bibfnamefont {W.}~\bibnamefont {van~der Schee}},
  \bibinfo {author} {\bibfnamefont {U.}~\bibnamefont {G{\"u}rsoy}},\ and\
  \bibinfo {author} {\bibfnamefont {R.}~\bibnamefont {Snellings}},\ }\bibfield
  {title} {\bibinfo {title} {{Bayesian analysis of heavy ion collisions with
  the heavy ion computational framework Trajectum}},\ }\href
  {https://doi.org/10.1103/PhysRevC.103.054909} {\bibfield  {journal} {\bibinfo
   {journal} {Phys. Rev. C}\ }\textbf {\bibinfo {volume} {103}},\ \bibinfo
  {pages} {054909} (\bibinfo {year} {2021})},\ \Eprint
  {https://arxiv.org/abs/2010.15134} {arXiv:2010.15134 [nucl-th]} \BibitemShut
  {NoStop}%
\bibitem [{\citenamefont {Bierlich}\ \emph {et~al.}(2018)\citenamefont
  {Bierlich}, \citenamefont {Gustafson}, \citenamefont {L\"onnblad},\ and\
  \citenamefont {Shah}}]{Bierlich:2018xfw}%
  \BibitemOpen
  \bibfield  {author} {\bibinfo {author} {\bibfnamefont {C.}~\bibnamefont
  {Bierlich}}, \bibinfo {author} {\bibfnamefont {G.}~\bibnamefont {Gustafson}},
  \bibinfo {author} {\bibfnamefont {L.}~\bibnamefont {L\"onnblad}},\ and\
  \bibinfo {author} {\bibfnamefont {H.}~\bibnamefont {Shah}},\ }\bibfield
  {title} {\bibinfo {title} {{The Angantyr model for Heavy-Ion Collisions in
  PYTHIA8}},\ }\href {https://doi.org/10.1007/JHEP10(2018)134} {\bibfield
  {journal} {\bibinfo  {journal} {JHEP}\ }\textbf {\bibinfo {volume} {10}},\
  \bibinfo {pages} {134}},\ \Eprint {https://arxiv.org/abs/1806.10820}
  {arXiv:1806.10820 [hep-ph]} \BibitemShut {NoStop}%
\bibitem [{\citenamefont {Kurkela}\ \emph {et~al.}(2021)\citenamefont
  {Kurkela}, \citenamefont {Mazeliauskas},\ and\ \citenamefont
  {T\"ornkvist}}]{Kurkela:2021ctp}%
  \BibitemOpen
  \bibfield  {author} {\bibinfo {author} {\bibfnamefont {A.}~\bibnamefont
  {Kurkela}}, \bibinfo {author} {\bibfnamefont {A.}~\bibnamefont
  {Mazeliauskas}},\ and\ \bibinfo {author} {\bibfnamefont {R.}~\bibnamefont
  {T\"ornkvist}},\ }\bibfield  {title} {\bibinfo {title} {{Collective flow in
  single-hit QCD kinetic theory}},\ }\href
  {https://doi.org/10.1007/JHEP11(2021)216} {\bibfield  {journal} {\bibinfo
  {journal} {JHEP}\ }\textbf {\bibinfo {volume} {11}},\ \bibinfo {pages}
  {216}},\ \Eprint {https://arxiv.org/abs/2104.08179} {arXiv:2104.08179
  [hep-ph]} \BibitemShut {NoStop}%
\bibitem [{\citenamefont {Bierlich}\ \emph {et~al.}(2024)\citenamefont
  {Bierlich}, \citenamefont {Christiansen}, \citenamefont {Gustafson},
  \citenamefont {L\"onnblad}, \citenamefont {T\"ornkvist},\ and\ \citenamefont
  {Zapp}}]{Bierlich:2024lmb}%
  \BibitemOpen
  \bibfield  {author} {\bibinfo {author} {\bibfnamefont {C.}~\bibnamefont
  {Bierlich}}, \bibinfo {author} {\bibfnamefont {P.}~\bibnamefont
  {Christiansen}}, \bibinfo {author} {\bibfnamefont {G.}~\bibnamefont
  {Gustafson}}, \bibinfo {author} {\bibfnamefont {L.}~\bibnamefont
  {L\"onnblad}}, \bibinfo {author} {\bibfnamefont {R.}~\bibnamefont
  {T\"ornkvist}},\ and\ \bibinfo {author} {\bibfnamefont {K.}~\bibnamefont
  {Zapp}},\ }\bibfield  {title} {\bibinfo {title} {{Going against the flow:
  Revealing the QCD degrees of freedom in hadronic collisions}},\ }\href@noop
  {} {\  (\bibinfo {year} {2024})},\ \Eprint {https://arxiv.org/abs/2409.16093}
  {arXiv:2409.16093 [hep-ph]} \BibitemShut {NoStop}%
\bibitem [{\citenamefont {Torres}\ \emph {et~al.}(2024)\citenamefont {Torres},
  \citenamefont {Feng},\ and\ \citenamefont {Wang}}]{Torres:2024rgw}%
  \BibitemOpen
  \bibfield  {author} {\bibinfo {author} {\bibfnamefont {M.~S.}\ \bibnamefont
  {Torres}}, \bibinfo {author} {\bibfnamefont {Y.}~\bibnamefont {Feng}},\ and\
  \bibinfo {author} {\bibfnamefont {F.}~\bibnamefont {Wang}},\ }\bibfield
  {title} {\bibinfo {title} {{Azimuthal Correlation Anisotropies in $p + p$
  Collisions Simulated by Pythia}},\ }\href@noop {} {\  (\bibinfo {year}
  {2024})},\ \Eprint {https://arxiv.org/abs/2410.13143} {arXiv:2410.13143
  [nucl-th]} \BibitemShut {NoStop}%
\bibitem [{\citenamefont {Soudi}\ and\ \citenamefont
  {Majumder}(2025)}]{Soudi:2024slz}%
  \BibitemOpen
  \bibfield  {author} {\bibinfo {author} {\bibfnamefont {I.}~\bibnamefont
  {Soudi}}\ and\ \bibinfo {author} {\bibfnamefont {A.}~\bibnamefont
  {Majumder}},\ }\bibfield  {title} {\bibinfo {title} {{T-odd parton
  distribution functions and azimuthal anisotropy at high transverse momentum
  in p-p and p-A collisions}},\ }\href
  {https://doi.org/10.1103/PhysRevC.111.024901} {\bibfield  {journal} {\bibinfo
   {journal} {Phys. Rev. C}\ }\textbf {\bibinfo {volume} {111}},\ \bibinfo
  {pages} {024901} (\bibinfo {year} {2025})},\ \Eprint
  {https://arxiv.org/abs/2404.05287} {arXiv:2404.05287 [hep-ph]} \BibitemShut
  {NoStop}%
\bibitem [{\citenamefont {Soudi}\ \emph {et~al.}(2025)\citenamefont {Soudi}
  \emph {et~al.}}]{JETSCAPE:2024dgu}%
  \BibitemOpen
  \bibfield  {author} {\bibinfo {author} {\bibfnamefont {I.}~\bibnamefont
  {Soudi}} \emph {et~al.} (\bibinfo {collaboration} {JETSCAPE}),\ }\bibfield
  {title} {\bibinfo {title} {{Soft-hard framework with exact four-momentum
  conservation for small systems}},\ }\href {https://doi.org/10.1103/r8jt-1xpk}
  {\bibfield  {journal} {\bibinfo  {journal} {Phys. Rev. C}\ }\textbf {\bibinfo
  {volume} {112}},\ \bibinfo {pages} {014905} (\bibinfo {year} {2025})},\
  \Eprint {https://arxiv.org/abs/2407.17443} {arXiv:2407.17443 [hep-ph]}
  \BibitemShut {NoStop}%
\bibitem [{\citenamefont {Armesto}\ and\ \citenamefont
  {Scomparin}(2016)}]{Armesto:2015ioy}%
  \BibitemOpen
  \bibfield  {author} {\bibinfo {author} {\bibfnamefont {N.}~\bibnamefont
  {Armesto}}\ and\ \bibinfo {author} {\bibfnamefont {E.}~\bibnamefont
  {Scomparin}},\ }\bibfield  {title} {\bibinfo {title} {{Heavy-ion collisions
  at the Large Hadron Collider: a review of the results from Run 1}},\ }\href
  {https://doi.org/10.1140/epjp/i2016-16052-4} {\bibfield  {journal} {\bibinfo
  {journal} {Eur. Phys. J. Plus}\ }\textbf {\bibinfo {volume} {131}},\ \bibinfo
  {pages} {52} (\bibinfo {year} {2016})},\ \Eprint
  {https://arxiv.org/abs/1511.02151} {arXiv:1511.02151 [nucl-ex]} \BibitemShut
  {NoStop}%
\bibitem [{\citenamefont {Connors}\ \emph {et~al.}(2018)\citenamefont
  {Connors}, \citenamefont {Nattrass}, \citenamefont {Reed},\ and\
  \citenamefont {Salur}}]{Connors:2017ptx}%
  \BibitemOpen
  \bibfield  {author} {\bibinfo {author} {\bibfnamefont {M.}~\bibnamefont
  {Connors}}, \bibinfo {author} {\bibfnamefont {C.}~\bibnamefont {Nattrass}},
  \bibinfo {author} {\bibfnamefont {R.}~\bibnamefont {Reed}},\ and\ \bibinfo
  {author} {\bibfnamefont {S.}~\bibnamefont {Salur}},\ }\bibfield  {title}
  {\bibinfo {title} {{Jet measurements in heavy ion physics}},\ }\href
  {https://doi.org/10.1103/RevModPhys.90.025005} {\bibfield  {journal}
  {\bibinfo  {journal} {Rev. Mod. Phys.}\ }\textbf {\bibinfo {volume} {90}},\
  \bibinfo {pages} {025005} (\bibinfo {year} {2018})},\ \Eprint
  {https://arxiv.org/abs/1705.01974} {arXiv:1705.01974 [nucl-ex]} \BibitemShut
  {NoStop}%
\bibitem [{\citenamefont {Cunqueiro}\ and\ \citenamefont
  {Sickles}(2022)}]{Cunqueiro:2021wls}%
  \BibitemOpen
  \bibfield  {author} {\bibinfo {author} {\bibfnamefont {L.}~\bibnamefont
  {Cunqueiro}}\ and\ \bibinfo {author} {\bibfnamefont {A.~M.}\ \bibnamefont
  {Sickles}},\ }\bibfield  {title} {\bibinfo {title} {{Studying the QGP with
  Jets at the LHC and RHIC}},\ }\href
  {https://doi.org/10.1016/j.ppnp.2022.103940} {\bibfield  {journal} {\bibinfo
  {journal} {Prog. Part. Nucl. Phys.}\ }\textbf {\bibinfo {volume} {124}},\
  \bibinfo {pages} {103940} (\bibinfo {year} {2022})},\ \Eprint
  {https://arxiv.org/abs/2110.14490} {arXiv:2110.14490 [nucl-ex]} \BibitemShut
  {NoStop}%
\bibitem [{\citenamefont {Casalderrey-Solana}\ and\ \citenamefont
  {Salgado}(2007)}]{Casalderrey-Solana:2007knd}%
  \BibitemOpen
  \bibfield  {author} {\bibinfo {author} {\bibfnamefont {J.}~\bibnamefont
  {Casalderrey-Solana}}\ and\ \bibinfo {author} {\bibfnamefont {C.~A.}\
  \bibnamefont {Salgado}},\ }\bibfield  {title} {\bibinfo {title}
  {{Introductory lectures on jet quenching in heavy ion collisions}},\
  }\href@noop {} {\bibfield  {journal} {\bibinfo  {journal} {Acta Phys. Polon.
  B}\ }\textbf {\bibinfo {volume} {38}},\ \bibinfo {pages} {3731} (\bibinfo
  {year} {2007})},\ \Eprint {https://arxiv.org/abs/0712.3443} {arXiv:0712.3443
  [hep-ph]} \BibitemShut {NoStop}%
\bibitem [{\citenamefont {d'Enterria}(2010)}]{dEnterria:2009xfs}%
  \BibitemOpen
  \bibfield  {author} {\bibinfo {author} {\bibfnamefont {D.}~\bibnamefont
  {d'Enterria}},\ }\bibfield  {title} {\bibinfo {title} {{Jet quenching}},\
  }\href {https://doi.org/10.1007/978-3-642-01539-7_16} {\bibfield  {journal}
  {\bibinfo  {journal} {Landolt-Bornstein}\ }\textbf {\bibinfo {volume} {23}},\
  \bibinfo {pages} {471} (\bibinfo {year} {2010})},\ \Eprint
  {https://arxiv.org/abs/0902.2011} {arXiv:0902.2011 [nucl-ex]} \BibitemShut
  {NoStop}%
\bibitem [{\citenamefont {Mehtar-Tani}\ \emph {et~al.}(2013)\citenamefont
  {Mehtar-Tani}, \citenamefont {Milhano},\ and\ \citenamefont
  {Tywoniuk}}]{Mehtar-Tani:2013pia}%
  \BibitemOpen
  \bibfield  {author} {\bibinfo {author} {\bibfnamefont {Y.}~\bibnamefont
  {Mehtar-Tani}}, \bibinfo {author} {\bibfnamefont {J.~G.}\ \bibnamefont
  {Milhano}},\ and\ \bibinfo {author} {\bibfnamefont {K.}~\bibnamefont
  {Tywoniuk}},\ }\bibfield  {title} {\bibinfo {title} {{Jet physics in
  heavy-ion collisions}},\ }\href {https://doi.org/10.1142/S0217751X13400137}
  {\bibfield  {journal} {\bibinfo  {journal} {Int. J. Mod. Phys. A}\ }\textbf
  {\bibinfo {volume} {28}},\ \bibinfo {pages} {1340013} (\bibinfo {year}
  {2013})},\ \Eprint {https://arxiv.org/abs/1302.2579} {arXiv:1302.2579
  [hep-ph]} \BibitemShut {NoStop}%
\bibitem [{\citenamefont {Cao}\ \emph {et~al.}(2024)\citenamefont {Cao},
  \citenamefont {Majumder}, \citenamefont {Modarresi-Yazdi}, \citenamefont
  {Soudi},\ and\ \citenamefont {Tachibana}}]{Cao:2024pxc}%
  \BibitemOpen
  \bibfield  {author} {\bibinfo {author} {\bibfnamefont {S.}~\bibnamefont
  {Cao}}, \bibinfo {author} {\bibfnamefont {A.}~\bibnamefont {Majumder}},
  \bibinfo {author} {\bibfnamefont {R.}~\bibnamefont {Modarresi-Yazdi}},
  \bibinfo {author} {\bibfnamefont {I.}~\bibnamefont {Soudi}},\ and\ \bibinfo
  {author} {\bibfnamefont {Y.}~\bibnamefont {Tachibana}},\ }\bibfield  {title}
  {\bibinfo {title} {{Jet Quenching: From Theory to Simulation}},\ }\href
  {https://doi.org/10.1142/S0218301324300029} {\bibfield  {journal} {\bibinfo
  {journal} {Int. J. Mod. Phys. E}\ }\textbf {\bibinfo {volume} {33}},\
  \bibinfo {pages} {2430002} (\bibinfo {year} {2024})},\ \Eprint
  {https://arxiv.org/abs/2401.10026} {arXiv:2401.10026 [hep-ph]} \BibitemShut
  {NoStop}%
\bibitem [{\citenamefont {Aad}\ \emph {et~al.}(2015)\citenamefont {Aad} \emph
  {et~al.}}]{ATLAS:2014cpa}%
  \BibitemOpen
  \bibfield  {author} {\bibinfo {author} {\bibfnamefont {G.}~\bibnamefont
  {Aad}} \emph {et~al.} (\bibinfo {collaboration} {ATLAS}),\ }\bibfield
  {title} {\bibinfo {title} {{Centrality and rapidity dependence of inclusive
  jet production in $\sqrt{s_\mathrm{NN}} = 5.02$ TeV proton-lead collisions
  with the ATLAS detector}},\ }\href
  {https://doi.org/10.1016/j.physletb.2015.07.023} {\bibfield  {journal}
  {\bibinfo  {journal} {Phys. Lett. B}\ }\textbf {\bibinfo {volume} {748}},\
  \bibinfo {pages} {392} (\bibinfo {year} {2015})},\ \Eprint
  {https://arxiv.org/abs/1412.4092} {arXiv:1412.4092 [hep-ex]} \BibitemShut
  {NoStop}%
\bibitem [{\citenamefont {Adam}\ \emph {et~al.}(2016)\citenamefont {Adam} \emph
  {et~al.}}]{ALICE:2016faw}%
  \BibitemOpen
  \bibfield  {author} {\bibinfo {author} {\bibfnamefont {J.}~\bibnamefont
  {Adam}} \emph {et~al.} (\bibinfo {collaboration} {ALICE}),\ }\bibfield
  {title} {\bibinfo {title} {{Centrality dependence of charged jet production
  in p-Pb collisions at $\sqrt{s_\mathrm{NN}}$ = 5.02 TeV}}\ }\href
  {https://doi.org/10.1140/epjc/s10052-016-4107-8}
  {10.1140/epjc/s10052-016-4107-8} (\bibinfo {year} {2016}),\ \Eprint
  {https://arxiv.org/abs/1603.03402} {arXiv:1603.03402 [nucl-ex]} \BibitemShut
  {NoStop}%
\bibitem [{\citenamefont {Acharya}\ \emph {et~al.}(2022)\citenamefont {Acharya}
  \emph {et~al.}}]{ALICE:2021wct}%
  \BibitemOpen
  \bibfield  {author} {\bibinfo {author} {\bibfnamefont {S.}~\bibnamefont
  {Acharya}} \emph {et~al.} (\bibinfo {collaboration} {ALICE}),\ }\bibfield
  {title} {\bibinfo {title} {{Measurement of inclusive charged-particle b-jet
  production in pp and p-Pb collisions at $ \sqrt{s_{\mathrm{NN}}} $ = 5.02
  TeV}},\ }\href {https://doi.org/10.1007/JHEP01(2022)178} {\bibfield
  {journal} {\bibinfo  {journal} {JHEP}\ }\textbf {\bibinfo {volume} {01}},\
  \bibinfo {pages} {178}},\ \Eprint {https://arxiv.org/abs/2110.06104}
  {arXiv:2110.06104 [nucl-ex]} \BibitemShut {NoStop}%
\bibitem [{\citenamefont {Acharya}\ \emph {et~al.}(2024)\citenamefont {Acharya}
  \emph {et~al.}}]{ALICE:2023ama}%
  \BibitemOpen
  \bibfield  {author} {\bibinfo {author} {\bibfnamefont {S.}~\bibnamefont
  {Acharya}} \emph {et~al.} (\bibinfo {collaboration} {ALICE}),\ }\bibfield
  {title} {\bibinfo {title} {{Measurement of inclusive charged-particle jet
  production in pp and p-Pb collisions at $ \sqrt{s_{\textrm{NN}}} $ = 5.02
  TeV}},\ }\href {https://doi.org/10.1007/JHEP05(2024)041} {\bibfield
  {journal} {\bibinfo  {journal} {JHEP}\ }\textbf {\bibinfo {volume} {05}},\
  \bibinfo {pages} {041}},\ \Eprint {https://arxiv.org/abs/2307.10860}
  {arXiv:2307.10860 [nucl-ex]} \BibitemShut {NoStop}%
\bibitem [{\citenamefont {Aad}\ \emph {et~al.}(2020)\citenamefont {Aad} \emph
  {et~al.}}]{ATLAS:2019vcm}%
  \BibitemOpen
  \bibfield  {author} {\bibinfo {author} {\bibfnamefont {G.}~\bibnamefont
  {Aad}} \emph {et~al.} (\bibinfo {collaboration} {ATLAS}),\ }\bibfield
  {title} {\bibinfo {title} {{Transverse momentum and process dependent
  azimuthal anisotropies in $\sqrt{s_{\mathrm{NN}}}=8.16$ TeV $p$+Pb collisions
  with the ATLAS detector}},\ }\href
  {https://doi.org/10.1140/epjc/s10052-020-7624-4} {\bibfield  {journal}
  {\bibinfo  {journal} {Eur. Phys. J. C}\ }\textbf {\bibinfo {volume} {80}},\
  \bibinfo {pages} {73} (\bibinfo {year} {2020})},\ \Eprint
  {https://arxiv.org/abs/1910.13978} {arXiv:1910.13978 [nucl-ex]} \BibitemShut
  {NoStop}%
\bibitem [{\citenamefont {Aad}\ \emph {et~al.}(2023{\natexlab{a}})\citenamefont
  {Aad} \emph {et~al.}}]{ATLAS:2023bmp}%
  \BibitemOpen
  \bibfield  {author} {\bibinfo {author} {\bibfnamefont {G.}~\bibnamefont
  {Aad}} \emph {et~al.} (\bibinfo {collaboration} {ATLAS}),\ }\bibfield
  {title} {\bibinfo {title} {{Measurement of the Sensitivity of Two-Particle
  Correlations in pp Collisions to the Presence of Hard Scatterings}},\ }\href
  {https://doi.org/10.1103/PhysRevLett.131.162301} {\bibfield  {journal}
  {\bibinfo  {journal} {Phys. Rev. Lett.}\ }\textbf {\bibinfo {volume} {131}},\
  \bibinfo {pages} {162301} (\bibinfo {year} {2023}{\natexlab{a}})},\ \Eprint
  {https://arxiv.org/abs/2303.17357} {arXiv:2303.17357 [nucl-ex]} \BibitemShut
  {NoStop}%
\bibitem [{\citenamefont {Chekhovsky}\ \emph {et~al.}(2025)\citenamefont
  {Chekhovsky} \emph {et~al.}}]{CMS:2025kzg}%
  \BibitemOpen
  \bibfield  {author} {\bibinfo {author} {\bibfnamefont {V.}~\bibnamefont
  {Chekhovsky}} \emph {et~al.} (\bibinfo {collaboration} {CMS}),\ }\bibfield
  {title} {\bibinfo {title} {{Evidence for similar collectivity of high
  transverse momentum particles in pPb and PbPb collisions}},\ }\href@noop {}
  {\  (\bibinfo {year} {2025})},\ \Eprint {https://arxiv.org/abs/2502.07525}
  {arXiv:2502.07525 [nucl-ex]} \BibitemShut {NoStop}%
\bibitem [{\citenamefont {Gyulassy}\ \emph {et~al.}(2001)\citenamefont
  {Gyulassy}, \citenamefont {Vitev},\ and\ \citenamefont
  {Wang}}]{Gyulassy:2000gk}%
  \BibitemOpen
  \bibfield  {author} {\bibinfo {author} {\bibfnamefont {M.}~\bibnamefont
  {Gyulassy}}, \bibinfo {author} {\bibfnamefont {I.}~\bibnamefont {Vitev}},\
  and\ \bibinfo {author} {\bibfnamefont {X.~N.}\ \bibnamefont {Wang}},\
  }\bibfield  {title} {\bibinfo {title} {{High p(T) azimuthal asymmetry in
  noncentral A+A at RHIC}},\ }\href
  {https://doi.org/10.1103/PhysRevLett.86.2537} {\bibfield  {journal} {\bibinfo
   {journal} {Phys. Rev. Lett.}\ }\textbf {\bibinfo {volume} {86}},\ \bibinfo
  {pages} {2537} (\bibinfo {year} {2001})},\ \Eprint
  {https://arxiv.org/abs/nucl-th/0012092} {arXiv:nucl-th/0012092} \BibitemShut
  {NoStop}%
\bibitem [{\citenamefont {Wang}(2001)}]{Wang:2000fq}%
  \BibitemOpen
  \bibfield  {author} {\bibinfo {author} {\bibfnamefont {X.-N.}\ \bibnamefont
  {Wang}},\ }\bibfield  {title} {\bibinfo {title} {{Jet quenching and azimuthal
  anisotropy of large p(T) spectra in noncentral high-energy heavy ion
  collisions}},\ }\href {https://doi.org/10.1103/PhysRevC.63.054902} {\bibfield
   {journal} {\bibinfo  {journal} {Phys. Rev. C}\ }\textbf {\bibinfo {volume}
  {63}},\ \bibinfo {pages} {054902} (\bibinfo {year} {2001})},\ \Eprint
  {https://arxiv.org/abs/nucl-th/0009019} {arXiv:nucl-th/0009019} \BibitemShut
  {NoStop}%
\bibitem [{\citenamefont {Citron}\ \emph {et~al.}(2019)\citenamefont {Citron}
  \emph {et~al.}}]{Citron:2018lsq}%
  \BibitemOpen
  \bibfield  {author} {\bibinfo {author} {\bibfnamefont {Z.}~\bibnamefont
  {Citron}} \emph {et~al.},\ }\bibfield  {title} {\bibinfo {title} {{Report
  from Working Group 5}: {Future physics opportunities for high-density QCD at
  the LHC with heavy-ion and proton beams}},\ }\href
  {https://doi.org/10.23731/CYRM-2019-007.1159} {\bibfield  {journal} {\bibinfo
   {journal} {CERN Yellow Rep. Monogr.}\ }\textbf {\bibinfo {volume} {7}},\
  \bibinfo {pages} {1159} (\bibinfo {year} {2019})},\ \Eprint
  {https://arxiv.org/abs/1812.06772} {arXiv:1812.06772 [hep-ph]} \BibitemShut
  {NoStop}%
\bibitem [{\citenamefont {Brewer}\ \emph {et~al.}(2021)\citenamefont {Brewer},
  \citenamefont {Mazeliauskas},\ and\ \citenamefont {van~der
  Schee}}]{Brewer:2021kiv}%
  \BibitemOpen
  \bibfield  {author} {\bibinfo {author} {\bibfnamefont {J.}~\bibnamefont
  {Brewer}}, \bibinfo {author} {\bibfnamefont {A.}~\bibnamefont
  {Mazeliauskas}},\ and\ \bibinfo {author} {\bibfnamefont {W.}~\bibnamefont
  {van~der Schee}},\ }\bibfield  {title} {\bibinfo {title} {{Opportunities of
  OO and $p$O collisions at the LHC}},\ }in\ \href@noop {} {\emph {\bibinfo
  {booktitle} {{Opportunities of OO and pO collisions at the LHC}}}}\ (\bibinfo
  {year} {2021})\ \Eprint {https://arxiv.org/abs/2103.01939} {arXiv:2103.01939
  [hep-ph]} \BibitemShut {NoStop}%
\bibitem [{\citenamefont {Loizides}\ and\ \citenamefont
  {Morsch}(2017)}]{Loizides:2017sqq}%
  \BibitemOpen
  \bibfield  {author} {\bibinfo {author} {\bibfnamefont {C.}~\bibnamefont
  {Loizides}}\ and\ \bibinfo {author} {\bibfnamefont {A.}~\bibnamefont
  {Morsch}},\ }\bibfield  {title} {\bibinfo {title} {{Absence of jet quenching
  in peripheral nucleus\textendash{}nucleus collisions}},\ }\href
  {https://doi.org/10.1016/j.physletb.2017.09.002} {\bibfield  {journal}
  {\bibinfo  {journal} {Phys. Lett. B}\ }\textbf {\bibinfo {volume} {773}},\
  \bibinfo {pages} {408} (\bibinfo {year} {2017})},\ \Eprint
  {https://arxiv.org/abs/1705.08856} {arXiv:1705.08856 [nucl-ex]} \BibitemShut
  {NoStop}%
\bibitem [{\citenamefont {Park}\ \emph {et~al.}(2025)\citenamefont {Park},
  \citenamefont {Nagle}, \citenamefont {Perepelitsa}, \citenamefont {Lim},\
  and\ \citenamefont {Loizides}}]{Park:2025mbt}%
  \BibitemOpen
  \bibfield  {author} {\bibinfo {author} {\bibfnamefont {J.}~\bibnamefont
  {Park}}, \bibinfo {author} {\bibfnamefont {J.~L.}\ \bibnamefont {Nagle}},
  \bibinfo {author} {\bibfnamefont {D.~V.}\ \bibnamefont {Perepelitsa}},
  \bibinfo {author} {\bibfnamefont {S.}~\bibnamefont {Lim}},\ and\ \bibinfo
  {author} {\bibfnamefont {C.}~\bibnamefont {Loizides}},\ }\bibfield  {title}
  {\bibinfo {title} {{Selection bias effects on high-$p_\mathrm{T}$ yield and
  correlation measurements in Oxygen+Oxygen collisions}},\ }\href@noop {} {\
  (\bibinfo {year} {2025})},\ \Eprint {https://arxiv.org/abs/2507.03603}
  {arXiv:2507.03603 [nucl-ex]} \BibitemShut {NoStop}%
\bibitem [{\citenamefont {Heinz}\ and\ \citenamefont
  {Snellings}(2013)}]{Heinz:2013th}%
  \BibitemOpen
  \bibfield  {author} {\bibinfo {author} {\bibfnamefont {U.}~\bibnamefont
  {Heinz}}\ and\ \bibinfo {author} {\bibfnamefont {R.}~\bibnamefont
  {Snellings}},\ }\bibfield  {title} {\bibinfo {title} {{Collective flow and
  viscosity in relativistic heavy-ion collisions}},\ }\href
  {https://doi.org/10.1146/annurev-nucl-102212-170540} {\bibfield  {journal}
  {\bibinfo  {journal} {Ann. Rev. Nucl. Part. Sci.}\ }\textbf {\bibinfo
  {volume} {63}},\ \bibinfo {pages} {123} (\bibinfo {year} {2013})},\ \Eprint
  {https://arxiv.org/abs/1301.2826} {arXiv:1301.2826 [nucl-th]} \BibitemShut
  {NoStop}%
\bibitem [{\citenamefont {Gale}\ \emph {et~al.}(2013)\citenamefont {Gale},
  \citenamefont {Jeon},\ and\ \citenamefont {Schenke}}]{Gale:2013da}%
  \BibitemOpen
  \bibfield  {author} {\bibinfo {author} {\bibfnamefont {C.}~\bibnamefont
  {Gale}}, \bibinfo {author} {\bibfnamefont {S.}~\bibnamefont {Jeon}},\ and\
  \bibinfo {author} {\bibfnamefont {B.}~\bibnamefont {Schenke}},\ }\bibfield
  {title} {\bibinfo {title} {{Hydrodynamic Modeling of Heavy-Ion Collisions}},\
  }\href {https://doi.org/10.1142/S0217751X13400113} {\bibfield  {journal}
  {\bibinfo  {journal} {Int. J. Mod. Phys. A}\ }\textbf {\bibinfo {volume}
  {28}},\ \bibinfo {pages} {1340011} (\bibinfo {year} {2013})},\ \Eprint
  {https://arxiv.org/abs/1301.5893} {arXiv:1301.5893 [nucl-th]} \BibitemShut
  {NoStop}%
\bibitem [{\citenamefont {M{\"a}ntysaari}\ \emph {et~al.}(2017)\citenamefont
  {M{\"a}ntysaari}, \citenamefont {Schenke}, \citenamefont {Shen},\ and\
  \citenamefont {Tribedy}}]{Mantysaari:2017cni}%
  \BibitemOpen
  \bibfield  {author} {\bibinfo {author} {\bibfnamefont {H.}~\bibnamefont
  {M{\"a}ntysaari}}, \bibinfo {author} {\bibfnamefont {B.}~\bibnamefont
  {Schenke}}, \bibinfo {author} {\bibfnamefont {C.}~\bibnamefont {Shen}},\ and\
  \bibinfo {author} {\bibfnamefont {P.}~\bibnamefont {Tribedy}},\ }\bibfield
  {title} {\bibinfo {title} {{Imprints of fluctuating proton shapes on flow in
  proton-lead collisions at the LHC}},\ }\href
  {https://doi.org/10.1016/j.physletb.2017.07.038} {\bibfield  {journal}
  {\bibinfo  {journal} {Phys. Lett. B}\ }\textbf {\bibinfo {volume} {772}},\
  \bibinfo {pages} {681} (\bibinfo {year} {2017})},\ \Eprint
  {https://arxiv.org/abs/1705.03177} {arXiv:1705.03177 [nucl-th]} \BibitemShut
  {NoStop}%
\bibitem [{\citenamefont {Romatschke}\ and\ \citenamefont
  {Romatschke}(2019)}]{Romatschke:2017ejr}%
  \BibitemOpen
  \bibfield  {author} {\bibinfo {author} {\bibfnamefont {P.}~\bibnamefont
  {Romatschke}}\ and\ \bibinfo {author} {\bibfnamefont {U.}~\bibnamefont
  {Romatschke}},\ }\href {https://doi.org/10.1017/9781108651998} {\emph
  {\bibinfo {title} {{Relativistic Fluid Dynamics In and Out of
  Equilibrium}}}},\ Cambridge Monographs on Mathematical Physics\ (\bibinfo
  {publisher} {Cambridge University Press},\ \bibinfo {year} {2019})\ \Eprint
  {https://arxiv.org/abs/1712.05815} {arXiv:1712.05815 [nucl-th]} \BibitemShut
  {NoStop}%
\bibitem [{\citenamefont {Mehtar-Tani}\ \emph {et~al.}(2021)\citenamefont
  {Mehtar-Tani}, \citenamefont {Pablos},\ and\ \citenamefont
  {Tywoniuk}}]{Mehtar-Tani:2021fud}%
  \BibitemOpen
  \bibfield  {author} {\bibinfo {author} {\bibfnamefont {Y.}~\bibnamefont
  {Mehtar-Tani}}, \bibinfo {author} {\bibfnamefont {D.}~\bibnamefont
  {Pablos}},\ and\ \bibinfo {author} {\bibfnamefont {K.}~\bibnamefont
  {Tywoniuk}},\ }\bibfield  {title} {\bibinfo {title} {{Cone-Size Dependence of
  Jet Suppression in Heavy-Ion Collisions}},\ }\href
  {https://doi.org/10.1103/PhysRevLett.127.252301} {\bibfield  {journal}
  {\bibinfo  {journal} {Phys. Rev. Lett.}\ }\textbf {\bibinfo {volume} {127}},\
  \bibinfo {pages} {252301} (\bibinfo {year} {2021})},\ \Eprint
  {https://arxiv.org/abs/2101.01742} {arXiv:2101.01742 [hep-ph]} \BibitemShut
  {NoStop}%
\bibitem [{\citenamefont {Takacs}\ and\ \citenamefont
  {Tywoniuk}(2021)}]{Takacs:2021bpv}%
  \BibitemOpen
  \bibfield  {author} {\bibinfo {author} {\bibfnamefont {A.}~\bibnamefont
  {Takacs}}\ and\ \bibinfo {author} {\bibfnamefont {K.}~\bibnamefont
  {Tywoniuk}},\ }\bibfield  {title} {\bibinfo {title} {{Quenching effects in
  the cumulative jet spectrum}},\ }\href
  {https://doi.org/10.1007/JHEP10(2021)038} {\bibfield  {journal} {\bibinfo
  {journal} {JHEP}\ }\textbf {\bibinfo {volume} {10}},\ \bibinfo {pages}
  {038}},\ \Eprint {https://arxiv.org/abs/2103.14676} {arXiv:2103.14676
  [hep-ph]} \BibitemShut {NoStop}%
\bibitem [{\citenamefont {Mehtar-Tani}\ \emph {et~al.}(2024)\citenamefont
  {Mehtar-Tani}, \citenamefont {Pablos},\ and\ \citenamefont
  {Tywoniuk}}]{Mehtar-Tani:2024jtd}%
  \BibitemOpen
  \bibfield  {author} {\bibinfo {author} {\bibfnamefont {Y.}~\bibnamefont
  {Mehtar-Tani}}, \bibinfo {author} {\bibfnamefont {D.}~\bibnamefont
  {Pablos}},\ and\ \bibinfo {author} {\bibfnamefont {K.}~\bibnamefont
  {Tywoniuk}},\ }\bibfield  {title} {\bibinfo {title} {{Jet suppression and
  azimuthal anisotropy from RHIC to LHC}},\ }\href
  {https://doi.org/10.1103/PhysRevD.110.014009} {\bibfield  {journal} {\bibinfo
   {journal} {Phys. Rev. D}\ }\textbf {\bibinfo {volume} {110}},\ \bibinfo
  {pages} {014009} (\bibinfo {year} {2024})},\ \Eprint
  {https://arxiv.org/abs/2402.07869} {arXiv:2402.07869 [hep-ph]} \BibitemShut
  {NoStop}%
\bibitem [{\citenamefont {Alwall}\ \emph {et~al.}(2014)\citenamefont {Alwall},
  \citenamefont {Frederix}, \citenamefont {Frixione}, \citenamefont {Hirschi},
  \citenamefont {Maltoni}, \citenamefont {Mattelaer}, \citenamefont {Shao},
  \citenamefont {Stelzer}, \citenamefont {Torrielli},\ and\ \citenamefont
  {Zaro}}]{Alwall:2014hca}%
  \BibitemOpen
  \bibfield  {author} {\bibinfo {author} {\bibfnamefont {J.}~\bibnamefont
  {Alwall}}, \bibinfo {author} {\bibfnamefont {R.}~\bibnamefont {Frederix}},
  \bibinfo {author} {\bibfnamefont {S.}~\bibnamefont {Frixione}}, \bibinfo
  {author} {\bibfnamefont {V.}~\bibnamefont {Hirschi}}, \bibinfo {author}
  {\bibfnamefont {F.}~\bibnamefont {Maltoni}}, \bibinfo {author} {\bibfnamefont
  {O.}~\bibnamefont {Mattelaer}}, \bibinfo {author} {\bibfnamefont {H.~S.}\
  \bibnamefont {Shao}}, \bibinfo {author} {\bibfnamefont {T.}~\bibnamefont
  {Stelzer}}, \bibinfo {author} {\bibfnamefont {P.}~\bibnamefont {Torrielli}},\
  and\ \bibinfo {author} {\bibfnamefont {M.}~\bibnamefont {Zaro}},\ }\bibfield
  {title} {\bibinfo {title} {{The automated computation of tree-level and
  next-to-leading order differential cross sections, and their matching to
  parton shower simulations}},\ }\href
  {https://doi.org/10.1007/JHEP07(2014)079} {\bibfield  {journal} {\bibinfo
  {journal} {JHEP}\ }\textbf {\bibinfo {volume} {07}},\ \bibinfo {pages}
  {079}},\ \Eprint {https://arxiv.org/abs/1405.0301} {arXiv:1405.0301 [hep-ph]}
  \BibitemShut {NoStop}%
\bibitem [{\citenamefont {Bierlich}\ \emph {et~al.}(2022)\citenamefont
  {Bierlich} \emph {et~al.}}]{Bierlich:2022pfr}%
  \BibitemOpen
  \bibfield  {author} {\bibinfo {author} {\bibfnamefont {C.}~\bibnamefont
  {Bierlich}} \emph {et~al.},\ }\bibfield  {title} {\bibinfo {title} {{A
  comprehensive guide to the physics and usage of PYTHIA 8.3}},\ }\href
  {https://doi.org/10.21468/SciPostPhysCodeb.8} {\bibfield  {journal} {\bibinfo
   {journal} {SciPost Phys. Codeb.}\ }\textbf {\bibinfo {volume} {2022}},\
  \bibinfo {pages} {8} (\bibinfo {year} {2022})},\ \Eprint
  {https://arxiv.org/abs/2203.11601} {arXiv:2203.11601 [hep-ph]} \BibitemShut
  {NoStop}%
\bibitem [{\citenamefont {Skands}\ \emph {et~al.}(2014)\citenamefont {Skands},
  \citenamefont {Carrazza},\ and\ \citenamefont {Rojo}}]{Skands:2014pea}%
  \BibitemOpen
  \bibfield  {author} {\bibinfo {author} {\bibfnamefont {P.}~\bibnamefont
  {Skands}}, \bibinfo {author} {\bibfnamefont {S.}~\bibnamefont {Carrazza}},\
  and\ \bibinfo {author} {\bibfnamefont {J.}~\bibnamefont {Rojo}},\ }\bibfield
  {title} {\bibinfo {title} {{Tuning PYTHIA 8.1: the Monash 2013 Tune}},\
  }\href {https://doi.org/10.1140/epjc/s10052-014-3024-y} {\bibfield  {journal}
  {\bibinfo  {journal} {Eur. Phys. J. C}\ }\textbf {\bibinfo {volume} {74}},\
  \bibinfo {pages} {3024} (\bibinfo {year} {2014})},\ \Eprint
  {https://arxiv.org/abs/1404.5630} {arXiv:1404.5630 [hep-ph]} \BibitemShut
  {NoStop}%
\bibitem [{\citenamefont {Banfi}\ \emph {et~al.}(2006)\citenamefont {Banfi},
  \citenamefont {Salam},\ and\ \citenamefont {Zanderighi}}]{Banfi:2006hf}%
  \BibitemOpen
  \bibfield  {author} {\bibinfo {author} {\bibfnamefont {A.}~\bibnamefont
  {Banfi}}, \bibinfo {author} {\bibfnamefont {G.~P.}\ \bibnamefont {Salam}},\
  and\ \bibinfo {author} {\bibfnamefont {G.}~\bibnamefont {Zanderighi}},\
  }\bibfield  {title} {\bibinfo {title} {{Infrared safe definition of jet
  flavor}},\ }\href {https://doi.org/10.1140/epjc/s2006-02552-4} {\bibfield
  {journal} {\bibinfo  {journal} {Eur. Phys. J. C}\ }\textbf {\bibinfo {volume}
  {47}},\ \bibinfo {pages} {113} (\bibinfo {year} {2006})},\ \Eprint
  {https://arxiv.org/abs/hep-ph/0601139} {arXiv:hep-ph/0601139} \BibitemShut
  {NoStop}%
\bibitem [{\citenamefont {Eskola}\ \emph {et~al.}(2022)\citenamefont {Eskola},
  \citenamefont {Paakkinen}, \citenamefont {Paukkunen},\ and\ \citenamefont
  {Salgado}}]{Eskola:2021nhw}%
  \BibitemOpen
  \bibfield  {author} {\bibinfo {author} {\bibfnamefont {K.~J.}\ \bibnamefont
  {Eskola}}, \bibinfo {author} {\bibfnamefont {P.}~\bibnamefont {Paakkinen}},
  \bibinfo {author} {\bibfnamefont {H.}~\bibnamefont {Paukkunen}},\ and\
  \bibinfo {author} {\bibfnamefont {C.~A.}\ \bibnamefont {Salgado}},\
  }\bibfield  {title} {\bibinfo {title} {{EPPS21: a global QCD analysis of
  nuclear PDFs}},\ }\href {https://doi.org/10.1140/epjc/s10052-022-10359-0}
  {\bibfield  {journal} {\bibinfo  {journal} {Eur. Phys. J. C}\ }\textbf
  {\bibinfo {volume} {82}},\ \bibinfo {pages} {413} (\bibinfo {year} {2022})},\
  \Eprint {https://arxiv.org/abs/2112.12462} {arXiv:2112.12462 [hep-ph]}
  \BibitemShut {NoStop}%
\bibitem [{\citenamefont {Buckley}\ \emph {et~al.}(2015)\citenamefont
  {Buckley}, \citenamefont {Ferrando}, \citenamefont {Lloyd}, \citenamefont
  {Nordstr\"om}, \citenamefont {Page}, \citenamefont {R\"ufenacht},
  \citenamefont {Sch\"onherr},\ and\ \citenamefont {Watt}}]{Buckley:2014ana}%
  \BibitemOpen
  \bibfield  {author} {\bibinfo {author} {\bibfnamefont {A.}~\bibnamefont
  {Buckley}}, \bibinfo {author} {\bibfnamefont {J.}~\bibnamefont {Ferrando}},
  \bibinfo {author} {\bibfnamefont {S.}~\bibnamefont {Lloyd}}, \bibinfo
  {author} {\bibfnamefont {K.}~\bibnamefont {Nordstr\"om}}, \bibinfo {author}
  {\bibfnamefont {B.}~\bibnamefont {Page}}, \bibinfo {author} {\bibfnamefont
  {M.}~\bibnamefont {R\"ufenacht}}, \bibinfo {author} {\bibfnamefont
  {M.}~\bibnamefont {Sch\"onherr}},\ and\ \bibinfo {author} {\bibfnamefont
  {G.}~\bibnamefont {Watt}},\ }\bibfield  {title} {\bibinfo {title} {{LHAPDF6:
  parton density access in the LHC precision era}},\ }\href
  {https://doi.org/10.1140/epjc/s10052-015-3318-8} {\bibfield  {journal}
  {\bibinfo  {journal} {Eur. Phys. J. C}\ }\textbf {\bibinfo {volume} {75}},\
  \bibinfo {pages} {132} (\bibinfo {year} {2015})},\ \Eprint
  {https://arxiv.org/abs/1412.7420} {arXiv:1412.7420 [hep-ph]} \BibitemShut
  {NoStop}%
\bibitem [{\citenamefont {Gao}\ \emph {et~al.}(2024)\citenamefont {Gao},
  \citenamefont {Liu}, \citenamefont {Shen}, \citenamefont {Xing},\ and\
  \citenamefont {Zhao}}]{Gao:2024dbv}%
  \BibitemOpen
  \bibfield  {author} {\bibinfo {author} {\bibfnamefont {J.}~\bibnamefont
  {Gao}}, \bibinfo {author} {\bibfnamefont {C.}~\bibnamefont {Liu}}, \bibinfo
  {author} {\bibfnamefont {X.}~\bibnamefont {Shen}}, \bibinfo {author}
  {\bibfnamefont {H.}~\bibnamefont {Xing}},\ and\ \bibinfo {author}
  {\bibfnamefont {Y.}~\bibnamefont {Zhao}},\ }\bibfield  {title} {\bibinfo
  {title} {{Global analysis of fragmentation functions to charged hadrons with
  high-precision data from the LHC}},\ }\href
  {https://doi.org/10.1103/PhysRevD.110.114019} {\bibfield  {journal} {\bibinfo
   {journal} {Phys. Rev. D}\ }\textbf {\bibinfo {volume} {110}},\ \bibinfo
  {pages} {114019} (\bibinfo {year} {2024})},\ \Eprint
  {https://arxiv.org/abs/2407.04422} {arXiv:2407.04422 [hep-ph]} \BibitemShut
  {NoStop}%
\bibitem [{\citenamefont {Mehtar-Tani}(2019)}]{Mehtar-Tani:2019tvy}%
  \BibitemOpen
  \bibfield  {author} {\bibinfo {author} {\bibfnamefont {Y.}~\bibnamefont
  {Mehtar-Tani}},\ }\bibfield  {title} {\bibinfo {title} {{Gluon bremsstrahlung
  in finite media beyond multiple soft scattering approximation}},\ }\href
  {https://doi.org/10.1007/JHEP07(2019)057} {\bibfield  {journal} {\bibinfo
  {journal} {JHEP}\ }\textbf {\bibinfo {volume} {07}},\ \bibinfo {pages}
  {057}},\ \Eprint {https://arxiv.org/abs/1903.00506} {arXiv:1903.00506
  [hep-ph]} \BibitemShut {NoStop}%
\bibitem [{\citenamefont {Barata}\ \emph {et~al.}(2021)\citenamefont {Barata},
  \citenamefont {Mehtar-Tani}, \citenamefont {Soto-Ontoso},\ and\ \citenamefont
  {Tywoniuk}}]{Barata:2020rdn}%
  \BibitemOpen
  \bibfield  {author} {\bibinfo {author} {\bibfnamefont {J.~a.}\ \bibnamefont
  {Barata}}, \bibinfo {author} {\bibfnamefont {Y.}~\bibnamefont {Mehtar-Tani}},
  \bibinfo {author} {\bibfnamefont {A.}~\bibnamefont {Soto-Ontoso}},\ and\
  \bibinfo {author} {\bibfnamefont {K.}~\bibnamefont {Tywoniuk}},\ }\bibfield
  {title} {\bibinfo {title} {{Revisiting transverse momentum broadening in
  dense QCD media}},\ }\href {https://doi.org/10.1103/PhysRevD.104.054047}
  {\bibfield  {journal} {\bibinfo  {journal} {Phys. Rev. D}\ }\textbf {\bibinfo
  {volume} {104}},\ \bibinfo {pages} {054047} (\bibinfo {year} {2021})},\
  \Eprint {https://arxiv.org/abs/2009.13667} {arXiv:2009.13667 [hep-ph]}
  \BibitemShut {NoStop}%
\bibitem [{\citenamefont {Adhya}\ \emph {et~al.}(2020)\citenamefont {Adhya},
  \citenamefont {Salgado}, \citenamefont {Spousta},\ and\ \citenamefont
  {Tywoniuk}}]{Adhya:2019qse}%
  \BibitemOpen
  \bibfield  {author} {\bibinfo {author} {\bibfnamefont {S.~P.}\ \bibnamefont
  {Adhya}}, \bibinfo {author} {\bibfnamefont {C.~A.}\ \bibnamefont {Salgado}},
  \bibinfo {author} {\bibfnamefont {M.}~\bibnamefont {Spousta}},\ and\ \bibinfo
  {author} {\bibfnamefont {K.}~\bibnamefont {Tywoniuk}},\ }\bibfield  {title}
  {\bibinfo {title} {{Medium-induced cascade in expanding media}},\ }\href
  {https://doi.org/10.1007/JHEP07(2020)150} {\bibfield  {journal} {\bibinfo
  {journal} {JHEP}\ }\textbf {\bibinfo {volume} {07}},\ \bibinfo {pages}
  {150}},\ \Eprint {https://arxiv.org/abs/1911.12193} {arXiv:1911.12193
  [hep-ph]} \BibitemShut {NoStop}%
\bibitem [{\citenamefont {Caucal}\ \emph {et~al.}(2020)\citenamefont {Caucal},
  \citenamefont {Iancu}, \citenamefont {Mueller},\ and\ \citenamefont
  {Soyez}}]{Caucal:2020xad}%
  \BibitemOpen
  \bibfield  {author} {\bibinfo {author} {\bibfnamefont {P.}~\bibnamefont
  {Caucal}}, \bibinfo {author} {\bibfnamefont {E.}~\bibnamefont {Iancu}},
  \bibinfo {author} {\bibfnamefont {A.~H.}\ \bibnamefont {Mueller}},\ and\
  \bibinfo {author} {\bibfnamefont {G.}~\bibnamefont {Soyez}},\ }\bibfield
  {title} {\bibinfo {title} {{Nuclear modification factors for jet
  fragmentation}},\ }\href {https://doi.org/10.1007/JHEP10(2020)204} {\bibfield
   {journal} {\bibinfo  {journal} {JHEP}\ }\textbf {\bibinfo {volume} {10}},\
  \bibinfo {pages} {204}},\ \Eprint {https://arxiv.org/abs/2005.05852}
  {arXiv:2005.05852 [hep-ph]} \BibitemShut {NoStop}%
\bibitem [{\citenamefont {Blaizot}\ and\ \citenamefont
  {Mehtar-Tani}(2015)}]{Blaizot:2015lma}%
  \BibitemOpen
  \bibfield  {author} {\bibinfo {author} {\bibfnamefont {J.-P.}\ \bibnamefont
  {Blaizot}}\ and\ \bibinfo {author} {\bibfnamefont {Y.}~\bibnamefont
  {Mehtar-Tani}},\ }\bibfield  {title} {\bibinfo {title} {{Jet Structure in
  Heavy Ion Collisions}},\ }\href {https://doi.org/10.1142/S021830131530012X}
  {\bibfield  {journal} {\bibinfo  {journal} {Int. J. Mod. Phys. E}\ }\textbf
  {\bibinfo {volume} {24}},\ \bibinfo {pages} {1530012} (\bibinfo {year}
  {2015})},\ \Eprint {https://arxiv.org/abs/1503.05958} {arXiv:1503.05958
  [hep-ph]} \BibitemShut {NoStop}%
\bibitem [{\citenamefont {Casalderrey-Solana}\ \emph
  {et~al.}(2005)\citenamefont {Casalderrey-Solana}, \citenamefont {Shuryak},\
  and\ \citenamefont {Teaney}}]{Casalderrey-Solana:2004fdk}%
  \BibitemOpen
  \bibfield  {author} {\bibinfo {author} {\bibfnamefont {J.}~\bibnamefont
  {Casalderrey-Solana}}, \bibinfo {author} {\bibfnamefont {E.~V.}\ \bibnamefont
  {Shuryak}},\ and\ \bibinfo {author} {\bibfnamefont {D.}~\bibnamefont
  {Teaney}},\ }\bibfield  {title} {\bibinfo {title} {{Conical flow induced by
  quenched QCD jets}},\ }\href {https://doi.org/10.1088/1742-6596/27/1/003}
  {\bibfield  {journal} {\bibinfo  {journal} {J. Phys. Conf. Ser.}\ }\textbf
  {\bibinfo {volume} {27}},\ \bibinfo {pages} {22} (\bibinfo {year} {2005})},\
  \Eprint {https://arxiv.org/abs/hep-ph/0411315} {arXiv:hep-ph/0411315}
  \BibitemShut {NoStop}%
\bibitem [{\citenamefont {Chesler}\ and\ \citenamefont
  {Yaffe}(2007)}]{Chesler:2007an}%
  \BibitemOpen
  \bibfield  {author} {\bibinfo {author} {\bibfnamefont {P.~M.}\ \bibnamefont
  {Chesler}}\ and\ \bibinfo {author} {\bibfnamefont {L.~G.}\ \bibnamefont
  {Yaffe}},\ }\bibfield  {title} {\bibinfo {title} {{The Wake of a quark moving
  through a strongly-coupled plasma}},\ }\href
  {https://doi.org/10.1103/PhysRevLett.99.152001} {\bibfield  {journal}
  {\bibinfo  {journal} {Phys. Rev. Lett.}\ }\textbf {\bibinfo {volume} {99}},\
  \bibinfo {pages} {152001} (\bibinfo {year} {2007})},\ \Eprint
  {https://arxiv.org/abs/0706.0368} {arXiv:0706.0368 [hep-th]} \BibitemShut
  {NoStop}%
\bibitem [{\citenamefont {Caucal}\ \emph {et~al.}(2018)\citenamefont {Caucal},
  \citenamefont {Iancu}, \citenamefont {Mueller},\ and\ \citenamefont
  {Soyez}}]{Caucal:2018dla}%
  \BibitemOpen
  \bibfield  {author} {\bibinfo {author} {\bibfnamefont {P.}~\bibnamefont
  {Caucal}}, \bibinfo {author} {\bibfnamefont {E.}~\bibnamefont {Iancu}},
  \bibinfo {author} {\bibfnamefont {A.~H.}\ \bibnamefont {Mueller}},\ and\
  \bibinfo {author} {\bibfnamefont {G.}~\bibnamefont {Soyez}},\ }\bibfield
  {title} {\bibinfo {title} {{Vacuum-like jet fragmentation in a dense QCD
  medium}},\ }\href {https://doi.org/10.1103/PhysRevLett.120.232001} {\bibfield
   {journal} {\bibinfo  {journal} {Phys. Rev. Lett.}\ }\textbf {\bibinfo
  {volume} {120}},\ \bibinfo {pages} {232001} (\bibinfo {year} {2018})},\
  \Eprint {https://arxiv.org/abs/1801.09703} {arXiv:1801.09703 [hep-ph]}
  \BibitemShut {NoStop}%
\bibitem [{\citenamefont {Mehtar-Tani}\ and\ \citenamefont
  {Tywoniuk}(2018)}]{Mehtar-Tani:2017web}%
  \BibitemOpen
  \bibfield  {author} {\bibinfo {author} {\bibfnamefont {Y.}~\bibnamefont
  {Mehtar-Tani}}\ and\ \bibinfo {author} {\bibfnamefont {K.}~\bibnamefont
  {Tywoniuk}},\ }\bibfield  {title} {\bibinfo {title} {{Sudakov suppression of
  jets in QCD media}},\ }\href {https://doi.org/10.1103/PhysRevD.98.051501}
  {\bibfield  {journal} {\bibinfo  {journal} {Phys. Rev. D}\ }\textbf {\bibinfo
  {volume} {98}},\ \bibinfo {pages} {051501} (\bibinfo {year} {2018})},\
  \Eprint {https://arxiv.org/abs/1707.07361} {arXiv:1707.07361 [hep-ph]}
  \BibitemShut {NoStop}%
\bibitem [{\citenamefont {Mehtar-Tani}\ \emph {et~al.}(2011)\citenamefont
  {Mehtar-Tani}, \citenamefont {Salgado},\ and\ \citenamefont
  {Tywoniuk}}]{Mehtar-Tani:2010ebp}%
  \BibitemOpen
  \bibfield  {author} {\bibinfo {author} {\bibfnamefont {Y.}~\bibnamefont
  {Mehtar-Tani}}, \bibinfo {author} {\bibfnamefont {C.~A.}\ \bibnamefont
  {Salgado}},\ and\ \bibinfo {author} {\bibfnamefont {K.}~\bibnamefont
  {Tywoniuk}},\ }\bibfield  {title} {\bibinfo {title} {{Anti-angular ordering
  of gluon radiation in QCD media}},\ }\href
  {https://doi.org/10.1103/PhysRevLett.106.122002} {\bibfield  {journal}
  {\bibinfo  {journal} {Phys. Rev. Lett.}\ }\textbf {\bibinfo {volume} {106}},\
  \bibinfo {pages} {122002} (\bibinfo {year} {2011})},\ \Eprint
  {https://arxiv.org/abs/1009.2965} {arXiv:1009.2965 [hep-ph]} \BibitemShut
  {NoStop}%
\bibitem [{\citenamefont {Mehtar-Tani}\ \emph {et~al.}(2012)\citenamefont
  {Mehtar-Tani}, \citenamefont {Salgado},\ and\ \citenamefont
  {Tywoniuk}}]{Mehtar-Tani:2011hma}%
  \BibitemOpen
  \bibfield  {author} {\bibinfo {author} {\bibfnamefont {Y.}~\bibnamefont
  {Mehtar-Tani}}, \bibinfo {author} {\bibfnamefont {C.~A.}\ \bibnamefont
  {Salgado}},\ and\ \bibinfo {author} {\bibfnamefont {K.}~\bibnamefont
  {Tywoniuk}},\ }\bibfield  {title} {\bibinfo {title} {{Jets in QCD Media: From
  Color Coherence to Decoherence}},\ }\href
  {https://doi.org/10.1016/j.physletb.2011.12.042} {\bibfield  {journal}
  {\bibinfo  {journal} {Phys. Lett. B}\ }\textbf {\bibinfo {volume} {707}},\
  \bibinfo {pages} {156} (\bibinfo {year} {2012})},\ \Eprint
  {https://arxiv.org/abs/1102.4317} {arXiv:1102.4317 [hep-ph]} \BibitemShut
  {NoStop}%
\bibitem [{\citenamefont {Casalderrey-Solana}\ and\ \citenamefont
  {Iancu}(2011)}]{Casalderrey-Solana:2011ule}%
  \BibitemOpen
  \bibfield  {author} {\bibinfo {author} {\bibfnamefont {J.}~\bibnamefont
  {Casalderrey-Solana}}\ and\ \bibinfo {author} {\bibfnamefont
  {E.}~\bibnamefont {Iancu}},\ }\bibfield  {title} {\bibinfo {title}
  {{Interference effects in medium-induced gluon radiation}},\ }\href
  {https://doi.org/10.1007/JHEP08(2011)015} {\bibfield  {journal} {\bibinfo
  {journal} {JHEP}\ }\textbf {\bibinfo {volume} {08}},\ \bibinfo {pages}
  {015}},\ \Eprint {https://arxiv.org/abs/1105.1760} {arXiv:1105.1760 [hep-ph]}
  \BibitemShut {NoStop}%
\bibitem [{\citenamefont {M{\"a}ntysaari}\ \emph {et~al.}(2025)\citenamefont
  {M{\"a}ntysaari}, \citenamefont {Schenke}, \citenamefont {Shen},\ and\
  \citenamefont {Zhao}}]{Mantysaari:2025tcg}%
  \BibitemOpen
  \bibfield  {author} {\bibinfo {author} {\bibfnamefont {H.}~\bibnamefont
  {M{\"a}ntysaari}}, \bibinfo {author} {\bibfnamefont {B.}~\bibnamefont
  {Schenke}}, \bibinfo {author} {\bibfnamefont {C.}~\bibnamefont {Shen}},\ and\
  \bibinfo {author} {\bibfnamefont {W.}~\bibnamefont {Zhao}},\ }\bibfield
  {title} {\bibinfo {title} {{Collision-Energy Dependence in Heavy-Ion
  Collisions from Nonlinear QCD Evolution}},\ }\href
  {https://doi.org/10.1103/gf4y-p5j7} {\bibfield  {journal} {\bibinfo
  {journal} {Phys. Rev. Lett.}\ }\textbf {\bibinfo {volume} {135}},\ \bibinfo
  {pages} {022302} (\bibinfo {year} {2025})},\ \Eprint
  {https://arxiv.org/abs/2502.05138} {arXiv:2502.05138 [nucl-th]} \BibitemShut
  {NoStop}%
\bibitem [{\citenamefont {Kowalski}\ and\ \citenamefont
  {Teaney}(2003)}]{Kowalski:2003hm}%
  \BibitemOpen
  \bibfield  {author} {\bibinfo {author} {\bibfnamefont {H.}~\bibnamefont
  {Kowalski}}\ and\ \bibinfo {author} {\bibfnamefont {D.}~\bibnamefont
  {Teaney}},\ }\bibfield  {title} {\bibinfo {title} {{An Impact parameter
  dipole saturation model}},\ }\href
  {https://doi.org/10.1103/PhysRevD.68.114005} {\bibfield  {journal} {\bibinfo
  {journal} {Phys. Rev. D}\ }\textbf {\bibinfo {volume} {68}},\ \bibinfo
  {pages} {114005} (\bibinfo {year} {2003})},\ \Eprint
  {https://arxiv.org/abs/hep-ph/0304189} {arXiv:hep-ph/0304189} \BibitemShut
  {NoStop}%
\bibitem [{\citenamefont {Schlichting}\ and\ \citenamefont
  {Teaney}(2019)}]{Schlichting:2019abc}%
  \BibitemOpen
  \bibfield  {author} {\bibinfo {author} {\bibfnamefont {S.}~\bibnamefont
  {Schlichting}}\ and\ \bibinfo {author} {\bibfnamefont {D.}~\bibnamefont
  {Teaney}},\ }\bibfield  {title} {\bibinfo {title} {{The First fm/c of
  Heavy-Ion Collisions}},\ }\href
  {https://doi.org/10.1146/annurev-nucl-101918-023825} {\bibfield  {journal}
  {\bibinfo  {journal} {Ann. Rev. Nucl. Part. Sci.}\ }\textbf {\bibinfo
  {volume} {69}},\ \bibinfo {pages} {447} (\bibinfo {year} {2019})},\ \Eprint
  {https://arxiv.org/abs/1908.02113} {arXiv:1908.02113 [nucl-th]} \BibitemShut
  {NoStop}%
\bibitem [{\citenamefont {Berges}\ \emph {et~al.}(2021)\citenamefont {Berges},
  \citenamefont {Heller}, \citenamefont {Mazeliauskas},\ and\ \citenamefont
  {Venugopalan}}]{Berges:2020fwq}%
  \BibitemOpen
  \bibfield  {author} {\bibinfo {author} {\bibfnamefont {J.}~\bibnamefont
  {Berges}}, \bibinfo {author} {\bibfnamefont {M.~P.}\ \bibnamefont {Heller}},
  \bibinfo {author} {\bibfnamefont {A.}~\bibnamefont {Mazeliauskas}},\ and\
  \bibinfo {author} {\bibfnamefont {R.}~\bibnamefont {Venugopalan}},\
  }\bibfield  {title} {\bibinfo {title} {{QCD thermalization: Ab initio
  approaches and interdisciplinary connections}},\ }\href
  {https://doi.org/10.1103/RevModPhys.93.035003} {\bibfield  {journal}
  {\bibinfo  {journal} {Rev. Mod. Phys.}\ }\textbf {\bibinfo {volume} {93}},\
  \bibinfo {pages} {035003} (\bibinfo {year} {2021})},\ \Eprint
  {https://arxiv.org/abs/2005.12299} {arXiv:2005.12299 [hep-th]} \BibitemShut
  {NoStop}%
\bibitem [{\citenamefont {Andres}\ \emph {et~al.}(2020)\citenamefont {Andres},
  \citenamefont {Armesto}, \citenamefont {Niemi}, \citenamefont {Paatelainen},\
  and\ \citenamefont {Salgado}}]{Andres:2019eus}%
  \BibitemOpen
  \bibfield  {author} {\bibinfo {author} {\bibfnamefont {C.}~\bibnamefont
  {Andres}}, \bibinfo {author} {\bibfnamefont {N.}~\bibnamefont {Armesto}},
  \bibinfo {author} {\bibfnamefont {H.}~\bibnamefont {Niemi}}, \bibinfo
  {author} {\bibfnamefont {R.}~\bibnamefont {Paatelainen}},\ and\ \bibinfo
  {author} {\bibfnamefont {C.~A.}\ \bibnamefont {Salgado}},\ }\bibfield
  {title} {\bibinfo {title} {{Jet quenching as a probe of the initial stages in
  heavy-ion collisions}},\ }\href
  {https://doi.org/10.1016/j.physletb.2020.135318} {\bibfield  {journal}
  {\bibinfo  {journal} {Phys. Lett. B}\ }\textbf {\bibinfo {volume} {803}},\
  \bibinfo {pages} {135318} (\bibinfo {year} {2020})},\ \Eprint
  {https://arxiv.org/abs/1902.03231} {arXiv:1902.03231 [hep-ph]} \BibitemShut
  {NoStop}%
\bibitem [{\citenamefont {Stojku}\ \emph {et~al.}(2022)\citenamefont {Stojku},
  \citenamefont {Auvinen}, \citenamefont {Djordjevic}, \citenamefont
  {Huovinen},\ and\ \citenamefont {Djordjevic}}]{Stojku:2020wkh}%
  \BibitemOpen
  \bibfield  {author} {\bibinfo {author} {\bibfnamefont {S.}~\bibnamefont
  {Stojku}}, \bibinfo {author} {\bibfnamefont {J.}~\bibnamefont {Auvinen}},
  \bibinfo {author} {\bibfnamefont {M.}~\bibnamefont {Djordjevic}}, \bibinfo
  {author} {\bibfnamefont {P.}~\bibnamefont {Huovinen}},\ and\ \bibinfo
  {author} {\bibfnamefont {M.}~\bibnamefont {Djordjevic}},\ }\bibfield  {title}
  {\bibinfo {title} {{Early evolution constrained by high-p{\ensuremath{\perp}}
  quark-gluon plasma tomography}},\ }\href
  {https://doi.org/10.1103/PhysRevC.105.L021901} {\bibfield  {journal}
  {\bibinfo  {journal} {Phys. Rev. C}\ }\textbf {\bibinfo {volume} {105}},\
  \bibinfo {pages} {L021901} (\bibinfo {year} {2022})},\ \Eprint
  {https://arxiv.org/abs/2008.08987} {arXiv:2008.08987 [nucl-th]} \BibitemShut
  {NoStop}%
\bibitem [{\citenamefont {Ipp}\ \emph {et~al.}(2020)\citenamefont {Ipp},
  \citenamefont {M{\"u}ller},\ and\ \citenamefont {Schuh}}]{Ipp:2020nfu}%
  \BibitemOpen
  \bibfield  {author} {\bibinfo {author} {\bibfnamefont {A.}~\bibnamefont
  {Ipp}}, \bibinfo {author} {\bibfnamefont {D.~I.}\ \bibnamefont
  {M{\"u}ller}},\ and\ \bibinfo {author} {\bibfnamefont {D.}~\bibnamefont
  {Schuh}},\ }\bibfield  {title} {\bibinfo {title} {{Jet momentum broadening in
  the pre-equilibrium Glasma}},\ }\href
  {https://doi.org/10.1016/j.physletb.2020.135810} {\bibfield  {journal}
  {\bibinfo  {journal} {Phys. Lett. B}\ }\textbf {\bibinfo {volume} {810}},\
  \bibinfo {pages} {135810} (\bibinfo {year} {2020})},\ \Eprint
  {https://arxiv.org/abs/2009.14206} {arXiv:2009.14206 [hep-ph]} \BibitemShut
  {NoStop}%
\bibitem [{\citenamefont {Avramescu}\ \emph {et~al.}(2023)\citenamefont
  {Avramescu}, \citenamefont {B\u{a}ran}, \citenamefont {Greco}, \citenamefont
  {Ipp}, \citenamefont {M\"uller},\ and\ \citenamefont
  {Ruggieri}}]{Avramescu:2023qvv}%
  \BibitemOpen
  \bibfield  {author} {\bibinfo {author} {\bibfnamefont {D.}~\bibnamefont
  {Avramescu}}, \bibinfo {author} {\bibfnamefont {V.}~\bibnamefont
  {B\u{a}ran}}, \bibinfo {author} {\bibfnamefont {V.}~\bibnamefont {Greco}},
  \bibinfo {author} {\bibfnamefont {A.}~\bibnamefont {Ipp}}, \bibinfo {author}
  {\bibfnamefont {D.~I.}\ \bibnamefont {M\"uller}},\ and\ \bibinfo {author}
  {\bibfnamefont {M.}~\bibnamefont {Ruggieri}},\ }\bibfield  {title} {\bibinfo
  {title} {{Simulating jets and heavy quarks in the glasma using the colored
  particle-in-cell method}},\ }\href
  {https://doi.org/10.1103/PhysRevD.107.114021} {\bibfield  {journal} {\bibinfo
   {journal} {Phys. Rev. D}\ }\textbf {\bibinfo {volume} {107}},\ \bibinfo
  {pages} {114021} (\bibinfo {year} {2023})},\ \Eprint
  {https://arxiv.org/abs/2303.05599} {arXiv:2303.05599 [hep-ph]} \BibitemShut
  {NoStop}%
\bibitem [{\citenamefont {Boguslavski}\ \emph
  {et~al.}(2024{\natexlab{a}})\citenamefont {Boguslavski}, \citenamefont
  {Kurkela}, \citenamefont {Lappi}, \citenamefont {Lindenbauer},\ and\
  \citenamefont {Peuron}}]{Boguslavski:2023alu}%
  \BibitemOpen
  \bibfield  {author} {\bibinfo {author} {\bibfnamefont {K.}~\bibnamefont
  {Boguslavski}}, \bibinfo {author} {\bibfnamefont {A.}~\bibnamefont
  {Kurkela}}, \bibinfo {author} {\bibfnamefont {T.}~\bibnamefont {Lappi}},
  \bibinfo {author} {\bibfnamefont {F.}~\bibnamefont {Lindenbauer}},\ and\
  \bibinfo {author} {\bibfnamefont {J.}~\bibnamefont {Peuron}},\ }\bibfield
  {title} {\bibinfo {title} {{Jet momentum broadening during initial stages in
  heavy-ion collisions}},\ }\href
  {https://doi.org/10.1016/j.physletb.2024.138525} {\bibfield  {journal}
  {\bibinfo  {journal} {Phys. Lett. B}\ }\textbf {\bibinfo {volume} {850}},\
  \bibinfo {pages} {138525} (\bibinfo {year} {2024}{\natexlab{a}})},\ \Eprint
  {https://arxiv.org/abs/2303.12595} {arXiv:2303.12595 [hep-ph]} \BibitemShut
  {NoStop}%
\bibitem [{\citenamefont {Boguslavski}\ \emph
  {et~al.}(2024{\natexlab{b}})\citenamefont {Boguslavski}, \citenamefont
  {Kurkela}, \citenamefont {Lappi}, \citenamefont {Lindenbauer},\ and\
  \citenamefont {Peuron}}]{Boguslavski:2023waw}%
  \BibitemOpen
  \bibfield  {author} {\bibinfo {author} {\bibfnamefont {K.}~\bibnamefont
  {Boguslavski}}, \bibinfo {author} {\bibfnamefont {A.}~\bibnamefont
  {Kurkela}}, \bibinfo {author} {\bibfnamefont {T.}~\bibnamefont {Lappi}},
  \bibinfo {author} {\bibfnamefont {F.}~\bibnamefont {Lindenbauer}},\ and\
  \bibinfo {author} {\bibfnamefont {J.}~\bibnamefont {Peuron}},\ }\bibfield
  {title} {\bibinfo {title} {{Jet quenching parameter in QCD kinetic theory}},\
  }\href {https://doi.org/10.1103/PhysRevD.110.034019} {\bibfield  {journal}
  {\bibinfo  {journal} {Phys. Rev. D}\ }\textbf {\bibinfo {volume} {110}},\
  \bibinfo {pages} {034019} (\bibinfo {year} {2024}{\natexlab{b}})},\ \Eprint
  {https://arxiv.org/abs/2312.00447} {arXiv:2312.00447 [hep-ph]} \BibitemShut
  {NoStop}%
\bibitem [{\citenamefont {Kurkela}\ \emph
  {et~al.}(2019{\natexlab{a}})\citenamefont {Kurkela}, \citenamefont
  {Mazeliauskas}, \citenamefont {Paquet}, \citenamefont {Schlichting},\ and\
  \citenamefont {Teaney}}]{Kurkela:2018wud}%
  \BibitemOpen
  \bibfield  {author} {\bibinfo {author} {\bibfnamefont {A.}~\bibnamefont
  {Kurkela}}, \bibinfo {author} {\bibfnamefont {A.}~\bibnamefont
  {Mazeliauskas}}, \bibinfo {author} {\bibfnamefont {J.-F.}\ \bibnamefont
  {Paquet}}, \bibinfo {author} {\bibfnamefont {S.}~\bibnamefont
  {Schlichting}},\ and\ \bibinfo {author} {\bibfnamefont {D.}~\bibnamefont
  {Teaney}},\ }\bibfield  {title} {\bibinfo {title} {{Matching the
  Nonequilibrium Initial Stage of Heavy Ion Collisions to Hydrodynamics with
  QCD Kinetic Theory}},\ }\href
  {https://doi.org/10.1103/PhysRevLett.122.122302} {\bibfield  {journal}
  {\bibinfo  {journal} {Phys. Rev. Lett.}\ }\textbf {\bibinfo {volume} {122}},\
  \bibinfo {pages} {122302} (\bibinfo {year} {2019}{\natexlab{a}})},\ \Eprint
  {https://arxiv.org/abs/1805.01604} {arXiv:1805.01604 [hep-ph]} \BibitemShut
  {NoStop}%
\bibitem [{\citenamefont {Kurkela}\ \emph
  {et~al.}(2019{\natexlab{b}})\citenamefont {Kurkela}, \citenamefont
  {Mazeliauskas}, \citenamefont {Paquet}, \citenamefont {Schlichting},\ and\
  \citenamefont {Teaney}}]{Kurkela:2018vqr}%
  \BibitemOpen
  \bibfield  {author} {\bibinfo {author} {\bibfnamefont {A.}~\bibnamefont
  {Kurkela}}, \bibinfo {author} {\bibfnamefont {A.}~\bibnamefont
  {Mazeliauskas}}, \bibinfo {author} {\bibfnamefont {J.-F.}\ \bibnamefont
  {Paquet}}, \bibinfo {author} {\bibfnamefont {S.}~\bibnamefont
  {Schlichting}},\ and\ \bibinfo {author} {\bibfnamefont {D.}~\bibnamefont
  {Teaney}},\ }\bibfield  {title} {\bibinfo {title} {{Effective kinetic
  description of event-by-event pre-equilibrium dynamics in high-energy
  heavy-ion collisions}},\ }\href {https://doi.org/10.1103/PhysRevC.99.034910}
  {\bibfield  {journal} {\bibinfo  {journal} {Phys. Rev. C}\ }\textbf {\bibinfo
  {volume} {99}},\ \bibinfo {pages} {034910} (\bibinfo {year}
  {2019}{\natexlab{b}})},\ \Eprint {https://arxiv.org/abs/1805.00961}
  {arXiv:1805.00961 [hep-ph]} \BibitemShut {NoStop}%
\bibitem [{\citenamefont {Mazeliauskas}\ and\ \citenamefont
  {Berges}(2019)}]{Mazeliauskas:2018yef}%
  \BibitemOpen
  \bibfield  {author} {\bibinfo {author} {\bibfnamefont {A.}~\bibnamefont
  {Mazeliauskas}}\ and\ \bibinfo {author} {\bibfnamefont {J.}~\bibnamefont
  {Berges}},\ }\bibfield  {title} {\bibinfo {title} {{Prescaling and
  far-from-equilibrium hydrodynamics in the quark-gluon plasma}},\ }\href
  {https://doi.org/10.1103/PhysRevLett.122.122301} {\bibfield  {journal}
  {\bibinfo  {journal} {Phys. Rev. Lett.}\ }\textbf {\bibinfo {volume} {122}},\
  \bibinfo {pages} {122301} (\bibinfo {year} {2019})},\ \Eprint
  {https://arxiv.org/abs/1810.10554} {arXiv:1810.10554 [hep-ph]} \BibitemShut
  {NoStop}%
\bibitem [{\citenamefont {Kurkela}\ \emph {et~al.}(2020)\citenamefont
  {Kurkela}, \citenamefont {van~der Schee}, \citenamefont {Wiedemann},\ and\
  \citenamefont {Wu}}]{Kurkela:2019set}%
  \BibitemOpen
  \bibfield  {author} {\bibinfo {author} {\bibfnamefont {A.}~\bibnamefont
  {Kurkela}}, \bibinfo {author} {\bibfnamefont {W.}~\bibnamefont {van~der
  Schee}}, \bibinfo {author} {\bibfnamefont {U.~A.}\ \bibnamefont
  {Wiedemann}},\ and\ \bibinfo {author} {\bibfnamefont {B.}~\bibnamefont
  {Wu}},\ }\bibfield  {title} {\bibinfo {title} {{Early- and Late-Time Behavior
  of Attractors in Heavy-Ion Collisions}},\ }\href
  {https://doi.org/10.1103/PhysRevLett.124.102301} {\bibfield  {journal}
  {\bibinfo  {journal} {Phys. Rev. Lett.}\ }\textbf {\bibinfo {volume} {124}},\
  \bibinfo {pages} {102301} (\bibinfo {year} {2020})},\ \Eprint
  {https://arxiv.org/abs/1907.08101} {arXiv:1907.08101 [hep-ph]} \BibitemShut
  {NoStop}%
\bibitem [{\citenamefont {Almaalol}\ \emph {et~al.}(2020)\citenamefont
  {Almaalol}, \citenamefont {Kurkela},\ and\ \citenamefont
  {Strickland}}]{Almaalol:2020rnu}%
  \BibitemOpen
  \bibfield  {author} {\bibinfo {author} {\bibfnamefont {D.}~\bibnamefont
  {Almaalol}}, \bibinfo {author} {\bibfnamefont {A.}~\bibnamefont {Kurkela}},\
  and\ \bibinfo {author} {\bibfnamefont {M.}~\bibnamefont {Strickland}},\
  }\bibfield  {title} {\bibinfo {title} {{Nonequilibrium Attractor in
  High-Temperature QCD Plasmas}},\ }\href
  {https://doi.org/10.1103/PhysRevLett.125.122302} {\bibfield  {journal}
  {\bibinfo  {journal} {Phys. Rev. Lett.}\ }\textbf {\bibinfo {volume} {125}},\
  \bibinfo {pages} {122302} (\bibinfo {year} {2020})},\ \Eprint
  {https://arxiv.org/abs/2004.05195} {arXiv:2004.05195 [hep-ph]} \BibitemShut
  {NoStop}%
\bibitem [{\citenamefont {Soloviev}(2022)}]{Soloviev:2021lhs}%
  \BibitemOpen
  \bibfield  {author} {\bibinfo {author} {\bibfnamefont {A.}~\bibnamefont
  {Soloviev}},\ }\bibfield  {title} {\bibinfo {title} {{Hydrodynamic attractors
  in heavy ion collisions: a review}},\ }\href
  {https://doi.org/10.1140/epjc/s10052-022-10282-4} {\bibfield  {journal}
  {\bibinfo  {journal} {Eur. Phys. J. C}\ }\textbf {\bibinfo {volume} {82}},\
  \bibinfo {pages} {319} (\bibinfo {year} {2022})},\ \Eprint
  {https://arxiv.org/abs/2109.15081} {arXiv:2109.15081 [hep-th]} \BibitemShut
  {NoStop}%
\bibitem [{\citenamefont {Giacalone}\ \emph {et~al.}(2019)\citenamefont
  {Giacalone}, \citenamefont {Mazeliauskas},\ and\ \citenamefont
  {Schlichting}}]{Giacalone:2019ldn}%
  \BibitemOpen
  \bibfield  {author} {\bibinfo {author} {\bibfnamefont {G.}~\bibnamefont
  {Giacalone}}, \bibinfo {author} {\bibfnamefont {A.}~\bibnamefont
  {Mazeliauskas}},\ and\ \bibinfo {author} {\bibfnamefont {S.}~\bibnamefont
  {Schlichting}},\ }\bibfield  {title} {\bibinfo {title} {{Hydrodynamic
  attractors, initial state energy and particle production in relativistic
  nuclear collisions}},\ }\href
  {https://doi.org/10.1103/PhysRevLett.123.262301} {\bibfield  {journal}
  {\bibinfo  {journal} {Phys. Rev. Lett.}\ }\textbf {\bibinfo {volume} {123}},\
  \bibinfo {pages} {262301} (\bibinfo {year} {2019})},\ \Eprint
  {https://arxiv.org/abs/1908.02866} {arXiv:1908.02866 [hep-ph]} \BibitemShut
  {NoStop}%
\bibitem [{\citenamefont {Garcia-Montero}\ \emph {et~al.}(2024)\citenamefont
  {Garcia-Montero}, \citenamefont {Mazeliauskas}, \citenamefont {Plaschke},\
  and\ \citenamefont {Schlichting}}]{Garcia-Montero:2023lrd}%
  \BibitemOpen
  \bibfield  {author} {\bibinfo {author} {\bibfnamefont {O.}~\bibnamefont
  {Garcia-Montero}}, \bibinfo {author} {\bibfnamefont {A.}~\bibnamefont
  {Mazeliauskas}}, \bibinfo {author} {\bibfnamefont {P.}~\bibnamefont
  {Plaschke}},\ and\ \bibinfo {author} {\bibfnamefont {S.}~\bibnamefont
  {Schlichting}},\ }\bibfield  {title} {\bibinfo {title} {{Pre-equilibrium
  photons from the early stages of heavy-ion collisions}},\ }\href
  {https://doi.org/10.1007/JHEP03(2024)053} {\bibfield  {journal} {\bibinfo
  {journal} {JHEP}\ }\textbf {\bibinfo {volume} {03}},\ \bibinfo {pages}
  {053}},\ \Eprint {https://arxiv.org/abs/2308.09747} {arXiv:2308.09747
  [hep-ph]} \BibitemShut {NoStop}%
\bibitem [{\citenamefont {Keegan}\ \emph {et~al.}(2016)\citenamefont {Keegan},
  \citenamefont {Kurkela}, \citenamefont {Mazeliauskas},\ and\ \citenamefont
  {Teaney}}]{Keegan:2016cpi}%
  \BibitemOpen
  \bibfield  {author} {\bibinfo {author} {\bibfnamefont {L.}~\bibnamefont
  {Keegan}}, \bibinfo {author} {\bibfnamefont {A.}~\bibnamefont {Kurkela}},
  \bibinfo {author} {\bibfnamefont {A.}~\bibnamefont {Mazeliauskas}},\ and\
  \bibinfo {author} {\bibfnamefont {D.}~\bibnamefont {Teaney}},\ }\bibfield
  {title} {\bibinfo {title} {{Initial conditions for hydrodynamics from weakly
  coupled pre-equilibrium evolution}},\ }\href
  {https://doi.org/10.1007/JHEP08(2016)171} {\bibfield  {journal} {\bibinfo
  {journal} {JHEP}\ }\textbf {\bibinfo {volume} {08}},\ \bibinfo {pages}
  {171}},\ \Eprint {https://arxiv.org/abs/1605.04287} {arXiv:1605.04287
  [hep-ph]} \BibitemShut {NoStop}%
\bibitem [{\citenamefont {Aad}\ \emph {et~al.}(2022)\citenamefont {Aad} \emph
  {et~al.}}]{ATLAS:2021ktw}%
  \BibitemOpen
  \bibfield  {author} {\bibinfo {author} {\bibfnamefont {G.}~\bibnamefont
  {Aad}} \emph {et~al.} (\bibinfo {collaboration} {ATLAS}),\ }\bibfield
  {title} {\bibinfo {title} {{Measurements of azimuthal anisotropies of jet
  production in Pb+Pb collisions at $\sqrt{s_{NN}} =$ 5.02 TeV with the ATLAS
  detector}},\ }\href {https://doi.org/10.1103/PhysRevC.105.064903} {\bibfield
  {journal} {\bibinfo  {journal} {Phys. Rev. C}\ }\textbf {\bibinfo {volume}
  {105}},\ \bibinfo {pages} {064903} (\bibinfo {year} {2022})},\ \Eprint
  {https://arxiv.org/abs/2111.06606} {arXiv:2111.06606 [nucl-ex]} \BibitemShut
  {NoStop}%
\bibitem [{\citenamefont {Noronha-Hostler}\ \emph {et~al.}(2016)\citenamefont
  {Noronha-Hostler}, \citenamefont {Betz}, \citenamefont {Noronha},\ and\
  \citenamefont {Gyulassy}}]{Noronha-Hostler:2016eow}%
  \BibitemOpen
  \bibfield  {author} {\bibinfo {author} {\bibfnamefont {J.}~\bibnamefont
  {Noronha-Hostler}}, \bibinfo {author} {\bibfnamefont {B.}~\bibnamefont
  {Betz}}, \bibinfo {author} {\bibfnamefont {J.}~\bibnamefont {Noronha}},\ and\
  \bibinfo {author} {\bibfnamefont {M.}~\bibnamefont {Gyulassy}},\ }\bibfield
  {title} {\bibinfo {title} {{Event-by-event hydrodynamics $+$ jet energy loss:
  A solution to the $R_{AA} \otimes v_2$ puzzle}},\ }\href
  {https://doi.org/10.1103/PhysRevLett.116.252301} {\bibfield  {journal}
  {\bibinfo  {journal} {Phys. Rev. Lett.}\ }\textbf {\bibinfo {volume} {116}},\
  \bibinfo {pages} {252301} (\bibinfo {year} {2016})},\ \Eprint
  {https://arxiv.org/abs/1602.03788} {arXiv:1602.03788 [nucl-th]} \BibitemShut
  {NoStop}%
\bibitem [{\citenamefont {Aaboud}\ \emph {et~al.}(2018)\citenamefont {Aaboud}
  \emph {et~al.}}]{ATLAS:2018ezv}%
  \BibitemOpen
  \bibfield  {author} {\bibinfo {author} {\bibfnamefont {M.}~\bibnamefont
  {Aaboud}} \emph {et~al.} (\bibinfo {collaboration} {ATLAS}),\ }\bibfield
  {title} {\bibinfo {title} {{Measurement of the azimuthal anisotropy of
  charged particles produced in $\sqrt{s_{_\text {NN}}}$ = 5.02 TeV Pb+Pb
  collisions with the ATLAS detector}},\ }\href
  {https://doi.org/10.1140/epjc/s10052-018-6468-7} {\bibfield  {journal}
  {\bibinfo  {journal} {Eur. Phys. J. C}\ }\textbf {\bibinfo {volume} {78}},\
  \bibinfo {pages} {997} (\bibinfo {year} {2018})},\ \Eprint
  {https://arxiv.org/abs/1808.03951} {arXiv:1808.03951 [nucl-ex]} \BibitemShut
  {NoStop}%
\bibitem [{\citenamefont {Betz}\ and\ \citenamefont
  {Gyulassy}(2014)}]{Betz:2014cza}%
  \BibitemOpen
  \bibfield  {author} {\bibinfo {author} {\bibfnamefont {B.}~\bibnamefont
  {Betz}}\ and\ \bibinfo {author} {\bibfnamefont {M.}~\bibnamefont
  {Gyulassy}},\ }\bibfield  {title} {\bibinfo {title} {{Constraints on the
  Path-Length Dependence of Jet Quenching in Nuclear Collisions at RHIC and
  LHC}},\ }\href {https://doi.org/10.1007/JHEP10(2014)043} {\bibfield
  {journal} {\bibinfo  {journal} {JHEP}\ }\textbf {\bibinfo {volume} {08}},\
  \bibinfo {pages} {090}},\ \bibinfo {note} {[Erratum: JHEP 10, 043 (2014)]},\
  \Eprint {https://arxiv.org/abs/1404.6378} {arXiv:1404.6378 [hep-ph]}
  \BibitemShut {NoStop}%
\bibitem [{\citenamefont {Andres}\ \emph {et~al.}(2023)\citenamefont {Andres},
  \citenamefont {Apolin{\'a}rio}, \citenamefont {Dominguez}, \citenamefont
  {Martinez},\ and\ \citenamefont {Salgado}}]{Andres:2022bql}%
  \BibitemOpen
  \bibfield  {author} {\bibinfo {author} {\bibfnamefont {C.}~\bibnamefont
  {Andres}}, \bibinfo {author} {\bibfnamefont {L.}~\bibnamefont
  {Apolin{\'a}rio}}, \bibinfo {author} {\bibfnamefont {F.}~\bibnamefont
  {Dominguez}}, \bibinfo {author} {\bibfnamefont {M.~G.}\ \bibnamefont
  {Martinez}},\ and\ \bibinfo {author} {\bibfnamefont {C.~A.}\ \bibnamefont
  {Salgado}},\ }\bibfield  {title} {\bibinfo {title} {{Medium-induced radiation
  with vacuum propagation in the pre-hydrodynamics phase}},\ }\href
  {https://doi.org/10.1007/JHEP03(2023)189} {\bibfield  {journal} {\bibinfo
  {journal} {JHEP}\ }\textbf {\bibinfo {volume} {03}},\ \bibinfo {pages}
  {189}},\ \Eprint {https://arxiv.org/abs/2211.10161} {arXiv:2211.10161
  [hep-ph]} \BibitemShut {NoStop}%
\bibitem [{\citenamefont {Arleo}\ and\ \citenamefont
  {Falmagne}(2024)}]{Arleo:2022shs}%
  \BibitemOpen
  \bibfield  {author} {\bibinfo {author} {\bibfnamefont {F.}~\bibnamefont
  {Arleo}}\ and\ \bibinfo {author} {\bibfnamefont {G.}~\bibnamefont
  {Falmagne}},\ }\bibfield  {title} {\bibinfo {title} {{Probing the path-length
  dependence of parton energy loss via scaling properties in heavy ion
  collisions}},\ }\href {https://doi.org/10.1103/PhysRevD.109.L051503}
  {\bibfield  {journal} {\bibinfo  {journal} {Phys. Rev. D}\ }\textbf {\bibinfo
  {volume} {109}},\ \bibinfo {pages} {L051503} (\bibinfo {year} {2024})},\
  \Eprint {https://arxiv.org/abs/2212.01324} {arXiv:2212.01324 [hep-ph]}
  \BibitemShut {NoStop}%
\bibitem [{\citenamefont {Faraday}\ and\ \citenamefont
  {Horowitz}(2025)}]{Faraday:2024qtl}%
  \BibitemOpen
  \bibfield  {author} {\bibinfo {author} {\bibfnamefont {C.}~\bibnamefont
  {Faraday}}\ and\ \bibinfo {author} {\bibfnamefont {W.~A.}\ \bibnamefont
  {Horowitz}},\ }\bibfield  {title} {\bibinfo {title} {{A unified description
  of small, peripheral, and large system suppression data from pQCD}},\ }\href
  {https://doi.org/10.1016/j.physletb.2025.139437} {\bibfield  {journal}
  {\bibinfo  {journal} {Phys. Lett. B}\ }\textbf {\bibinfo {volume} {864}},\
  \bibinfo {pages} {139437} (\bibinfo {year} {2025})},\ \Eprint
  {https://arxiv.org/abs/2411.09647} {arXiv:2411.09647 [hep-ph]} \BibitemShut
  {NoStop}%
\bibitem [{\citenamefont {Zigic}\ \emph
  {et~al.}(2019{\natexlab{a}})\citenamefont {Zigic}, \citenamefont {Salom},
  \citenamefont {Auvinen}, \citenamefont {Djordjevic},\ and\ \citenamefont
  {Djordjevic}}]{Zigic:2018smz}%
  \BibitemOpen
  \bibfield  {author} {\bibinfo {author} {\bibfnamefont {D.}~\bibnamefont
  {Zigic}}, \bibinfo {author} {\bibfnamefont {I.}~\bibnamefont {Salom}},
  \bibinfo {author} {\bibfnamefont {J.}~\bibnamefont {Auvinen}}, \bibinfo
  {author} {\bibfnamefont {M.}~\bibnamefont {Djordjevic}},\ and\ \bibinfo
  {author} {\bibfnamefont {M.}~\bibnamefont {Djordjevic}},\ }\bibfield  {title}
  {\bibinfo {title} {{DREENA-C framework: joint $R_{AA}$ and $v_2$ predictions
  and implications to QGP tomography}},\ }\href
  {https://doi.org/10.1088/1361-6471/ab2356} {\bibfield  {journal} {\bibinfo
  {journal} {J. Phys. G}\ }\textbf {\bibinfo {volume} {46}},\ \bibinfo {pages}
  {085101} (\bibinfo {year} {2019}{\natexlab{a}})},\ \Eprint
  {https://arxiv.org/abs/1805.03494} {arXiv:1805.03494 [nucl-th]} \BibitemShut
  {NoStop}%
\bibitem [{\citenamefont {Zigic}\ \emph
  {et~al.}(2019{\natexlab{b}})\citenamefont {Zigic}, \citenamefont {Salom},
  \citenamefont {Auvinen}, \citenamefont {Djordjevic},\ and\ \citenamefont
  {Djordjevic}}]{Zigic:2018ovr}%
  \BibitemOpen
  \bibfield  {author} {\bibinfo {author} {\bibfnamefont {D.}~\bibnamefont
  {Zigic}}, \bibinfo {author} {\bibfnamefont {I.}~\bibnamefont {Salom}},
  \bibinfo {author} {\bibfnamefont {J.}~\bibnamefont {Auvinen}}, \bibinfo
  {author} {\bibfnamefont {M.}~\bibnamefont {Djordjevic}},\ and\ \bibinfo
  {author} {\bibfnamefont {M.}~\bibnamefont {Djordjevic}},\ }\bibfield  {title}
  {\bibinfo {title} {{DREENA-B framework: first predictions of $R_{AA}$ and
  $v_2$ within dynamical energy loss formalism in evolving QCD medium}},\
  }\href {https://doi.org/10.1016/j.physletb.2019.02.020} {\bibfield  {journal}
  {\bibinfo  {journal} {Phys. Lett. B}\ }\textbf {\bibinfo {volume} {791}},\
  \bibinfo {pages} {236} (\bibinfo {year} {2019}{\natexlab{b}})},\ \Eprint
  {https://arxiv.org/abs/1805.04786} {arXiv:1805.04786 [nucl-th]} \BibitemShut
  {NoStop}%
\bibitem [{\citenamefont {Zigic}\ \emph {et~al.}(2020)\citenamefont {Zigic},
  \citenamefont {Ilic}, \citenamefont {Djordjevic},\ and\ \citenamefont
  {Djordjevic}}]{Zigic:2019sth}%
  \BibitemOpen
  \bibfield  {author} {\bibinfo {author} {\bibfnamefont {D.}~\bibnamefont
  {Zigic}}, \bibinfo {author} {\bibfnamefont {B.}~\bibnamefont {Ilic}},
  \bibinfo {author} {\bibfnamefont {M.}~\bibnamefont {Djordjevic}},\ and\
  \bibinfo {author} {\bibfnamefont {M.}~\bibnamefont {Djordjevic}},\ }\bibfield
   {title} {\bibinfo {title} {{Exploring the initial stages in heavy-ion
  collisions with high-$p_\bot$ $R_{AA}$ and $v_2$ theory and data}},\ }\href
  {https://doi.org/10.1103/PhysRevC.101.064909} {\bibfield  {journal} {\bibinfo
   {journal} {Phys. Rev. C}\ }\textbf {\bibinfo {volume} {101}},\ \bibinfo
  {pages} {064909} (\bibinfo {year} {2020})},\ \Eprint
  {https://arxiv.org/abs/1908.11866} {arXiv:1908.11866 [hep-ph]} \BibitemShut
  {NoStop}%
\bibitem [{\citenamefont {Zigic}\ \emph
  {et~al.}(2022{\natexlab{a}})\citenamefont {Zigic}, \citenamefont {Salom},
  \citenamefont {Auvinen}, \citenamefont {Huovinen},\ and\ \citenamefont
  {Djordjevic}}]{Zigic:2021rku}%
  \BibitemOpen
  \bibfield  {author} {\bibinfo {author} {\bibfnamefont {D.}~\bibnamefont
  {Zigic}}, \bibinfo {author} {\bibfnamefont {I.}~\bibnamefont {Salom}},
  \bibinfo {author} {\bibfnamefont {J.}~\bibnamefont {Auvinen}}, \bibinfo
  {author} {\bibfnamefont {P.}~\bibnamefont {Huovinen}},\ and\ \bibinfo
  {author} {\bibfnamefont {M.}~\bibnamefont {Djordjevic}},\ }\bibfield  {title}
  {\bibinfo {title} {{DREENA-A framework as a QGP tomography tool}},\ }\href
  {https://doi.org/10.3389/fphy.2022.957019} {\bibfield  {journal} {\bibinfo
  {journal} {Front. in Phys.}\ }\textbf {\bibinfo {volume} {10}},\ \bibinfo
  {pages} {957019} (\bibinfo {year} {2022}{\natexlab{a}})},\ \Eprint
  {https://arxiv.org/abs/2110.01544} {arXiv:2110.01544 [nucl-th]} \BibitemShut
  {NoStop}%
\bibitem [{\citenamefont {Zigic}\ \emph
  {et~al.}(2022{\natexlab{b}})\citenamefont {Zigic}, \citenamefont {Auvinen},
  \citenamefont {Salom}, \citenamefont {Djordjevic},\ and\ \citenamefont
  {Huovinen}}]{Zigic:2022xks}%
  \BibitemOpen
  \bibfield  {author} {\bibinfo {author} {\bibfnamefont {D.}~\bibnamefont
  {Zigic}}, \bibinfo {author} {\bibfnamefont {J.}~\bibnamefont {Auvinen}},
  \bibinfo {author} {\bibfnamefont {I.}~\bibnamefont {Salom}}, \bibinfo
  {author} {\bibfnamefont {M.}~\bibnamefont {Djordjevic}},\ and\ \bibinfo
  {author} {\bibfnamefont {P.}~\bibnamefont {Huovinen}},\ }\bibfield  {title}
  {\bibinfo {title} {{Importance of higher harmonics and v4 puzzle in
  quark-gluon plasma tomography}},\ }\href
  {https://doi.org/10.1103/PhysRevC.106.044909} {\bibfield  {journal} {\bibinfo
   {journal} {Phys. Rev. C}\ }\textbf {\bibinfo {volume} {106}},\ \bibinfo
  {pages} {044909} (\bibinfo {year} {2022}{\natexlab{b}})},\ \Eprint
  {https://arxiv.org/abs/2208.09886} {arXiv:2208.09886 [hep-ph]} \BibitemShut
  {NoStop}%
\bibitem [{\citenamefont {Karmakar}\ \emph {et~al.}(2023)\citenamefont
  {Karmakar}, \citenamefont {Zigic}, \citenamefont {Salom}, \citenamefont
  {Auvinen}, \citenamefont {Huovinen}, \citenamefont {Djordjevic},\ and\
  \citenamefont {Djordjevic}}]{Karmakar:2023ity}%
  \BibitemOpen
  \bibfield  {author} {\bibinfo {author} {\bibfnamefont {B.}~\bibnamefont
  {Karmakar}}, \bibinfo {author} {\bibfnamefont {D.}~\bibnamefont {Zigic}},
  \bibinfo {author} {\bibfnamefont {I.}~\bibnamefont {Salom}}, \bibinfo
  {author} {\bibfnamefont {J.}~\bibnamefont {Auvinen}}, \bibinfo {author}
  {\bibfnamefont {P.}~\bibnamefont {Huovinen}}, \bibinfo {author}
  {\bibfnamefont {M.}~\bibnamefont {Djordjevic}},\ and\ \bibinfo {author}
  {\bibfnamefont {M.}~\bibnamefont {Djordjevic}},\ }\bibfield  {title}
  {\bibinfo {title} {{Constraining \ensuremath{\eta}/s through
  high-p\ensuremath{\perp} theory and data}},\ }\href
  {https://doi.org/10.1103/PhysRevC.108.044907} {\bibfield  {journal} {\bibinfo
   {journal} {Phys. Rev. C}\ }\textbf {\bibinfo {volume} {108}},\ \bibinfo
  {pages} {044907} (\bibinfo {year} {2023})},\ \Eprint
  {https://arxiv.org/abs/2305.11318} {arXiv:2305.11318 [hep-ph]} \BibitemShut
  {NoStop}%
\bibitem [{\citenamefont {Karmakar}\ \emph {et~al.}(2024)\citenamefont
  {Karmakar}, \citenamefont {Zigic}, \citenamefont {Djordjevic}, \citenamefont
  {Huovinen}, \citenamefont {Djordjevic},\ and\ \citenamefont
  {Auvinen}}]{Karmakar:2024jak}%
  \BibitemOpen
  \bibfield  {author} {\bibinfo {author} {\bibfnamefont {B.}~\bibnamefont
  {Karmakar}}, \bibinfo {author} {\bibfnamefont {D.}~\bibnamefont {Zigic}},
  \bibinfo {author} {\bibfnamefont {M.}~\bibnamefont {Djordjevic}}, \bibinfo
  {author} {\bibfnamefont {P.}~\bibnamefont {Huovinen}}, \bibinfo {author}
  {\bibfnamefont {M.}~\bibnamefont {Djordjevic}},\ and\ \bibinfo {author}
  {\bibfnamefont {J.}~\bibnamefont {Auvinen}},\ }\bibfield  {title} {\bibinfo
  {title} {{Probing the shape of the quark-gluon plasma droplet via
  event-by-event quark-gluon plasma tomography}},\ }\href
  {https://doi.org/10.1103/PhysRevC.110.044906} {\bibfield  {journal} {\bibinfo
   {journal} {Phys. Rev. C}\ }\textbf {\bibinfo {volume} {110}},\ \bibinfo
  {pages} {044906} (\bibinfo {year} {2024})},\ \Eprint
  {https://arxiv.org/abs/2403.17817} {arXiv:2403.17817 [hep-ph]} \BibitemShut
  {NoStop}%
\bibitem [{\citenamefont {Zhao}\ \emph {et~al.}(2022)\citenamefont {Zhao},
  \citenamefont {Ke}, \citenamefont {Chen}, \citenamefont {Luo},\ and\
  \citenamefont {Wang}}]{Zhao:2021vmu}%
  \BibitemOpen
  \bibfield  {author} {\bibinfo {author} {\bibfnamefont {W.}~\bibnamefont
  {Zhao}}, \bibinfo {author} {\bibfnamefont {W.}~\bibnamefont {Ke}}, \bibinfo
  {author} {\bibfnamefont {W.}~\bibnamefont {Chen}}, \bibinfo {author}
  {\bibfnamefont {T.}~\bibnamefont {Luo}},\ and\ \bibinfo {author}
  {\bibfnamefont {X.-N.}\ \bibnamefont {Wang}},\ }\bibfield  {title} {\bibinfo
  {title} {{From Hydrodynamics to Jet Quenching, Coalescence, and Hadron
  Cascade: A Coupled Approach to Solving the RAA{\ensuremath{\otimes}}v2
  Puzzle}},\ }\href {https://doi.org/10.1103/PhysRevLett.128.022302} {\bibfield
   {journal} {\bibinfo  {journal} {Phys. Rev. Lett.}\ }\textbf {\bibinfo
  {volume} {128}},\ \bibinfo {pages} {022302} (\bibinfo {year} {2022})},\
  \Eprint {https://arxiv.org/abs/2103.14657} {arXiv:2103.14657 [hep-ph]}
  \BibitemShut {NoStop}%
\bibitem [{\citenamefont {Barreto}\ \emph {et~al.}(2025)\citenamefont
  {Barreto}, \citenamefont {Canedo}, \citenamefont {Munhoz}, \citenamefont
  {Noronha},\ and\ \citenamefont {Noronha-Hostler}}]{Barreto:2022ulg}%
  \BibitemOpen
  \bibfield  {author} {\bibinfo {author} {\bibfnamefont {L.}~\bibnamefont
  {Barreto}}, \bibinfo {author} {\bibfnamefont {F.~M.}\ \bibnamefont {Canedo}},
  \bibinfo {author} {\bibfnamefont {M.~G.}\ \bibnamefont {Munhoz}}, \bibinfo
  {author} {\bibfnamefont {J.}~\bibnamefont {Noronha}},\ and\ \bibinfo {author}
  {\bibfnamefont {J.}~\bibnamefont {Noronha-Hostler}},\ }\bibfield  {title}
  {\bibinfo {title} {{Jet cone radius dependence of RAA and v2 at PbPb 5.02 TeV
  from JEWEL+TRENTo+v-USPhydro}},\ }\href
  {https://doi.org/10.1016/j.physletb.2024.139217} {\bibfield  {journal}
  {\bibinfo  {journal} {Phys. Lett. B}\ }\textbf {\bibinfo {volume} {860}},\
  \bibinfo {pages} {139217} (\bibinfo {year} {2025})},\ \Eprint
  {https://arxiv.org/abs/2208.02061} {arXiv:2208.02061 [nucl-th]} \BibitemShut
  {NoStop}%
\bibitem [{\citenamefont {He}\ \emph {et~al.}(2022)\citenamefont {He},
  \citenamefont {Chen}, \citenamefont {Luo}, \citenamefont {Cao}, \citenamefont
  {Pang},\ and\ \citenamefont {Wang}}]{He:2022evt}%
  \BibitemOpen
  \bibfield  {author} {\bibinfo {author} {\bibfnamefont {Y.}~\bibnamefont
  {He}}, \bibinfo {author} {\bibfnamefont {W.}~\bibnamefont {Chen}}, \bibinfo
  {author} {\bibfnamefont {T.}~\bibnamefont {Luo}}, \bibinfo {author}
  {\bibfnamefont {S.}~\bibnamefont {Cao}}, \bibinfo {author} {\bibfnamefont
  {L.-G.}\ \bibnamefont {Pang}},\ and\ \bibinfo {author} {\bibfnamefont
  {X.-N.}\ \bibnamefont {Wang}},\ }\bibfield  {title} {\bibinfo {title}
  {{Event-by-event jet anisotropy and hard-soft tomography of the quark-gluon
  plasma}},\ }\href {https://doi.org/10.1103/PhysRevC.106.044904} {\bibfield
  {journal} {\bibinfo  {journal} {Phys. Rev. C}\ }\textbf {\bibinfo {volume}
  {106}},\ \bibinfo {pages} {044904} (\bibinfo {year} {2022})},\ \Eprint
  {https://arxiv.org/abs/2201.08408} {arXiv:2201.08408 [hep-ph]} \BibitemShut
  {NoStop}%
\bibitem [{\citenamefont {Ellis}\ \emph {et~al.}(2003)\citenamefont {Ellis},
  \citenamefont {Stirling},\ and\ \citenamefont {Webber}}]{Ellis:318585}%
  \BibitemOpen
  \bibfield  {author} {\bibinfo {author} {\bibfnamefont {R.~K.}\ \bibnamefont
  {Ellis}}, \bibinfo {author} {\bibfnamefont {W.~J.}\ \bibnamefont
  {Stirling}},\ and\ \bibinfo {author} {\bibfnamefont {B.~R.}\ \bibnamefont
  {Webber}},\ }\href {https://doi.org/10.1017/CBO9780511628788} {\emph
  {\bibinfo {title} {{QCD and collider physics}}}},\ Cambridge monographs on
  particle physics, nuclear physics, and cosmology\ (\bibinfo  {publisher}
  {Cambridge University Press},\ \bibinfo {address} {Cambridge},\ \bibinfo
  {year} {2003})\ \bibinfo {note} {photography by S. Vascotto}\BibitemShut
  {NoStop}%
\bibitem [{\citenamefont {Salam}(2010)}]{Salam:2010zt}%
  \BibitemOpen
  \bibfield  {author} {\bibinfo {author} {\bibfnamefont {G.~P.}\ \bibnamefont
  {Salam}},\ }\bibfield  {title} {\bibinfo {title} {{Elements of QCD for hadron
  colliders}},\ }in\ \href@noop {} {\emph {\bibinfo {booktitle} {{2009 European
  School of High-Energy Physics}}}}\ (\bibinfo {year} {2010})\ \Eprint
  {https://arxiv.org/abs/1011.5131} {arXiv:1011.5131 [hep-ph]} \BibitemShut
  {NoStop}%
\bibitem [{\citenamefont {Luisoni}\ and\ \citenamefont
  {Marzani}(2015)}]{Luisoni:2015xha}%
  \BibitemOpen
  \bibfield  {author} {\bibinfo {author} {\bibfnamefont {G.}~\bibnamefont
  {Luisoni}}\ and\ \bibinfo {author} {\bibfnamefont {S.}~\bibnamefont
  {Marzani}},\ }\bibfield  {title} {\bibinfo {title} {{QCD resummation for
  hadronic final states}},\ }\href
  {https://doi.org/10.1088/0954-3899/42/10/103101} {\bibfield  {journal}
  {\bibinfo  {journal} {J. Phys. G}\ }\textbf {\bibinfo {volume} {42}},\
  \bibinfo {pages} {103101} (\bibinfo {year} {2015})},\ \Eprint
  {https://arxiv.org/abs/1505.04084} {arXiv:1505.04084 [hep-ph]} \BibitemShut
  {NoStop}%
\bibitem [{\citenamefont {Marzani}\ \emph {et~al.}(2019)\citenamefont
  {Marzani}, \citenamefont {Soyez},\ and\ \citenamefont
  {Spannowsky}}]{Marzani:2019hun}%
  \BibitemOpen
  \bibfield  {author} {\bibinfo {author} {\bibfnamefont {S.}~\bibnamefont
  {Marzani}}, \bibinfo {author} {\bibfnamefont {G.}~\bibnamefont {Soyez}},\
  and\ \bibinfo {author} {\bibfnamefont {M.}~\bibnamefont {Spannowsky}},\
  }\href {https://doi.org/10.1007/978-3-030-15709-8} {\emph {\bibinfo {title}
  {{Looking inside jets: an introduction to jet substructure and boosted-object
  phenomenology}}}},\ Vol.\ \bibinfo {volume} {958}\ (\bibinfo  {publisher}
  {Springer},\ \bibinfo {year} {2019})\ \Eprint
  {https://arxiv.org/abs/1901.10342} {arXiv:1901.10342 [hep-ph]} \BibitemShut
  {NoStop}%
\bibitem [{\citenamefont {Collaboration}(2025)}]{JETSCAPE/STAT_GitHub}%
  \BibitemOpen
  \bibfield  {author} {\bibinfo {author} {\bibfnamefont {J.}~\bibnamefont
  {Collaboration}},\ }\href@noop {} {\bibinfo {title} {Jetscape/stat}},\
  \bibinfo {howpublished} {\url{https://github.com/JETSCAPE/STAT}} (\bibinfo
  {year} {2025}),\ \bibinfo {note} {accessed: (04.08.2025)}\BibitemShut
  {NoStop}%
\bibitem [{\citenamefont {Cao}\ \emph {et~al.}(2021)\citenamefont {Cao} \emph
  {et~al.}}]{JETSCAPE:2021ehl}%
  \BibitemOpen
  \bibfield  {author} {\bibinfo {author} {\bibfnamefont {S.}~\bibnamefont
  {Cao}} \emph {et~al.} (\bibinfo {collaboration} {JETSCAPE}),\ }\bibfield
  {title} {\bibinfo {title} {{Determining the jet transport coefficient
  $\hat{q}$ from inclusive hadron suppression measurements using Bayesian
  parameter estimation}},\ }\href {https://doi.org/10.1103/PhysRevC.104.024905}
  {\bibfield  {journal} {\bibinfo  {journal} {Phys. Rev. C}\ }\textbf {\bibinfo
  {volume} {104}},\ \bibinfo {pages} {024905} (\bibinfo {year} {2021})},\
  \Eprint {https://arxiv.org/abs/2102.11337} {arXiv:2102.11337 [nucl-th]}
  \BibitemShut {NoStop}%
\bibitem [{\citenamefont {Ehlers}\ \emph {et~al.}(2025)\citenamefont {Ehlers}
  \emph {et~al.}}]{JETSCAPE:2024cqe}%
  \BibitemOpen
  \bibfield  {author} {\bibinfo {author} {\bibfnamefont {R.}~\bibnamefont
  {Ehlers}} \emph {et~al.} (\bibinfo {collaboration} {JETSCAPE}),\ }\bibfield
  {title} {\bibinfo {title} {{Bayesian inference analysis of jet quenching
  using inclusive jet and hadron suppression measurements}},\ }\href
  {https://doi.org/10.1103/PhysRevC.111.054913} {\bibfield  {journal} {\bibinfo
   {journal} {Phys. Rev. C}\ }\textbf {\bibinfo {volume} {111}},\ \bibinfo
  {pages} {054913} (\bibinfo {year} {2025})},\ \Eprint
  {https://arxiv.org/abs/2408.08247} {arXiv:2408.08247 [hep-ph]} \BibitemShut
  {NoStop}%
\bibitem [{\citenamefont {Aaboud}\ \emph {et~al.}(2019)\citenamefont {Aaboud}
  \emph {et~al.}}]{ATLAS:2018gwx}%
  \BibitemOpen
  \bibfield  {author} {\bibinfo {author} {\bibfnamefont {M.}~\bibnamefont
  {Aaboud}} \emph {et~al.} (\bibinfo {collaboration} {ATLAS}),\ }\bibfield
  {title} {\bibinfo {title} {{Measurement of the nuclear modification factor
  for inclusive jets in Pb+Pb collisions at $\sqrt{s_\mathrm{NN}}=5.02$ TeV
  with the ATLAS detector}},\ }\href
  {https://doi.org/10.1016/j.physletb.2018.10.076} {\bibfield  {journal}
  {\bibinfo  {journal} {Phys. Lett. B}\ }\textbf {\bibinfo {volume} {790}},\
  \bibinfo {pages} {108} (\bibinfo {year} {2019})},\ \Eprint
  {https://arxiv.org/abs/1805.05635} {arXiv:1805.05635 [nucl-ex]} \BibitemShut
  {NoStop}%
\bibitem [{\citenamefont {Acharya}\ \emph {et~al.}(2020)\citenamefont {Acharya}
  \emph {et~al.}}]{ALICE:2019qyj}%
  \BibitemOpen
  \bibfield  {author} {\bibinfo {author} {\bibfnamefont {S.}~\bibnamefont
  {Acharya}} \emph {et~al.} (\bibinfo {collaboration} {ALICE}),\ }\bibfield
  {title} {\bibinfo {title} {{Measurements of inclusive jet spectra in pp and
  central Pb-Pb collisions at $\sqrt{s_{\rm{NN}}}$ = 5.02 TeV}},\ }\href
  {https://doi.org/10.1103/PhysRevC.101.034911} {\bibfield  {journal} {\bibinfo
   {journal} {Phys. Rev. C}\ }\textbf {\bibinfo {volume} {101}},\ \bibinfo
  {pages} {034911} (\bibinfo {year} {2020})},\ \Eprint
  {https://arxiv.org/abs/1909.09718} {arXiv:1909.09718 [nucl-ex]} \BibitemShut
  {NoStop}%
\bibitem [{\citenamefont {Sirunyan}\ \emph {et~al.}(2021)\citenamefont
  {Sirunyan} \emph {et~al.}}]{CMS:2021vui}%
  \BibitemOpen
  \bibfield  {author} {\bibinfo {author} {\bibfnamefont {A.~M.}\ \bibnamefont
  {Sirunyan}} \emph {et~al.} (\bibinfo {collaboration} {CMS}),\ }\bibfield
  {title} {\bibinfo {title} {{First measurement of large area jet transverse
  momentum spectra in heavy-ion collisions}},\ }\href
  {https://doi.org/10.1007/JHEP05(2021)284} {\bibfield  {journal} {\bibinfo
  {journal} {JHEP}\ }\textbf {\bibinfo {volume} {05}},\ \bibinfo {pages}
  {284}},\ \Eprint {https://arxiv.org/abs/2102.13080} {arXiv:2102.13080
  [hep-ex]} \BibitemShut {NoStop}%
\bibitem [{\citenamefont {Soltz}\ \emph {et~al.}(2025)\citenamefont {Soltz},
  \citenamefont {Hangal},\ and\ \citenamefont {Angerami}}]{Soltz:2024gkm}%
  \BibitemOpen
  \bibfield  {author} {\bibinfo {author} {\bibfnamefont {R.~A.}\ \bibnamefont
  {Soltz}}, \bibinfo {author} {\bibfnamefont {D.~A.}\ \bibnamefont {Hangal}},\
  and\ \bibinfo {author} {\bibfnamefont {A.}~\bibnamefont {Angerami}},\
  }\bibfield  {title} {\bibinfo {title} {{Simple model to investigate jet
  quenching and correlated errors for centrality-dependent nuclear modification
  factors in relativistic heavy-ion collisions}},\ }\href
  {https://doi.org/10.1103/PhysRevC.111.034911} {\bibfield  {journal} {\bibinfo
   {journal} {Phys. Rev. C}\ }\textbf {\bibinfo {volume} {111}},\ \bibinfo
  {pages} {034911} (\bibinfo {year} {2025})},\ \Eprint
  {https://arxiv.org/abs/2412.03724} {arXiv:2412.03724 [nucl-th]} \BibitemShut
  {NoStop}%
\bibitem [{\citenamefont {Adam}\ \emph {et~al.}(2020)\citenamefont {Adam} \emph
  {et~al.}}]{STAR:2020xiv}%
  \BibitemOpen
  \bibfield  {author} {\bibinfo {author} {\bibfnamefont {J.}~\bibnamefont
  {Adam}} \emph {et~al.} (\bibinfo {collaboration} {STAR}),\ }\bibfield
  {title} {\bibinfo {title} {{Measurement of inclusive charged-particle jet
  production in Au + Au collisions at $\sqrt{s_{NN}}=$200 GeV}},\ }\href
  {https://doi.org/10.1103/PhysRevC.102.054913} {\bibfield  {journal} {\bibinfo
   {journal} {Phys. Rev. C}\ }\textbf {\bibinfo {volume} {102}},\ \bibinfo
  {pages} {054913} (\bibinfo {year} {2020})},\ \Eprint
  {https://arxiv.org/abs/2006.00582} {arXiv:2006.00582 [nucl-ex]} \BibitemShut
  {NoStop}%
\bibitem [{\citenamefont {Khachatryan}\ \emph {et~al.}(2017)\citenamefont
  {Khachatryan} \emph {et~al.}}]{CMS:2016xef}%
  \BibitemOpen
  \bibfield  {author} {\bibinfo {author} {\bibfnamefont {V.}~\bibnamefont
  {Khachatryan}} \emph {et~al.} (\bibinfo {collaboration} {CMS}),\ }\bibfield
  {title} {\bibinfo {title} {{Charged-particle nuclear modification factors in
  PbPb and pPb collisions at $ \sqrt{s_{\mathrm{N}\;\mathrm{N}}}=5.02 $ TeV}},\
  }\href {https://doi.org/10.1007/JHEP04(2017)039} {\bibfield  {journal}
  {\bibinfo  {journal} {JHEP}\ }\textbf {\bibinfo {volume} {04}},\ \bibinfo
  {pages} {039}},\ \Eprint {https://arxiv.org/abs/1611.01664} {arXiv:1611.01664
  [nucl-ex]} \BibitemShut {NoStop}%
\bibitem [{\citenamefont {Acharya}\ \emph {et~al.}(2018)\citenamefont {Acharya}
  \emph {et~al.}}]{ALICE:2018vuu}%
  \BibitemOpen
  \bibfield  {author} {\bibinfo {author} {\bibfnamefont {S.}~\bibnamefont
  {Acharya}} \emph {et~al.} (\bibinfo {collaboration} {ALICE}),\ }\bibfield
  {title} {\bibinfo {title} {{Transverse momentum spectra and nuclear
  modification factors of charged particles in pp, p-Pb and Pb-Pb collisions at
  the LHC}},\ }\href {https://doi.org/10.1007/JHEP11(2018)013} {\bibfield
  {journal} {\bibinfo  {journal} {JHEP}\ }\textbf {\bibinfo {volume} {11}},\
  \bibinfo {pages} {013}},\ \Eprint {https://arxiv.org/abs/1802.09145}
  {arXiv:1802.09145 [nucl-ex]} \BibitemShut {NoStop}%
\bibitem [{\citenamefont {Aad}\ \emph {et~al.}(2023{\natexlab{b}})\citenamefont
  {Aad} \emph {et~al.}}]{ATLAS:2022kqu}%
  \BibitemOpen
  \bibfield  {author} {\bibinfo {author} {\bibfnamefont {G.}~\bibnamefont
  {Aad}} \emph {et~al.} (\bibinfo {collaboration} {ATLAS}),\ }\bibfield
  {title} {\bibinfo {title} {{Charged-hadron production in $pp$, $p$+Pb, Pb+Pb,
  and Xe+Xe collisions at $\sqrt{s_{_\text{NN}}}=5$ TeV with the ATLAS detector
  at the LHC}},\ }\href {https://doi.org/10.1007/JHEP07(2023)074} {\bibfield
  {journal} {\bibinfo  {journal} {JHEP}\ }\textbf {\bibinfo {volume} {07}},\
  \bibinfo {pages} {074}},\ \Eprint {https://arxiv.org/abs/2211.15257}
  {arXiv:2211.15257 [hep-ex]} \BibitemShut {NoStop}%
\bibitem [{\citenamefont {Brewer}\ \emph {et~al.}(2022)\citenamefont {Brewer},
  \citenamefont {Huss}, \citenamefont {Mazeliauskas},\ and\ \citenamefont
  {van~der Schee}}]{Brewer:2021tyv}%
  \BibitemOpen
  \bibfield  {author} {\bibinfo {author} {\bibfnamefont {J.}~\bibnamefont
  {Brewer}}, \bibinfo {author} {\bibfnamefont {A.}~\bibnamefont {Huss}},
  \bibinfo {author} {\bibfnamefont {A.}~\bibnamefont {Mazeliauskas}},\ and\
  \bibinfo {author} {\bibfnamefont {W.}~\bibnamefont {van~der Schee}},\
  }\bibfield  {title} {\bibinfo {title} {{Ratios of jet and hadron spectra at
  LHC energies: Measuring high-$p_T$ suppression without a pp reference}},\
  }\href {https://doi.org/10.1103/PhysRevD.105.074040} {\bibfield  {journal}
  {\bibinfo  {journal} {Phys. Rev. D}\ }\textbf {\bibinfo {volume} {105}},\
  \bibinfo {pages} {074040} (\bibinfo {year} {2022})},\ \Eprint
  {https://arxiv.org/abs/2108.13434} {arXiv:2108.13434 [hep-ph]} \BibitemShut
  {NoStop}%
\bibitem [{\citenamefont {Paakkinen}(2022)}]{Paakkinen:2021jjp}%
  \BibitemOpen
  \bibfield  {author} {\bibinfo {author} {\bibfnamefont {P.}~\bibnamefont
  {Paakkinen}},\ }\bibfield  {title} {\bibinfo {title} {{Light-nuclei gluons
  from dijet production in proton-oxygen collisions}},\ }\href
  {https://doi.org/10.1103/PhysRevD.105.L031504} {\bibfield  {journal}
  {\bibinfo  {journal} {Phys. Rev. D}\ }\textbf {\bibinfo {volume} {105}},\
  \bibinfo {pages} {L031504} (\bibinfo {year} {2022})},\ \Eprint
  {https://arxiv.org/abs/2111.05368} {arXiv:2111.05368 [hep-ph]} \BibitemShut
  {NoStop}%
\bibitem [{\citenamefont {Gebhard}\ \emph {et~al.}(2025)\citenamefont
  {Gebhard}, \citenamefont {Mazeliauskas},\ and\ \citenamefont
  {Takacs}}]{Gebhard:2024flv}%
  \BibitemOpen
  \bibfield  {author} {\bibinfo {author} {\bibfnamefont {J.}~\bibnamefont
  {Gebhard}}, \bibinfo {author} {\bibfnamefont {A.}~\bibnamefont
  {Mazeliauskas}},\ and\ \bibinfo {author} {\bibfnamefont {A.}~\bibnamefont
  {Takacs}},\ }\bibfield  {title} {\bibinfo {title} {{No-quenching baseline for
  energy loss signals in oxygen-oxygen collisions}},\ }\href
  {https://doi.org/10.1007/JHEP04(2025)034} {\bibfield  {journal} {\bibinfo
  {journal} {JHEP}\ }\textbf {\bibinfo {volume} {04}},\ \bibinfo {pages}
  {034}},\ \Eprint {https://arxiv.org/abs/2410.22405} {arXiv:2410.22405
  [hep-ph]} \BibitemShut {NoStop}%
\bibitem [{\citenamefont {Mazeliauskas}(2025)}]{Mazeliauskas:2025clt}%
  \BibitemOpen
  \bibfield  {author} {\bibinfo {author} {\bibfnamefont {A.}~\bibnamefont
  {Mazeliauskas}},\ }\bibfield  {title} {\bibinfo {title} {{Energy loss
  baseline for light hadrons in oxygen-oxygen collisions at
  $\sqrt{s_\mathrm{NN}}=5.36\,\text{TeV}$}},\ }\href@noop {} {\  (\bibinfo
  {year} {2025})},\ \Eprint {https://arxiv.org/abs/2509.07008}
  {arXiv:2509.07008 [hep-ph]} \BibitemShut {NoStop}%
\bibitem [{\citenamefont {Katz}\ \emph {et~al.}(2020)\citenamefont {Katz},
  \citenamefont {Prado}, \citenamefont {Noronha-Hostler},\ and\ \citenamefont
  {Suaide}}]{Katz:2019qwv}%
  \BibitemOpen
  \bibfield  {author} {\bibinfo {author} {\bibfnamefont {R.}~\bibnamefont
  {Katz}}, \bibinfo {author} {\bibfnamefont {C.~A.~G.}\ \bibnamefont {Prado}},
  \bibinfo {author} {\bibfnamefont {J.}~\bibnamefont {Noronha-Hostler}},\ and\
  \bibinfo {author} {\bibfnamefont {A.~A.~P.}\ \bibnamefont {Suaide}},\
  }\bibfield  {title} {\bibinfo {title} {{System-size scan of $D$ meson
  $R_{AA}$ and $v_n$ using PbPb, XeXe, ArAr, and OO collisions at energies
  available at the CERN Large Hadron Collider}},\ }\href
  {https://doi.org/10.1103/PhysRevC.102.041901} {\bibfield  {journal} {\bibinfo
   {journal} {Phys. Rev. C}\ }\textbf {\bibinfo {volume} {102}},\ \bibinfo
  {pages} {041901} (\bibinfo {year} {2020})},\ \Eprint
  {https://arxiv.org/abs/1907.03308} {arXiv:1907.03308 [nucl-th]} \BibitemShut
  {NoStop}%
\bibitem [{\citenamefont {Huss}\ \emph
  {et~al.}(2021{\natexlab{a}})\citenamefont {Huss}, \citenamefont {Kurkela},
  \citenamefont {Mazeliauskas}, \citenamefont {Paatelainen}, \citenamefont
  {van~der Schee},\ and\ \citenamefont {Wiedemann}}]{Huss:2020dwe}%
  \BibitemOpen
  \bibfield  {author} {\bibinfo {author} {\bibfnamefont {A.}~\bibnamefont
  {Huss}}, \bibinfo {author} {\bibfnamefont {A.}~\bibnamefont {Kurkela}},
  \bibinfo {author} {\bibfnamefont {A.}~\bibnamefont {Mazeliauskas}}, \bibinfo
  {author} {\bibfnamefont {R.}~\bibnamefont {Paatelainen}}, \bibinfo {author}
  {\bibfnamefont {W.}~\bibnamefont {van~der Schee}},\ and\ \bibinfo {author}
  {\bibfnamefont {U.~A.}\ \bibnamefont {Wiedemann}},\ }\bibfield  {title}
  {\bibinfo {title} {{Discovering Partonic Rescattering in Light Nucleus
  Collisions}},\ }\href {https://doi.org/10.1103/PhysRevLett.126.192301}
  {\bibfield  {journal} {\bibinfo  {journal} {Phys. Rev. Lett.}\ }\textbf
  {\bibinfo {volume} {126}},\ \bibinfo {pages} {192301} (\bibinfo {year}
  {2021}{\natexlab{a}})},\ \Eprint {https://arxiv.org/abs/2007.13754}
  {arXiv:2007.13754 [hep-ph]} \BibitemShut {NoStop}%
\bibitem [{\citenamefont {Huss}\ \emph
  {et~al.}(2021{\natexlab{b}})\citenamefont {Huss}, \citenamefont {Kurkela},
  \citenamefont {Mazeliauskas}, \citenamefont {Paatelainen}, \citenamefont
  {van~der Schee},\ and\ \citenamefont {Wiedemann}}]{Huss:2020whe}%
  \BibitemOpen
  \bibfield  {author} {\bibinfo {author} {\bibfnamefont {A.}~\bibnamefont
  {Huss}}, \bibinfo {author} {\bibfnamefont {A.}~\bibnamefont {Kurkela}},
  \bibinfo {author} {\bibfnamefont {A.}~\bibnamefont {Mazeliauskas}}, \bibinfo
  {author} {\bibfnamefont {R.}~\bibnamefont {Paatelainen}}, \bibinfo {author}
  {\bibfnamefont {W.}~\bibnamefont {van~der Schee}},\ and\ \bibinfo {author}
  {\bibfnamefont {U.~A.}\ \bibnamefont {Wiedemann}},\ }\bibfield  {title}
  {\bibinfo {title} {{Predicting parton energy loss in small collision
  systems}},\ }\href {https://doi.org/10.1103/PhysRevC.103.054903} {\bibfield
  {journal} {\bibinfo  {journal} {Phys. Rev. C}\ }\textbf {\bibinfo {volume}
  {103}},\ \bibinfo {pages} {054903} (\bibinfo {year} {2021}{\natexlab{b}})},\
  \Eprint {https://arxiv.org/abs/2007.13758} {arXiv:2007.13758 [hep-ph]}
  \BibitemShut {NoStop}%
\bibitem [{\citenamefont {Zakharov}(2021)}]{Zakharov:2021uza}%
  \BibitemOpen
  \bibfield  {author} {\bibinfo {author} {\bibfnamefont {B.~G.}\ \bibnamefont
  {Zakharov}},\ }\bibfield  {title} {\bibinfo {title} {{Jet quenching from
  heavy to light ion collisions}},\ }\href
  {https://doi.org/10.1007/JHEP09(2021)087} {\bibfield  {journal} {\bibinfo
  {journal} {JHEP}\ }\textbf {\bibinfo {volume} {09}},\ \bibinfo {pages}
  {087}},\ \Eprint {https://arxiv.org/abs/2105.09350} {arXiv:2105.09350
  [hep-ph]} \BibitemShut {NoStop}%
\bibitem [{\citenamefont {Xie}\ \emph {et~al.}(2024)\citenamefont {Xie},
  \citenamefont {Ke}, \citenamefont {Zhang},\ and\ \citenamefont
  {Wang}}]{Xie:2022fak}%
  \BibitemOpen
  \bibfield  {author} {\bibinfo {author} {\bibfnamefont {M.}~\bibnamefont
  {Xie}}, \bibinfo {author} {\bibfnamefont {W.}~\bibnamefont {Ke}}, \bibinfo
  {author} {\bibfnamefont {H.}~\bibnamefont {Zhang}},\ and\ \bibinfo {author}
  {\bibfnamefont {X.-N.}\ \bibnamefont {Wang}},\ }\bibfield  {title} {\bibinfo
  {title} {{Global constraint on the jet transport coefficient from
  single-hadron, dihadron, and \ensuremath{\gamma}-hadron spectra in
  high-energy heavy-ion collisions}},\ }\href
  {https://doi.org/10.1103/PhysRevC.109.064917} {\bibfield  {journal} {\bibinfo
   {journal} {Phys. Rev. C}\ }\textbf {\bibinfo {volume} {109}},\ \bibinfo
  {pages} {064917} (\bibinfo {year} {2024})},\ \Eprint
  {https://arxiv.org/abs/2208.14419} {arXiv:2208.14419 [hep-ph]} \BibitemShut
  {NoStop}%
\bibitem [{\citenamefont {Ke}\ and\ \citenamefont {Vitev}(2023)}]{Ke:2022gkq}%
  \BibitemOpen
  \bibfield  {author} {\bibinfo {author} {\bibfnamefont {W.}~\bibnamefont
  {Ke}}\ and\ \bibinfo {author} {\bibfnamefont {I.}~\bibnamefont {Vitev}},\
  }\bibfield  {title} {\bibinfo {title} {{Searching for QGP droplets with
  high-pT hadrons and heavy flavor}},\ }\href
  {https://doi.org/10.1103/PhysRevC.107.064903} {\bibfield  {journal} {\bibinfo
   {journal} {Phys. Rev. C}\ }\textbf {\bibinfo {volume} {107}},\ \bibinfo
  {pages} {064903} (\bibinfo {year} {2023})},\ \Eprint
  {https://arxiv.org/abs/2204.00634} {arXiv:2204.00634 [hep-ph]} \BibitemShut
  {NoStop}%
\bibitem [{\citenamefont {Vitev}\ and\ \citenamefont
  {Ke}(2024)}]{Vitev:2023nti}%
  \BibitemOpen
  \bibfield  {author} {\bibinfo {author} {\bibfnamefont {I.}~\bibnamefont
  {Vitev}}\ and\ \bibinfo {author} {\bibfnamefont {W.}~\bibnamefont {Ke}},\
  }\bibfield  {title} {\bibinfo {title} {{Initial-state and final-state effects
  on hadron production in small collision systems}},\ }\href
  {https://doi.org/10.1051/epjconf/202429615002} {\bibfield  {journal}
  {\bibinfo  {journal} {EPJ Web Conf.}\ }\textbf {\bibinfo {volume} {296}},\
  \bibinfo {pages} {15002} (\bibinfo {year} {2024})},\ \Eprint
  {https://arxiv.org/abs/2312.12580} {arXiv:2312.12580 [hep-ph]} \BibitemShut
  {NoStop}%
\bibitem [{\citenamefont {Ogrodnik}\ \emph {et~al.}(2025)\citenamefont
  {Ogrodnik}, \citenamefont {Ryb{\'a}{\v{r}}},\ and\ \citenamefont
  {Spousta}}]{Ogrodnik:2024qug}%
  \BibitemOpen
  \bibfield  {author} {\bibinfo {author} {\bibfnamefont {A.}~\bibnamefont
  {Ogrodnik}}, \bibinfo {author} {\bibfnamefont {M.}~\bibnamefont
  {Ryb{\'a}{\v{r}}}},\ and\ \bibinfo {author} {\bibfnamefont {M.}~\bibnamefont
  {Spousta}},\ }\bibfield  {title} {\bibinfo {title} {{Flavor and path-length
  dependence of jet quenching from inclusive jet and $\gamma $-jet
  suppression}},\ }\href {https://doi.org/10.1140/epjc/s10052-025-14629-5}
  {\bibfield  {journal} {\bibinfo  {journal} {Eur. Phys. J. C}\ }\textbf
  {\bibinfo {volume} {85}},\ \bibinfo {pages} {899} (\bibinfo {year} {2025})},\
  \Eprint {https://arxiv.org/abs/2407.11234} {arXiv:2407.11234 [hep-ph]}
  \BibitemShut {NoStop}%
\bibitem [{\citenamefont {van~der Schee}\ \emph {et~al.}(2025)\citenamefont
  {van~der Schee}, \citenamefont {Kolb{\'e}}, \citenamefont {Nijs},
  \citenamefont {Ruhani}, \citenamefont {Ahmed},\ and\ \citenamefont
  {Iqbal}}]{vanderSchee:2025hoe}%
  \BibitemOpen
  \bibfield  {author} {\bibinfo {author} {\bibfnamefont {W.}~\bibnamefont
  {van~der Schee}}, \bibinfo {author} {\bibfnamefont {I.}~\bibnamefont
  {Kolb{\'e}}}, \bibinfo {author} {\bibfnamefont {G.}~\bibnamefont {Nijs}},
  \bibinfo {author} {\bibfnamefont {K.}~\bibnamefont {Ruhani}}, \bibinfo
  {author} {\bibfnamefont {I.}~\bibnamefont {Ahmed}},\ and\ \bibinfo {author}
  {\bibfnamefont {S.}~\bibnamefont {Iqbal}},\ }\bibfield  {title} {\bibinfo
  {title} {{Three models for charged hadron nuclear modification from light to
  heavy ions}},\ }\href@noop {} {\  (\bibinfo {year} {2025})},\ \Eprint
  {https://arxiv.org/abs/2509.04299} {arXiv:2509.04299 [nucl-th]} \BibitemShut
  {NoStop}%
\bibitem [{\citenamefont {Rybczy{\'n}ski}\ and\ \citenamefont
  {Broniowski}(2019)}]{Rybczynski:2019adt}%
  \BibitemOpen
  \bibfield  {author} {\bibinfo {author} {\bibfnamefont {M.}~\bibnamefont
  {Rybczy{\'n}ski}}\ and\ \bibinfo {author} {\bibfnamefont {W.}~\bibnamefont
  {Broniowski}},\ }\bibfield  {title} {\bibinfo {title} {{Glauber Monte Carlo
  predictions for ultrarelativistic collisions with $^{16}$O}},\ }\href
  {https://doi.org/10.1103/PhysRevC.100.064912} {\bibfield  {journal} {\bibinfo
   {journal} {Phys. Rev. C}\ }\textbf {\bibinfo {volume} {100}},\ \bibinfo
  {pages} {064912} (\bibinfo {year} {2019})},\ \Eprint
  {https://arxiv.org/abs/1910.09489} {arXiv:1910.09489 [hep-ph]} \BibitemShut
  {NoStop}%
\bibitem [{\citenamefont {Casalderrey-Solana}\ \emph
  {et~al.}(2019)\citenamefont {Casalderrey-Solana}, \citenamefont {Hulcher},
  \citenamefont {Milhano}, \citenamefont {Pablos},\ and\ \citenamefont
  {Rajagopal}}]{Casalderrey-Solana:2018wrw}%
  \BibitemOpen
  \bibfield  {author} {\bibinfo {author} {\bibfnamefont {J.}~\bibnamefont
  {Casalderrey-Solana}}, \bibinfo {author} {\bibfnamefont {Z.}~\bibnamefont
  {Hulcher}}, \bibinfo {author} {\bibfnamefont {G.}~\bibnamefont {Milhano}},
  \bibinfo {author} {\bibfnamefont {D.}~\bibnamefont {Pablos}},\ and\ \bibinfo
  {author} {\bibfnamefont {K.}~\bibnamefont {Rajagopal}},\ }\bibfield  {title}
  {\bibinfo {title} {{Simultaneous description of hadron and jet suppression in
  heavy-ion collisions}},\ }\href {https://doi.org/10.1103/PhysRevC.99.051901}
  {\bibfield  {journal} {\bibinfo  {journal} {Phys. Rev. C}\ }\textbf {\bibinfo
  {volume} {99}},\ \bibinfo {pages} {051901} (\bibinfo {year} {2019})},\
  \Eprint {https://arxiv.org/abs/1808.07386} {arXiv:1808.07386 [hep-ph]}
  \BibitemShut {NoStop}%
\bibitem [{\citenamefont {Hayrapetyan}\ \emph {et~al.}(2025)\citenamefont
  {Hayrapetyan} \emph {et~al.}}]{CMS:2025bta}%
  \BibitemOpen
  \bibfield  {author} {\bibinfo {author} {\bibfnamefont {A.}~\bibnamefont
  {Hayrapetyan}} \emph {et~al.} (\bibinfo {collaboration} {CMS}),\ }\bibfield
  {title} {\bibinfo {title} {{Discovery of suppressed charged-particle
  production in ultrarelativistic oxygen-oxygen collisions}},\ }\href@noop {}
  {\  (\bibinfo {year} {2025})},\ \Eprint {https://arxiv.org/abs/2510.09864}
  {arXiv:2510.09864 [nucl-ex]} \BibitemShut {NoStop}%
\bibitem [{\citenamefont {Takacs}\ and\ \citenamefont
  {Pablos}(2026)}]{Takacs_Bayesian_constraints_on_2026}%
  \BibitemOpen
  \bibfield  {author} {\bibinfo {author} {\bibfnamefont {A.}~\bibnamefont
  {Takacs}}\ and\ \bibinfo {author} {\bibfnamefont {D.}~\bibnamefont
  {Pablos}},\ }\bibfield  {title} {\bibinfo {title} {{Bayesian constraints on
  pre-equilibrium jet quenching and predictions for oxygen collisions}},\
  }\href {https://github.com/adam-takacs/STAT}
  {https://github.com/adam-takacs/STAT} (\bibinfo {year} {2026})\BibitemShut
  {NoStop}%
\bibitem [{\citenamefont {Sadofyev}\ \emph {et~al.}(2021)\citenamefont
  {Sadofyev}, \citenamefont {Sievert},\ and\ \citenamefont
  {Vitev}}]{Sadofyev:2021ohn}%
  \BibitemOpen
  \bibfield  {author} {\bibinfo {author} {\bibfnamefont {A.~V.}\ \bibnamefont
  {Sadofyev}}, \bibinfo {author} {\bibfnamefont {M.~D.}\ \bibnamefont
  {Sievert}},\ and\ \bibinfo {author} {\bibfnamefont {I.}~\bibnamefont
  {Vitev}},\ }\bibfield  {title} {\bibinfo {title} {{Ab~initio coupling of jets
  to collective flow in the opacity expansion approach}},\ }\href
  {https://doi.org/10.1103/PhysRevD.104.094044} {\bibfield  {journal} {\bibinfo
   {journal} {Phys. Rev. D}\ }\textbf {\bibinfo {volume} {104}},\ \bibinfo
  {pages} {094044} (\bibinfo {year} {2021})},\ \Eprint
  {https://arxiv.org/abs/2104.09513} {arXiv:2104.09513 [hep-ph]} \BibitemShut
  {NoStop}%
\bibitem [{\citenamefont {Fu}\ \emph {et~al.}(2023)\citenamefont {Fu},
  \citenamefont {Casalderrey-Solana},\ and\ \citenamefont {Wang}}]{Fu:2022idl}%
  \BibitemOpen
  \bibfield  {author} {\bibinfo {author} {\bibfnamefont {Y.}~\bibnamefont
  {Fu}}, \bibinfo {author} {\bibfnamefont {J.}~\bibnamefont
  {Casalderrey-Solana}},\ and\ \bibinfo {author} {\bibfnamefont {X.-N.}\
  \bibnamefont {Wang}},\ }\bibfield  {title} {\bibinfo {title} {{Asymmetric
  transverse momentum broadening in an inhomogeneous medium}},\ }\href
  {https://doi.org/10.1103/PhysRevD.107.054038} {\bibfield  {journal} {\bibinfo
   {journal} {Phys. Rev. D}\ }\textbf {\bibinfo {volume} {107}},\ \bibinfo
  {pages} {054038} (\bibinfo {year} {2023})},\ \Eprint
  {https://arxiv.org/abs/2204.05323} {arXiv:2204.05323 [hep-ph]} \BibitemShut
  {NoStop}%
\bibitem [{\citenamefont {Barata}\ \emph {et~al.}(2022)\citenamefont {Barata},
  \citenamefont {Sadofyev},\ and\ \citenamefont {Salgado}}]{Barata:2022krd}%
  \BibitemOpen
  \bibfield  {author} {\bibinfo {author} {\bibfnamefont {J.~a.}\ \bibnamefont
  {Barata}}, \bibinfo {author} {\bibfnamefont {A.~V.}\ \bibnamefont
  {Sadofyev}},\ and\ \bibinfo {author} {\bibfnamefont {C.~A.}\ \bibnamefont
  {Salgado}},\ }\bibfield  {title} {\bibinfo {title} {{Jet broadening in dense
  inhomogeneous matter}},\ }\href {https://doi.org/10.1103/PhysRevD.105.114010}
  {\bibfield  {journal} {\bibinfo  {journal} {Phys. Rev. D}\ }\textbf {\bibinfo
  {volume} {105}},\ \bibinfo {pages} {114010} (\bibinfo {year} {2022})},\
  \Eprint {https://arxiv.org/abs/2202.08847} {arXiv:2202.08847 [hep-ph]}
  \BibitemShut {NoStop}%
\bibitem [{\citenamefont {Andres}\ \emph {et~al.}(2022)\citenamefont {Andres},
  \citenamefont {Dominguez}, \citenamefont {Sadofyev},\ and\ \citenamefont
  {Salgado}}]{Andres:2022ndd}%
  \BibitemOpen
  \bibfield  {author} {\bibinfo {author} {\bibfnamefont {C.}~\bibnamefont
  {Andres}}, \bibinfo {author} {\bibfnamefont {F.}~\bibnamefont {Dominguez}},
  \bibinfo {author} {\bibfnamefont {A.~V.}\ \bibnamefont {Sadofyev}},\ and\
  \bibinfo {author} {\bibfnamefont {C.~A.}\ \bibnamefont {Salgado}},\
  }\bibfield  {title} {\bibinfo {title} {{Jet broadening in flowing matter:
  Resummation}},\ }\href {https://doi.org/10.1103/PhysRevD.106.074023}
  {\bibfield  {journal} {\bibinfo  {journal} {Phys. Rev. D}\ }\textbf {\bibinfo
  {volume} {106}},\ \bibinfo {pages} {074023} (\bibinfo {year} {2022})},\
  \Eprint {https://arxiv.org/abs/2207.07141} {arXiv:2207.07141 [hep-ph]}
  \BibitemShut {NoStop}%
\bibitem [{\citenamefont {Hauksson}\ and\ \citenamefont
  {Iancu}(2023)}]{Hauksson:2023tze}%
  \BibitemOpen
  \bibfield  {author} {\bibinfo {author} {\bibfnamefont {S.}~\bibnamefont
  {Hauksson}}\ and\ \bibinfo {author} {\bibfnamefont {E.}~\bibnamefont
  {Iancu}},\ }\bibfield  {title} {\bibinfo {title} {{Jet polarisation in an
  anisotropic medium}},\ }\href {https://doi.org/10.1007/JHEP08(2023)027}
  {\bibfield  {journal} {\bibinfo  {journal} {JHEP}\ }\textbf {\bibinfo
  {volume} {08}},\ \bibinfo {pages} {027}},\ \Eprint
  {https://arxiv.org/abs/2303.03914} {arXiv:2303.03914 [hep-ph]} \BibitemShut
  {NoStop}%
\bibitem [{\citenamefont {Kuzmin}\ \emph {et~al.}(2024)\citenamefont {Kuzmin},
  \citenamefont {Mayo~L{\'o}pez}, \citenamefont {Reiten},\ and\ \citenamefont
  {Sadofyev}}]{Kuzmin:2023hko}%
  \BibitemOpen
  \bibfield  {author} {\bibinfo {author} {\bibfnamefont {M.~V.}\ \bibnamefont
  {Kuzmin}}, \bibinfo {author} {\bibfnamefont {X.}~\bibnamefont
  {Mayo~L{\'o}pez}}, \bibinfo {author} {\bibfnamefont {J.}~\bibnamefont
  {Reiten}},\ and\ \bibinfo {author} {\bibfnamefont {A.~V.}\ \bibnamefont
  {Sadofyev}},\ }\bibfield  {title} {\bibinfo {title} {{Jet quenching in
  anisotropic flowing matter}},\ }\href
  {https://doi.org/10.1103/PhysRevD.109.014036} {\bibfield  {journal} {\bibinfo
   {journal} {Phys. Rev. D}\ }\textbf {\bibinfo {volume} {109}},\ \bibinfo
  {pages} {014036} (\bibinfo {year} {2024})},\ \Eprint
  {https://arxiv.org/abs/2309.00683} {arXiv:2309.00683 [hep-ph]} \BibitemShut
  {NoStop}%
\bibitem [{\citenamefont {Adhya}\ and\ \citenamefont
  {Tywoniuk}(2024)}]{Adhya:2024nwx}%
  \BibitemOpen
  \bibfield  {author} {\bibinfo {author} {\bibfnamefont {S.~P.}\ \bibnamefont
  {Adhya}}\ and\ \bibinfo {author} {\bibfnamefont {K.}~\bibnamefont
  {Tywoniuk}},\ }\bibfield  {title} {\bibinfo {title} {{Sensitivity of jet
  quenching to the initial state in heavy-ion collisions}},\ }\href@noop {} {\
  (\bibinfo {year} {2024})},\ \Eprint {https://arxiv.org/abs/2409.04295}
  {arXiv:2409.04295 [hep-ph]} \BibitemShut {NoStop}%
\bibitem [{\citenamefont {Barata}\ \emph {et~al.}(2024)\citenamefont {Barata},
  \citenamefont {Hauksson}, \citenamefont {Mayo~L{\'o}pez},\ and\ \citenamefont
  {Sadofyev}}]{Barata:2024xwy}%
  \BibitemOpen
  \bibfield  {author} {\bibinfo {author} {\bibfnamefont {J.}~\bibnamefont
  {Barata}}, \bibinfo {author} {\bibfnamefont {S.}~\bibnamefont {Hauksson}},
  \bibinfo {author} {\bibfnamefont {X.}~\bibnamefont {Mayo~L{\'o}pez}},\ and\
  \bibinfo {author} {\bibfnamefont {A.~V.}\ \bibnamefont {Sadofyev}},\
  }\bibfield  {title} {\bibinfo {title} {{Jet quenching in the glasma phase:
  Medium-induced radiation}},\ }\href
  {https://doi.org/10.1103/PhysRevD.110.094055} {\bibfield  {journal} {\bibinfo
   {journal} {Phys. Rev. D}\ }\textbf {\bibinfo {volume} {110}},\ \bibinfo
  {pages} {094055} (\bibinfo {year} {2024})},\ \Eprint
  {https://arxiv.org/abs/2406.07615} {arXiv:2406.07615 [hep-ph]} \BibitemShut
  {NoStop}%
\bibitem [{\citenamefont {Altenburger}\ \emph {et~al.}(2025)\citenamefont
  {Altenburger}, \citenamefont {Boguslavski},\ and\ \citenamefont
  {Lindenbauer}}]{Altenburger:2025iqa}%
  \BibitemOpen
  \bibfield  {author} {\bibinfo {author} {\bibfnamefont {A.}~\bibnamefont
  {Altenburger}}, \bibinfo {author} {\bibfnamefont {K.}~\bibnamefont
  {Boguslavski}},\ and\ \bibinfo {author} {\bibfnamefont {F.}~\bibnamefont
  {Lindenbauer}},\ }\bibfield  {title} {\bibinfo {title} {{Jet broadening and
  radiation in the early anisotropic plasma in heavy-ion collisions}},\
  }\href@noop {} {\  (\bibinfo {year} {2025})},\ \Eprint
  {https://arxiv.org/abs/2509.03868} {arXiv:2509.03868 [hep-ph]} \BibitemShut
  {NoStop}%
\bibitem [{\citenamefont {Lindenbauer}(2025)}]{Lindenbauer:2025ctw}%
  \BibitemOpen
  \bibfield  {author} {\bibinfo {author} {\bibfnamefont {F.}~\bibnamefont
  {Lindenbauer}},\ }\bibfield  {title} {\bibinfo {title} {{Gluon splitting
  rates in an anisotropic plasma in the AMY formalism}},\ }\href@noop {} {\
  (\bibinfo {year} {2025})},\ \Eprint {https://arxiv.org/abs/2509.09897}
  {arXiv:2509.09897 [hep-ph]} \BibitemShut {NoStop}%
\end{thebibliography}%

\end{document}